\documentclass[12pt,letterpaper,twoside,openany]{book}
\usepackage{amsfonts,amssymb,amsmath,amscd,theorem,oldgerm,epsfig,units}
\usepackage[mathscr]{eucal}
\usepackage{mathrsfs}
\makeatletter

\def\diagram{\m@th\leftwidth=\z@ \rightwidth=\z@ \topheight=\z@
\botheight=\z@ \setbox\@picbox\hbox\bgroup}

\def\enddiagram{\egroup\wd\@picbox\rightwidth\unitlength
\ht\@picbox\topheight\unitlength \dp\@picbox\botheight\unitlength
\hskip\leftwidth\unitlength\box\@picbox}

\def\bfig{\begin{diagram}}
\def\efig{\end{diagram}}
\newcount\wideness \newcount\leftwidth \newcount\rightwidth
\newcount\highness \newcount\topheight \newcount\botheight

\def\ratchet#1#2{\ifnum#1<#2 \global #1=#2 \fi}

\def\putbox(#1,#2)#3{%
\horsize{\wideness}{#3} \divide\wideness by 2
{\advance\wideness by #1 \ratchet{\rightwidth}{\wideness}}
{\advance\wideness by -#1 \ratchet{\leftwidth}{\wideness}}
\vertsize{\highness}{#3} \divide\highness by 2
{\advance\highness by #2 \ratchet{\topheight}{\highness}}
{\advance\highness by -#2 \ratchet{\botheight}{\highness}}
\put(#1,#2){\makebox(0,0){$#3$}}}

\def\putlbox(#1,#2)#3{%
\horsize{\wideness}{#3}
{\advance\wideness by #1 \ratchet{\rightwidth}{\wideness}}
{\ratchet{\leftwidth}{-#1}}
\vertsize{\highness}{#3} \divide\highness by 2
{\advance\highness by #2 \ratchet{\topheight}{\highness}}
{\advance\highness by -#2 \ratchet{\botheight}{\highness}}
\put(#1,#2){\makebox(0,0)[l]{$#3$}}}

\def\putrbox(#1,#2)#3{%
\horsize{\wideness}{#3}
{\ratchet{\rightwidth}{#1}}
{\advance\wideness by -#1 \ratchet{\leftwidth}{\wideness}}
\vertsize{\highness}{#3} \divide\highness by 2
{\advance\highness by #2 \ratchet{\topheight}{\highness}}
{\advance\highness by -#2 \ratchet{\botheight}{\highness}}
\put(#1,#2){\makebox(0,0)[r]{$#3$}}}

\def\adjust[#1]{} 

\newcount \coefa
\newcount \coefb
\newcount \coefc
\newcount\tempcounta
\newcount\tempcountb
\newcount\tempcountc
\newcount\tempcountd
\newcount\xext
\newcount\yext
\newcount\xoff
\newcount\yoff
\newcount\gap%
\newcount\arrowtypea
\newcount\arrowtypeb
\newcount\arrowtypec
\newcount\arrowtyped
\newcount\arrowtypee
\newcount\height
\newcount\width
\newcount\xpos
\newcount\ypos
\newcount\run
\newcount\rise
\newcount\arrowlength
\newcount\halflength
\newcount\arrowtype
\newdimen\tempdimen
\newdimen\xlen
\newdimen\ylen
\newsavebox{\tempboxa}%
\newsavebox{\tempboxb}%
\newsavebox{\tempboxc}%

\newdimen\w@dth

\def\setw@dth#1#2{\setbox\z@\hbox{\m@th$#1$}\w@dth=\wd\z@
\setbox\@ne\hbox{\m@th$#2$}\ifnum\w@dth<\wd\@ne \w@dth=\wd\@ne \fi
\advance\w@dth by 1.2em}

\def\t@^#1_#2{\allowbreak\def\n@one{#1}\def\n@two{#2}\mathrel
{\setw@dth{#1}{#2}
\mathop{\hbox to \w@dth{\rightarrowfill}}\limits
\ifx\n@one\empty\else ^{\box\z@}\fi
\ifx\n@two\empty\else _{\box\@ne}\fi}}
\def\t@@^#1{\@ifnextchar_{\t@^{#1}}{\t@^{#1}_{}}}
\def\to{\@ifnextchar^{\t@@}{\t@@^{}}}

\def\t@left^#1_#2{\def\n@one{#1}\def\n@two{#2}\mathrel{\setw@dth{#1}{#2}
\mathop{\hbox to \w@dth{\leftarrowfill}}\limits
\ifx\n@one\empty\else ^{\box\z@}\fi
\ifx\n@two\empty\else _{\box\@ne}\fi}}
\def\t@@left^#1{\@ifnextchar_{\t@left^{#1}}{\t@left^{#1}_{}}}
\def\toleft{\@ifnextchar^{\t@@left}{\t@@left^{}}}

\def\two@^#1_#2{\allowbreak
\def\n@one{#1}\def\n@two{#2}\mathrel{\setw@dth{#1}{#2}
\mathop{\vcenter{\lineskip\z@\baselineskip\z@
                 \hbox to \w@dth{\rightarrowfill}%
                 \hbox to \w@dth{\rightarrowfill}}%
       }\limits
\ifx\n@one\empty\else ^{\box\z@}\fi
\ifx\n@two\empty\else _{\box\@ne}\fi}}
\def\tw@@^#1{\@ifnextchar _{\two@^{#1}}{\two@^{#1}_{}}}
\def\two{\@ifnextchar ^{\tw@@}{\tw@@^{}}}

\def\tofr@^#1_#2{\def\n@one{#1}\def\n@two{#2}\mathrel{\setw@dth{#1}{#2}
\mathop{\vcenter{\hbox to \w@dth{\rightarrowfill}\kern-1.7ex
                 \hbox to \w@dth{\leftarrowfill}}%
       }\limits
\ifx\n@one\empty\else ^{\box\z@}\fi
\ifx\n@two\empty\else _{\box\@ne}\fi}}
\def\t@fr@^#1{\@ifnextchar_ {\tofr@^{#1}}{\tofr@^{#1}_{}}}
\def\tofro{\@ifnextchar^ {\t@fr@}{\t@fr@^{}}}

\def\mon{\mathop{\m@th\hbox to
      14.6\P@{\lasyb\char'51\hskip-2.1\P@$\arrext$\hss
$\mathord\rightarrow$}}\limits} 
\def\leftmono{\mathrel{\m@th\hbox to
14.6\P@{$\mathord\leftarrow$\hss$\arrext$\hskip-2.1\P@\lasyb\char'50%
}}\limits} 
\mathchardef\arrext="0200       

\def\settypes(#1,#2,#3){\arrowtypea#1 \arrowtypeb#2 \arrowtypec#3}
\def\settoheight#1#2{\setbox\@tempboxa\hbox{#2}#1\ht\@tempboxa\relax}%
\def\settodepth#1#2{\setbox\@tempboxa\hbox{#2}#1\dp\@tempboxa\relax}%
\def\settokens`#1`#2`#3`#4`{%
     \def\tokena{#1}\def\tokenb{#2}\def\tokenc{#3}\def\tokend{#4}}
\def\setsqparms[#1`#2`#3`#4;#5`#6]{%
\arrowtypea #1
\arrowtypeb #2
\arrowtypec #3
\arrowtyped #4
\width #5
\height #6
}
\def\setpos(#1,#2){\xpos=#1 \ypos#2}

\def\settriparms[#1`#2`#3;#4]{\settripairparms[#1`#2`#3`1`1;#4]}%

\def\settripairparms[#1`#2`#3`#4`#5;#6]{%
\arrowtypea #1
\arrowtypeb #2
\arrowtypec #3
\arrowtyped #4
\arrowtypee #5
\width #6
\height #6
}

\def\resetparms{\settripairparms[1`1`1`1`1;500]\width 500}

\resetparms

\def\mvector(#1,#2)#3{
\put(0,0){\vector(#1,#2){#3}}%
\put(0,0){\vector(#1,#2){26}}%
}
\def\evector(#1,#2)#3{{
\arrowlength #3
\put(0,0){\vector(#1,#2){\arrowlength}}%
\advance \arrowlength by-30
\put(0,0){\vector(#1,#2){\arrowlength}}%
}}

\def\horsize#1#2{%
\settowidth{\tempdimen}{$#2$}%
#1=\tempdimen
\divide #1 by\unitlength
}

\def\vertsize#1#2{%
\settoheight{\tempdimen}{$#2$}%
#1=\tempdimen
\settodepth{\tempdimen}{$#2$}%
\advance #1 by\tempdimen
\divide #1 by\unitlength
}

\def\putvector(#1,#2)(#3,#4)#5#6{{%
\ifnum3<\arrowtype
\putdashvector(#1,#2)(#3,#4)#5\arrowtype
\else
\ifnum\arrowtype<-3
\putdashvector(#1,#2)(#3,#4)#5\arrowtype
\else
\xpos=#1
\ypos=#2
\run=#3
\rise=#4
\arrowlength=#5
\ifnum \arrowtype<0
    \ifnum \run=0
        \advance \ypos by-\arrowlength
    \else
        \tempcounta \arrowlength
        \multiply \tempcounta by\rise
        \divide \tempcounta by\run
        \ifnum\run>0
            \advance \xpos by\arrowlength
            \advance \ypos by\tempcounta
        \else
            \advance \xpos by-\arrowlength
            \advance \ypos by-\tempcounta
        \fi
    \fi
    \multiply \arrowtype by-1
    \multiply \rise by-1
    \multiply \run by-1
\fi
\ifcase \arrowtype
\or \put(\xpos,\ypos){\vector(\run,\rise){\arrowlength}}%
\or \put(\xpos,\ypos){\mvector(\run,\rise)\arrowlength}%
\or \put(\xpos,\ypos){\evector(\run,\rise){\arrowlength}}%
\fi\fi\fi
}}

\def\putsplitvector(#1,#2)#3#4{
\xpos #1
\ypos #2
\arrowtype #4
\halflength #3
\arrowlength #3
\gap 140
\advance \halflength by-\gap
\divide \halflength by2
\ifnum\arrowtype>0
   \ifcase \arrowtype
   \or \put(\xpos,\ypos){\line(0,-1){\halflength}}%
       \advance\ypos by-\halflength
       \advance\ypos by-\gap
       \put(\xpos,\ypos){\vector(0,-1){\halflength}}%
   \or \put(\xpos,\ypos){\line(0,-1)\halflength}%
       \put(\xpos,\ypos){\vector(0,-1)3}%
       \advance\ypos by-\halflength
       \advance\ypos by-\gap
       \put(\xpos,\ypos){\vector(0,-1){\halflength}}%
   \or \put(\xpos,\ypos){\line(0,-1)\halflength}%
       \advance\ypos by-\halflength
       \advance\ypos by-\gap
       \put(\xpos,\ypos){\evector(0,-1){\halflength}}%
   \fi
\else \arrowtype=-\arrowtype
   \ifcase\arrowtype
   \or \advance \ypos by-\arrowlength
       \put(\xpos,\ypos){\line(0,1){\halflength}}%
       \advance\ypos by\halflength
       \advance\ypos by\gap
       \put(\xpos,\ypos){\vector(0,1){\halflength}}%
   \or \advance \ypos by-\arrowlength
       \put(\xpos,\ypos){\line(0,1)\halflength}%
       \put(\xpos,\ypos){\vector(0,1)3}%
       \advance\ypos by\halflength
       \advance\ypos by\gap
       \put(\xpos,\ypos){\vector(0,1){\halflength}}%
   \or \advance \ypos by-\arrowlength
       \put(\xpos,\ypos){\line(0,1)\halflength}%
       \advance\ypos by\halflength
       \advance\ypos by\gap
       \put(\xpos,\ypos){\evector(0,1){\halflength}}%
   \fi
\fi
}

\def\putmorphism(#1)(#2,#3)[#4`#5`#6]#7#8#9{{%
\run #2
\rise #3
\ifnum\rise=0
  \puthmorphism(#1)[#4`#5`#6]{#7}{#8}#9%
\else\ifnum\run=0
  \putvmorphism(#1)[#4`#5`#6]{#7}{#8}#9%
\else
\setpos(#1)%
\arrowlength #7
\arrowtype #8
\ifnum\run=0
\else\ifnum\rise=0
\else
\ifnum\run>0
    \coefa=1
\else
   \coefa=-1
\fi
\ifnum\arrowtype>0
   \coefb=0
   \coefc=-1
\else
   \coefb=\coefa
   \coefc=1
   \arrowtype=-\arrowtype
\fi
\width=2
\multiply \width by\run
\divide \width by\rise
\ifnum \width<0  \width=-\width\fi
\advance\width by60
\if l#9 \width=-\width\fi
\putbox(\xpos,\ypos){#4}
{\multiply \coefa by\arrowlength
\advance\xpos by\coefa
\multiply \coefa by\rise
\divide \coefa by\run
\advance \ypos by\coefa
\putbox(\xpos,\ypos){#5} }%
{\multiply \coefa by\arrowlength
\divide \coefa by2
\advance \xpos by\coefa
\advance \xpos by\width
\multiply \coefa by\rise
\divide \coefa by\run
\advance \ypos by\coefa
\if l#9%
   \putrbox(\xpos,\ypos){#6}%
\else\if r#9%
   \putlbox(\xpos,\ypos){#6}%
\fi\fi }%
{\multiply \rise by-\coefc
\multiply \run by-\coefc
\multiply \coefb by\arrowlength
\advance \xpos by\coefb
\multiply \coefb by\rise
\divide \coefb by\run
\advance \ypos by\coefb
\multiply \coefc by70
\advance \ypos by\coefc
\multiply \coefc by\run
\divide \coefc by\rise
\advance \xpos by\coefc
\multiply \coefa by140
\multiply \coefa by\run
\divide \coefa by\rise
\advance \arrowlength by\coefa
\ifcase\arrowtype
\or \put(\xpos,\ypos){\vector(\run,\rise){\arrowlength}}%
\or \put(\xpos,\ypos){\mvector(\run,\rise){\arrowlength}}%
\or \put(\xpos,\ypos){\evector(\run,\rise){\arrowlength}}%
\fi}\fi\fi\fi\fi}}

\newcount\numbdashes \newcount\lengthdash \newcount\increment

\def\howmanydashes{
\numbdashes=\arrowlength \lengthdash=40
\divide\numbdashes by \lengthdash
\lengthdash=\arrowlength
\divide\lengthdash by \numbdashes
\increment=\lengthdash
\multiply\lengthdash by 3
\divide\lengthdash by 5
}

\def\putdashvector(#1)(#2,#3)#4#5{%
\ifnum#3=0 \putdashhvector(#1){#4}#5
\else
\ifnum#2=0
\putdashvvector(#1){#4}#5\fi\fi}

\def\putdashhvector(#1,#2)#3#4{{%
\arrowlength=#3 \howmanydashes
\multiput(#1,#2)(\increment,0){\numbdashes}%
{\vrule height .4pt width \lengthdash\unitlength}
\arrowtype=#4 \xpos=#1
\ifnum\arrowtype<0 \advance\arrowtype by 7 \fi
\ifcase\arrowtype
\or \advance\xpos by 10
    \put(\xpos,#2){\vector(-1,0){\lengthdash}}
    \advance\xpos by 40
    \put(\xpos,#2){\vector(-1,0){\lengthdash}}
\or \advance \xpos by 10
    \put(\xpos,#2){\vector(-1,0){\lengthdash}}
    \advance\xpos by  \arrowlength
    \advance\xpos by  -50
    \put(\xpos,#2){\vector(-1,0){\lengthdash}}
\or \advance\xpos by 10
    \put(\xpos,#2){\vector(-1,0){\lengthdash}}
\or \advance\xpos by \arrowlength
    \advance\xpos by -\lengthdash
    \put(\xpos,#2){\vector(1,0){\lengthdash}}
\or {\advance\xpos by 10
    \put(\xpos,#2){\vector(1,0){\lengthdash}}}
    \advance\xpos by \arrowlength
    \advance\xpos by -\lengthdash
    \put(\xpos,#2){\vector(1,0){\lengthdash}}
\or \advance\xpos by \arrowlength
    \advance\xpos by -\lengthdash
    \put(\xpos,#2){\vector(1,0){\lengthdash}}
    \advance\xpos by -40
    \put(\xpos,#2){\vector(1,0){\lengthdash}}
   \fi
}}

\def\putdashvvector(#1,#2)#3#4{{%
\arrowlength=#3 \howmanydashes
\ypos=#2 \advance\ypos by -\arrowlength
\multiput(#1,#2)(0,\increment){\numbdashes}%
    {\vrule width .4pt height \lengthdash\unitlength}
\arrowtype=#4 \ypos=#2
\ifnum\arrowtype<0 \advance\arrowtype by 7 \fi
\ifcase\arrowtype
\or \advance\ypos by \arrowlength \advance\ypos by -40
    \put(#1,\ypos){\vector(0,1){\lengthdash}}
    \advance\ypos by -40
    \put(#1,\ypos){\vector(0,1){\lengthdash}}
\or \advance\ypos by 10
    \put(#1,\ypos){\vector(0,1){\lengthdash}}
    \advance\ypos by \arrowlength \advance\ypos by -40
    \put(#1,\ypos){\vector(0,1){\lengthdash}}
\or \advance\ypos by \arrowlength \advance\ypos by -40
    \put(#1,\ypos){\vector(0,1){\lengthdash}}
\or \advance\ypos by 10
    \put(#1,\ypos){\vector(0,-1){\lengthdash}}
\or \advance\ypos by 10
    \put(#1,\ypos){\vector(0,-1){\lengthdash}}
    \advance\ypos by \arrowlength \advance\ypos by -40
    \put(#1,\ypos){\vector(0,-1){\lengthdash}}
\or \advance\ypos by 10
    \put(#1,\ypos){\vector(0,-1){\lengthdash}}
    \advance\ypos by 40
    \put(#1,\ypos){\vector(0,-1){\lengthdash}}
\fi
}}

\def\puthmorphism(#1,#2)[#3`#4`#5]#6#7#8{{%
\xpos #1
\ypos #2
\width #6
\arrowlength #6
\arrowtype=#7
\putbox(\xpos,\ypos){#3\vphantom{#4}}%
{\advance \xpos by\arrowlength
\putbox(\xpos,\ypos){\vphantom{#3}#4}}%
\horsize{\tempcounta}{#3}%
\horsize{\tempcountb}{#4}%
\divide \tempcounta by2
\divide \tempcountb by2
\advance \tempcounta by30
\advance \tempcountb by30
\advance \xpos by\tempcounta
\advance \arrowlength by-\tempcounta
\advance \arrowlength by-\tempcountb
\putvector(\xpos,\ypos)(1,0)\arrowlength\arrowtype
\divide \arrowlength by2
\advance \xpos by\arrowlength
\vertsize{\tempcounta}{#5}%
\divide\tempcounta by2
\advance \tempcounta by20
\if a#8 %
   \advance \ypos by\tempcounta
   \putbox(\xpos,\ypos){#5}%
\else
   \advance \ypos by-\tempcounta
   \putbox(\xpos,\ypos){#5}%
\fi}}

\def\putvmorphism(#1,#2)[#3`#4`#5]#6#7#8{{%
\xpos #1
\ypos #2
\arrowlength #6
\arrowtype #7
\settowidth{\xlen}{$#5$}%
\putbox(\xpos,\ypos){#3}%
{\advance \ypos by-\arrowlength
\putbox(\xpos,\ypos){#4}}%
{\advance\arrowlength by-140
\advance \ypos by-70
\ifdim\xlen>0pt
   \if m#8%
      \putsplitvector(\xpos,\ypos)\arrowlength\arrowtype
   \else
   \putvector(\xpos,\ypos)(0,-1)\arrowlength\arrowtype
   \fi
\else
   \putvector(\xpos,\ypos)(0,-1)\arrowlength\arrowtype
\fi}%
\ifdim\xlen>0pt
   \divide \arrowlength by2
   \advance\ypos by-\arrowlength
   \if l#8%
      \advance \xpos by-40
      \putrbox(\xpos,\ypos){#5}%
   \else\if r#8%
      \advance \xpos by40
      \putlbox(\xpos,\ypos){#5}%
   \else
      \putbox(\xpos,\ypos){#5}%
   \fi\fi
\fi
}}

\def\putsquarep<#1>(#2)[#3;#4`#5`#6`#7]{{%
\setsqparms[#1]%
\setpos(#2)%
\settokens`#3`%
\puthmorphism(\xpos,\ypos)[\tokenc`\tokend`{#7}]{\width}{\arrowtyped}b%
\advance\ypos by \height
\puthmorphism(\xpos,\ypos)[\tokena`\tokenb`{#4}]{\width}{\arrowtypea}a%
\putvmorphism(\xpos,\ypos)[``{#5}]{\height}{\arrowtypeb}l%
\advance\xpos by \width
\putvmorphism(\xpos,\ypos)[``{#6}]{\height}{\arrowtypec}r%
}}

\def\putsquare{\@ifnextchar <{\putsquarep}{\putsquarep%
   <\arrowtypea`\arrowtypeb`\arrowtypec`\arrowtyped;\width`\height>}}
\def\square{\@ifnextchar< {\squarep}{\squarep
   <\arrowtypea`\arrowtypeb`\arrowtypec`\arrowtyped;\width`\height>}}
\def\squarep<#1>[#2`#3`#4`#5;#6`#7`#8`#9]{{
\setsqparms[#1]
\diagram
\putsquarep<\arrowtypea`\arrowtypeb`\arrowtypec`
\arrowtyped;\width`\height>
(0,0)[#2`#3`#4`{#5};#6`#7`#8`{#9}]
\enddiagram
}}                                                 
\def\putptrianglep<#1>(#2,#3)[#4`#5`#6;#7`#8`#9]{{%
\settriparms[#1]%
\xpos=#2 \ypos=#3
\advance\ypos by \height
\puthmorphism(\xpos,\ypos)[#4`#5`{#7}]{\height}{\arrowtypea}a%
\putvmorphism(\xpos,\ypos)[`#6`{#8}]{\height}{\arrowtypeb}l%
\advance\xpos by\height
\putmorphism(\xpos,\ypos)(-1,-1)[``{#9}]{\height}{\arrowtypec}r%
}}

\def\putptriangle{\@ifnextchar <{\putptrianglep}{\putptrianglep
   <\arrowtypea`\arrowtypeb`\arrowtypec;\height>}}
\def\ptriangle{\@ifnextchar <{\ptrianglep}{\ptrianglep
   <\arrowtypea`\arrowtypeb`\arrowtypec;\height>}}
\def\ptrianglep<#1>[#2`#3`#4;#5`#6`#7]{{
\settriparms[#1]
\diagram
\putptrianglep<\arrowtypea`\arrowtypeb`
\arrowtypec;\height>
(0,0)[#2`#3`#4;#5`#6`{#7}]
\enddiagram
}}                                            

\def\putqtrianglep<#1>(#2,#3)[#4`#5`#6;#7`#8`#9]{{%
\settriparms[#1]%
\xpos=#2 \ypos=#3
\advance\ypos by\height
\puthmorphism(\xpos,\ypos)[#4`#5`{#7}]{\height}{\arrowtypea}a%
\putmorphism(\xpos,\ypos)(1,-1)[``{#8}]{\height}{\arrowtypeb}l%
\advance\xpos by\height
\putvmorphism(\xpos,\ypos)[`#6`{#9}]{\height}{\arrowtypec}r%
}}

\def\putqtriangle{\@ifnextchar <{\putqtrianglep}{\putqtrianglep
   <\arrowtypea`\arrowtypeb`\arrowtypec;\height>}}
\def\qtriangle{\@ifnextchar <{\qtrianglep}{\qtrianglep
   <\arrowtypea`\arrowtypeb`\arrowtypec;\height>}}
\def\qtrianglep<#1>[#2`#3`#4;#5`#6`#7]{{
\settriparms[#1]
\width=\height                                
\diagram
\putqtrianglep<\arrowtypea`\arrowtypeb`
\arrowtypec;\height>
(0,0)[#2`#3`#4;#5`#6`{#7}]
\enddiagram
}}

\def\putdtrianglep<#1>(#2,#3)[#4`#5`#6;#7`#8`#9]{{%
\settriparms[#1]%
\xpos=#2 \ypos=#3
\puthmorphism(\xpos,\ypos)[#5`#6`{#9}]{\height}{\arrowtypec}b%
\advance\xpos by \height \advance\ypos by\height
\putmorphism(\xpos,\ypos)(-1,-1)[``{#7}]{\height}{\arrowtypea}l%
\putvmorphism(\xpos,\ypos)[#4``{#8}]{\height}{\arrowtypeb}r%
}}

\def\putdtriangle{\@ifnextchar <{\putdtrianglep}{\putdtrianglep
   <\arrowtypea`\arrowtypeb`\arrowtypec;\height>}}
\def\dtriangle{\@ifnextchar <{\dtrianglep}{\dtrianglep
   <\arrowtypea`\arrowtypeb`\arrowtypec;\height>}}
\def\dtrianglep<#1>[#2`#3`#4;#5`#6`#7]{{
\settriparms[#1]
\width=\height                                
\diagram
\putdtrianglep<\arrowtypea`\arrowtypeb`
\arrowtypec;\height>
(0,0)[#2`#3`#4;#5`#6`{#7}]
\enddiagram
}}

\def\putbtrianglep<#1>(#2,#3)[#4`#5`#6;#7`#8`#9]{{%
\settriparms[#1]%
\xpos=#2 \ypos=#3
\puthmorphism(\xpos,\ypos)[#5`#6`{#9}]{\height}{\arrowtypec}b%
\advance\ypos by\height
\putmorphism(\xpos,\ypos)(1,-1)[``{#8}]{\height}{\arrowtypeb}r%
\putvmorphism(\xpos,\ypos)[#4``{#7}]{\height}{\arrowtypea}l%
}}

\def\putbtriangle{\@ifnextchar <{\putbtrianglep}{\putbtrianglep
   <\arrowtypea`\arrowtypeb`\arrowtypec;\height>}}
\def\btriangle{\@ifnextchar <{\btrianglep}{\btrianglep
   <\arrowtypea`\arrowtypeb`\arrowtypec;\height>}}
\def\btrianglep<#1>[#2`#3`#4;#5`#6`#7]{{
\settriparms[#1]
\width=\height                               
\diagram
\putbtrianglep<\arrowtypea`\arrowtypeb`
\arrowtypec;\height>
(0,0)[#2`#3`#4;#5`#6`{#7}]
\enddiagram
}}

\def\putAtrianglep<#1>(#2,#3)[#4`#5`#6;#7`#8`#9]{{%
\settriparms[#1]%
\xpos=#2 \ypos=#3
{\multiply \height by2
\puthmorphism(\xpos,\ypos)[#5`#6`{#9}]{\height}{\arrowtypec}b}%
\advance\xpos by\height \advance\ypos by\height
\putmorphism(\xpos,\ypos)(-1,-1)[#4``{#7}]{\height}{\arrowtypea}l%
\putmorphism(\xpos,\ypos)(1,-1)[``{#8}]{\height}{\arrowtypeb}r%
}}

\def\putAtriangle{\@ifnextchar <{\putAtrianglep}{\putAtrianglep
   <\arrowtypea`\arrowtypeb`\arrowtypec;\height>}}
\def\Atriangle{\@ifnextchar <{\Atrianglep}{\Atrianglep
   <\arrowtypea`\arrowtypeb`\arrowtypec;\height>}}
\def\Atrianglep<#1>[#2`#3`#4;#5`#6`#7]{{
\settriparms[#1]
\width=\height                                     
\diagram
\putAtrianglep<\arrowtypea`\arrowtypeb`
\arrowtypec;\height>
(0,0)[#2`#3`#4;#5`#6`{#7}]
\enddiagram
}}

\def\putAtrianglepairp<#1>(#2)[#3;#4`#5`#6`#7`#8]{{%
\settripairparms[#1]%
\setpos(#2)%
\settokens`#3`%
\puthmorphism(\xpos,\ypos)[\tokenb`\tokenc`{#7}]{\height}{\arrowtyped}b%
\advance\xpos by\height
\puthmorphism(\xpos,\ypos)[\phantom{\tokenc}`\tokend`{#8}]%
{\height}{\arrowtypee}b%
\advance\ypos by\height
\putmorphism(\xpos,\ypos)(-1,-1)[\tokena``{#4}]{\height}{\arrowtypea}l%
\putvmorphism(\xpos,\ypos)[``{#5}]{\height}{\arrowtypeb}m%
\putmorphism(\xpos,\ypos)(1,-1)[``{#6}]{\height}{\arrowtypec}r%
}}

\def\putAtrianglepair{\@ifnextchar <{\putAtrianglepairp}{\putAtrianglepairp%
   <\arrowtypea`\arrowtypeb`\arrowtypec`\arrowtyped`\arrowtypee;\height>}}
\def\Atrianglepair{\@ifnextchar <{\Atrianglepairp}{\Atrianglepairp%
   <\arrowtypea`\arrowtypeb`\arrowtypec`\arrowtyped`\arrowtypee;\height>}}

\def\Atrianglepairp<#1>[#2;#3`#4`#5`#6`#7]{{
\settripairparms[#1]
\settokens`#2`
\width=\height                                
\diagram
\putAtrianglepairp                            
<\arrowtypea`\arrowtypeb`\arrowtypec`
\arrowtyped`\arrowtypee;\height>
(0,0)[{#2};#3`#4`#5`#6`{#7}]
\enddiagram
}}

\def\putVtrianglep<#1>(#2,#3)[#4`#5`#6;#7`#8`#9]{{%
\settriparms[#1]%
\xpos=#2 \ypos=#3
\advance\ypos by\height
{\multiply\height by2
\puthmorphism(\xpos,\ypos)[#4`#5`{#7}]{\height}{\arrowtypea}a}%
\putmorphism(\xpos,\ypos)(1,-1)[`#6`{#8}]{\height}{\arrowtypeb}l%
\advance\xpos by\height
\advance\xpos by\height
\putmorphism(\xpos,\ypos)(-1,-1)[``{#9}]{\height}{\arrowtypec}r%
}}

\def\putVtriangle{\@ifnextchar <{\putVtrianglep}{\putVtrianglep
   <\arrowtypea`\arrowtypeb`\arrowtypec;\height>}}
\def\Vtriangle{\@ifnextchar <{\Vtrianglep}{\Vtrianglep
   <\arrowtypea`\arrowtypeb`\arrowtypec;\height>}}
\def\Vtrianglep<#1>[#2`#3`#4;#5`#6`#7]{{
\settriparms[#1]
\width=\height                                 
\diagram
\putVtrianglep<\arrowtypea`\arrowtypeb`
\arrowtypec;\height>
(0,0)[#2`#3`#4;#5`#6`{#7}]
\enddiagram
}}

\def\putVtrianglepairp<#1>(#2)[#3;#4`#5`#6`#7`#8]{{
\settripairparms[#1]%
\setpos(#2)%
\settokens`#3`%
\advance\ypos by\height
\putmorphism(\xpos,\ypos)(1,-1)[`\tokend`{#6}]{\height}{\arrowtypec}l%
\puthmorphism(\xpos,\ypos)[\tokena`\tokenb`{#4}]{\height}{\arrowtypea}a%
\advance\xpos by\height
\puthmorphism(\xpos,\ypos)[\phantom{\tokenb}`\tokenc`{#5}]%
{\height}{\arrowtypeb}a%
\putvmorphism(\xpos,\ypos)[``{#7}]{\height}{\arrowtyped}m%
\advance\xpos by\height
\putmorphism(\xpos,\ypos)(-1,-1)[``{#8}]{\height}{\arrowtypee}r%
}}

\def\putVtrianglepair{\@ifnextchar <{\putVtrianglepairp}{\putVtrianglepairp%
    <\arrowtypea`\arrowtypeb`\arrowtypec`\arrowtyped`\arrowtypee;\height>}}
\def\Vtrianglepair{\@ifnextchar <{\Vtrianglepairp}{\Vtrianglepairp%
    <\arrowtypea`\arrowtypeb`\arrowtypec`\arrowtyped`\arrowtypee;\height>}}
\def\Vtrianglepairp<#1>[#2;#3`#4`#5`#6`#7]{{
\settripairparms[#1]
\settokens`#2`
\diagram
\putVtrianglepairp                             
<\arrowtypea`\arrowtypeb`\arrowtypec`
\arrowtyped`\arrowtypee;\height>
(0,0)[{#2};#3`#4`#5`#6`{#7}]
\enddiagram
}}

\def\putCtrianglep<#1>(#2,#3)[#4`#5`#6;#7`#8`#9]{{%
\settriparms[#1]%
\xpos=#2 \ypos=#3
\advance\ypos by\height
\putmorphism(\xpos,\ypos)(1,-1)[``{#9}]{\height}{\arrowtypec}l%
\advance\xpos by\height
\advance\ypos by\height
\putmorphism(\xpos,\ypos)(-1,-1)[#4`#5`{#7}]{\height}{\arrowtypea}l%
{\multiply\height by 2
\putvmorphism(\xpos,\ypos)[`#6`{#8}]{\height}{\arrowtypeb}r}%
}}

\def\putCtriangle{\@ifnextchar <{\putCtrianglep}{\putCtrianglep
    <\arrowtypea`\arrowtypeb`\arrowtypec;\height>}}
\def\Ctriangle{\@ifnextchar <{\Ctrianglep}{\Ctrianglep
    <\arrowtypea`\arrowtypeb`\arrowtypec;\height>}}
\def\Ctrianglep<#1>[#2`#3`#4;#5`#6`#7]{{
\settriparms[#1]
\width=\height                               
\diagram
\putCtrianglep<\arrowtypea`\arrowtypeb`
\arrowtypec;\height>
(0,0)[#2`#3`#4;#5`#6`{#7}]
\enddiagram
}}                                           
\def\putDtrianglep<#1>(#2,#3)[#4`#5`#6;#7`#8`#9]{{%
\settriparms[#1]%
\xpos=#2 \ypos=#3
\advance\xpos by\height \advance\ypos by\height
\putmorphism(\xpos,\ypos)(-1,-1)[``{#9}]{\height}{\arrowtypec}r%
\advance\xpos by-\height \advance\ypos by\height
\putmorphism(\xpos,\ypos)(1,-1)[`#5`{#8}]{\height}{\arrowtypeb}r%
{\multiply\height by 2
\putvmorphism(\xpos,\ypos)[#4`#6`{#7}]{\height}{\arrowtypea}l}%
}}

\def\putDtriangle{\@ifnextchar <{\putDtrianglep}{\putDtrianglep
    <\arrowtypea`\arrowtypeb`\arrowtypec;\height>}}
\def\Dtriangle{\@ifnextchar <{\Dtrianglep}{\Dtrianglep
   <\arrowtypea`\arrowtypeb`\arrowtypec;\height>}}
\def\Dtrianglep<#1>[#2`#3`#4;#5`#6`#7]{{
\settriparms[#1]
\width=\height                              
\diagram
\putDtrianglep<\arrowtypea`\arrowtypeb`
\arrowtypec;\height>
(0,0)[#2`#3`#4;#5`#6`{#7}]
\enddiagram
}}                                          
\def\setrecparms[#1`#2]{\width=#1 \height=#2}%

\def\recursep<#1`#2>[#3;#4`#5`#6`#7`#8]{{\m@th
\width=#1 \height=#2
\settokens`#3`
\settowidth{\tempdimen}{$\tokena$}
\ifdim\tempdimen=0pt
  \savebox{\tempboxa}{\hbox{$\tokenb$}}%
  \savebox{\tempboxb}{\hbox{$\tokend$}}%
  \savebox{\tempboxc}{\hbox{$#6$}}%
\else
  \savebox{\tempboxa}{\hbox{$\hbox{$\tokena$}\times\hbox{$\tokenb$}$}}%
  \savebox{\tempboxb}{\hbox{$\hbox{$\tokena$}\times\hbox{$\tokend$}$}}%
  \savebox{\tempboxc}{\hbox{$\hbox{$\tokena$}\times\hbox{$#6$}$}}%
\fi
\ypos=\height
\divide\ypos by 2
\xpos=\ypos
\advance\xpos by \width
\bfig
\putCtrianglep<-1`1`1;\ypos>(0,0)[`\tokenc`;#5`#6`{#7}]%
\puthmorphism(\ypos,0)[\tokend`\usebox{\tempboxb}`{#8}]{\width}{-1}b%
\puthmorphism(\ypos,\height)[\tokenb`\usebox{\tempboxa}`{#4}]{\width}{-1}a%
\advance\ypos by \width
\putvmorphism(\ypos,\height)[``\usebox{\tempboxc}]{\height}1r%
\efig
}}

\def\recurse{\@ifnextchar <{\recursep}{\recursep<\width`\height>}}

\def\puttwohmorphisms(#1,#2)[#3`#4;#5`#6]#7#8#9{{%
\puthmorphism(#1,#2)[#3`#4`]{#7}0a
\ypos=#2
\advance\ypos by 20
\puthmorphism(#1,\ypos)[\phantom{#3}`\phantom{#4}`#5]{#7}{#8}a
\advance\ypos by -40
\puthmorphism(#1,\ypos)[\phantom{#3}`\phantom{#4}`#6]{#7}{#9}b
}}

\def\puttwovmorphisms(#1,#2)[#3`#4;#5`#6]#7#8#9{{%
\putvmorphism(#1,#2)[#3`#4`]{#7}0a
\xpos=#1
\advance\xpos by -20
\putvmorphism(\xpos,#2)[\phantom{#3}`\phantom{#4}`#5]{#7}{#8}l
\advance\xpos by 40
\putvmorphism(\xpos,#2)[\phantom{#3}`\phantom{#4}`#6]{#7}{#9}r
}}

\def\puthcoequalizer(#1)[#2`#3`#4;#5`#6`#7]#8#9{{%
\setpos(#1)%
\puttwohmorphisms(\xpos,\ypos)[#2`#3;#5`#6]{#8}11%
\advance\xpos by #8
\puthmorphism(\xpos,\ypos)[\phantom{#3}`#4`#7]{#8}1{#9}
}}

\def\putvcoequalizer(#1)[#2`#3`#4;#5`#6`#7]#8#9{{%
\setpos(#1)%
\puttwovmorphisms(\xpos,\ypos)[#2`#3;#5`#6]{#8}11%
\advance\ypos by -#8
\putvmorphism(\xpos,\ypos)[\phantom{#3}`#4`#7]{#8}1{#9}
}}

\def\putthreehmorphisms(#1)[#2`#3;#4`#5`#6]#7(#8)#9{{%
\setpos(#1) \settypes(#8)
\if a#9 %
     \vertsize{\tempcounta}{#5}%
     \vertsize{\tempcountb}{#6}%
     \ifnum \tempcounta<\tempcountb \tempcounta=\tempcountb \fi
\else
     \vertsize{\tempcounta}{#4}%
     \vertsize{\tempcountb}{#5}%
     \ifnum \tempcounta<\tempcountb \tempcounta=\tempcountb \fi
\fi
\advance \tempcounta by 60
\puthmorphism(\xpos,\ypos)[#2`#3`#5]{#7}{\arrowtypeb}{#9}
\advance\ypos by \tempcounta
\puthmorphism(\xpos,\ypos)[\phantom{#2}`\phantom{#3}`#4]{#7}{\arrowtypea}{#9}
\advance\ypos by -\tempcounta \advance\ypos by -\tempcounta
\puthmorphism(\xpos,\ypos)[\phantom{#2}`\phantom{#3}`#6]{#7}{\arrowtypec}{#9}
}}

\def\setarrowtoks[#1`#2`#3`#4`#5`#6]{%
\def\toka{#1}
\def\tokb{#2}
\def\tokc{#3}
\def\tokd{#4}
\def\toke{#5}
\def\tokf{#6}
}
\def\hex{\@ifnextchar <{\hexp}{\hexp<1000`400>}}
\def\hexp<#1`#2>[#3`#4`#5`#6`#7`#8;#9]{%
\setarrowtoks[#9]
\yext=#2 \advance \yext by #2
\xext=#1 \advance\xext by \yext
\bfig
\putCtriangle<-1`0`1;#2>(0,0)[`#5`;\tokb``\tokd]
\xext=#1 \yext=#2 \advance \yext by #2
\putsquare<1`0`0`1;\xext`\yext>(#2,0)[#3`#4`#7`#8;\toka```\tokf]
\advance \xext by #2
\putDtriangle<0`1`-1;#2>(\xext,0)[`#6`;`\tokc`\toke]
\efig
}
\makeatother

\newtheorem{lemma}{Lemma}[section]
\newtheorem{proposition}[lemma]{Proposition}
\newtheorem{remark}[lemma]{Remark}

\newtheorem{theorem}[lemma]{Theorem}

\newtheorem{definition}[lemma]{Definition}
\newtheorem{corollary}[lemma]{Corollary}
\newtheorem{claim}[lemma]{Claim}

\newcommand{\BProof}{{\it Proof.} \r}
\newcommand{\EProof}{\hfill \blacksquare}

\def\veps{\varepsilon}

\def\ome {\omega}

\def\XI{\xi}
\def\XII{\eta}
\def\CHI{\chi}
\def\lam{\lambda}
\def\lto{\longrightarrow}

\def\iso{\backsimeq}

\def\R{{\mathbb R}}
\def\Q{{\mathbb Q}}
\def\N{{\mathbb N}}
\def\Z{{\mathbb Z}}
\def\C{{\mathbb C}}
\def\Fq {\mathbb{F}_{q}}   
\def\Fqm{\Fq^*}
\def\lgn {\sigma}  
\def\Fp{\mathbb{F}_{p}}

\def\Fpm{\Fp^*}

\def\Lam{\Lambda}
\def\Lm{\Lambda^*}

\def\GL{\mathrm{GL}}
\def\PGL{\mathrm{PGL}}
\def\GZ{\mathrm{SL}_2(\Z)}
\def\G{\mathrm{\Gamma}}
\def\GG{\mathrm{SL}_2(\Fp)}
\def\GGG{\mathrm{Sp}(2n,\Fp)}
\def\GZZ{\mathrm{Sp}(2n,\Z)} 

\def\AGG{\mathbb{S} \mathbb{L}_2} 
\def\AT{\mathbb{T}}      
\def\GrA{\langle A \rangle}
\def\CA{{\mathrm T}_A}
\def\ACA{\mathbb{T}_A}
\def\CAD{{\mathrm T}^*_A} 
\def\Gf{\mathrm{\Gamma}_{p}}

\def\Sp {\mathrm{Sp}_{\omega}} 
\def\ASp {\mathbb{S} \mathrm{p}_\ome} 
\def\heiz {\mathrm{E}}      
\def\Aheiz{\mathbb{E}}     
\def \semi {\mathrm{D}} 
\def \Asemi{\mathbb{D}} 

\def\Ga {\mathbb{G}_a} 
\def\Gm {\mathbb{G}_m} 

\def\Br {\mathrm{B}} 
\def\ABr {\mathbb{B}} 

\def\oB{\Br} 
\def\AoB{\ABr} 

\def\AUn{{{\mathbb U}^\circ}} 

\def\AoU{\mathbb U} 

\def\ob{b} 
\def\ou{u}
\def\u{{u^\circ}}
\def\S0{S}

\def\FST{{\cal S}(\T)}
\def\F{C^{\infty}({\bf T})}
\def\T{{\bf T}}
\def\Td{\T^\vee} 

\def\Irr{\mathrm{Irr}({\cal A} _\hbar)}
\def\A{\cal A}
\def\Ad{ {\cal A} _\hbar}

\def\Adf{{\cal A}_{p}}
\def\hb{\hbar}
\def\h{ \hbar }
\def\Aeh{{\cal A} _{\varepsilon,\hbar}} 
\def\Irreh{\mathrm{Irr}(\Aeh)}

\def \Wigner {\mathcal{W}} 
\def \nWigner{\tilde{\mathcal{W}}} 

\def\H{{\cal H}}
\def\Hh{{\cal H} _\h}
\def\Hc {{\mathcal{H}_\V}} 
\def\V{\mathrm{V}} 
\def\AV {\mathbb{V}} 
\def\VI {\mathrm{V}_1} 
\def\AVI {\mathbb{V}_1}
\def\VII{\mathrm{V}_2} 
\def\W{{\mathrm W}}

\def\rhoh{\rho _{_\h}}
\def\pih{\pi i {\h}}
\def\Pih{\pi_{_\hbar}}
\def\Pi{\pi}
\def\rhop {\rho_{_{\mathrm p}}}
\def\rhof {\rho_{_p}}

\def \bL {{\bf L}}
\def\bM{{\bf M}}
\def\Pif{\pi_{p}}

\def\FSVI{{\cal S}(\VI)}
\def\FSFq{{\cal S}(\Fq)}

\def\Y{Y_{0}}
\def\YY{Y}
\def\AYY{\mathbb{Y}}

\def \Lmmf {\mathrm{V}}

\def \bA{\mathbb{A}}
\def\projI {\mathbb{P}^1}  
\def\P1{\mathbb{P}^1}

\def\AX {\mathbb{X}} 

\def\Sw {\mathrm{O}_w}  
\def\ASw {\mathbb{O}_w}  
\def\O{{\mathbb O}} 
\def\calU{{\cal U}} 

\def \AUnx {\AUn^{\times}}

\def\i_XI{i_{_{\XI}}}
\def\iXII{i_{_{\XII}}}
\def\p_XI{p_{_{\XI}}}

\def \mRSwE {\mathrm{R}}

\def\SE{\mathcal{E}}

\def\SF{\mathcal{F}}
\def\SG{\mathcal{G}}
\def\SL{\mathscr L}

\def \SI {\mathcal{I}}

\def\SK {\mathcal{K}} 

\def\SKoB {\SK_{\AoB}}  
\def\SKUx{\SK_{\AUn ^ \times}} 
\def\SKO{\SK_\O} 
\def\SKU{\SK_{\cal U}} 

\def\SKpi {\SK_{\Aheiz}} 

\def\SKn {\tilde{\SK}}  
\def\SKnSw {\SKn_{\ASw}}
\def\SKnoB {\SKn_{\AoB}}
\def\SKnSwE {\SKn_\O}
\def\SKnoU {\SKn_{\AoU}}
\def\SKnUx {\SKn_{\AUn ^ \times}}

\def\SKnoUxw{\SKn_{\AoU^\times w}}

\def\SKnE   {\SKn_{\Aheiz}}
\def\SKnw {\SKn_w}
\def\SKoU {\SK_{\AoU}} 
\def\SKUx {\SK_{\AUn ^ \times}} 
\def\SKoUxw {\SK_{{\AoU}^\times w}}

\def\SKob {\SK_{\ob}}

\def\SKw {\SK_w} 
\def\SKwI {\SK_{w^{-1}}}
\def\SAw  {\SA_w}

\def\SA {\mathcal{A}}  
\def\SASw {\SA_{\ASw}}
\def\SAoB {\SA_{\AoB}}
\def\SASwE  {\SA_{\mathbb O}}
\def\SAoU {\SA_{\AoU}}

\def \SAUx {\SA_{\AUn ^ \times}}
\def \SAoUxw {\SA_{\AoU^\times w}}

\def\Sartin{\SL_{\psi}} 
\def\Skummer {\SL_{\chi}} 

\def\Slegendre {\SL_{\lgn}}

\def\Tr{\mathrm{Tr}}
\def\Av{\mathrm{\bf Av}}
\def\ev{{\mathrm{e.v}}}

\def\lang{\langle}
\def\rang{\rangle}
\def\sub{\subset}
\def\End{\mathrm{End}}
\def\Hom{\mathrm{Hom}}
\def\dim{\mathrm{dim}}
\def\cent{Z}

\def\orn {\varrho}
\def\ornlag {\mathrm{Lag}^{\circ}} 
\def\lag {\mathrm{Lag}} 
\def\ornL {\mathrm{L^{^\circ}}} 
\def\ornLI {{\mathrm{L}^{^\circ}_1}} 
\def\ornLIm {{\mathrm{L}^{^{\overline{\circ}}}_1}} 
\def\ornLII {{\mathrm{L}^{^\circ}_2}} 
\def \ornLIII {{\mathrm{L}^{^\circ}_3}} 
\def\interLILII {\Theta_{\ornLII,\ornLI}} 
\def\L {\mathrm{L}} 
\def\LI {{\mathrm{L}_1}} 
\def\LII {{\mathrm{L}_2}} 
\def \LIII {{\mathrm{L}_3}} 
\def\thetaLILII {\theta_{_{\ornLII,\ornLI}}} 
\def \thetaLILIII  {\theta_{_{\ornLIII,\ornLI}}} 
\def \thetaLIILIII  {\theta_{_{\ornLIII,\ornLII}}} 
\def\thetagLIgLII {\theta_{_{g\ornLII,g\ornLI}}} 
\def\ornofL {\orn_{_\L}} 
\def\ornVII  {{\mathrm{\VII^{^\circ}}}} 
\def\thetagVIIVII {\theta_{_{\ornVII,g\ornVII}}} 
\def\thetaLIILI {\theta_{_{\ornLI,\ornLII}}} 
\def\thetaLIILIm {\theta_{_{\ornLIm,\ornLII}}} 

\def\thnLILII {\tilde{\theta}_{_{\ornLII,\ornLI}}} 
\def \thnLILIII  {\tilde{\theta}_{_{\ornLIII,\ornLI}}} 
\def \thnLIILIII  {\tilde{\theta}_{_{\ornLIII,\ornLII}}} 
\def \thnLIILI  {\tilde{\theta}_{_{\ornLI,\ornLII}}} 

\def\aLILII {\mathrm{a}_{\ornLII,\ornLI}} 
\def\aLILIII {\mathrm{a}_{\ornLIII,\ornLI}} 
\def\aLIILIII {\mathrm{a}_{\ornLIII,\ornLII}} 
\def\aLIILI {\mathrm{a}_{\ornLI,\ornLII}} 

\def\xiLI {\xi_{_\LI}} 
\def\xiLII {\xi_{_\LII}} 
\def\xiLIII {\xi_{_\LIII}} 

\def\rLIILIII {\mathrm{r}_{_{\LIII,\LII}}} 
\def\ri {\mathrm{r}_{_{\mathrm{L}_{i},\mathrm{L}_{k}}}} 

\def\Cconst{\mathrm{C}} 
\def\Dconst{\mathrm{D}} 

\def\cq {\mathrm{q}} 
\def\cp {\mathrm{p}} 

\def\Db{{\cal D}^{b}_{\mathrm{c},\mathrm{w}}}
\def\Fr{\mathrm{Fr}}
\def\chiFr{\chi_{_\Fr}}
\def\Qlb{\overline{\Q}_\ell}
\def\sint{\smallint}
\def\coH{{\mathrm H}}
\def\Swan{\mathrm{Swan}}

\def\rev{\quad}

\def\r{\;}
\def\above{\overset}

\def\half{\begin{smallmatrix} \frac{1}{2} \end{smallmatrix}}

\def\q{{\bf{q}}} 

\def\2n{\mathrm{2n}}

\def\Nd{\mathrm{N}^\mathrm{n}}



\begin{document}

\date{}

\frontmatter
\chapter*{Acknowledgements}
This work is for me a conclusion of four years of activity in a
research project under the direction of prof. Joseph Bernstein. I
would like to use this opportunity to thank him sincerely for
letting me join this intellectual voyage and having thought me as
much as the little that I know. Working with Joseph and learning
from him is an experience that I will cherish for all my life.
\\
\\
Large parts of this work are a result of a joint project with my
friend Ronny Hadani. I thank him and share my deep appreciation
for the long way we walked side by side and the thousands of hours
we spent working together here in Israel and far abroad the
states.
\\
\\
This thesis stands across the fine bridge between mathematics and
physics. This is a fascinating and exotic domain of knowledge,
which spreads its roots into multiple areas of research. I owe a
great deal to Dr. Par Kurlberg and Prof. Zeev Rudnick for
explaining me their ideas about "Quantum Chaos". I would like to
thank Prof. David Kazhdan for sharing his thoughts about the
possible existence of canonical Hilbert spaces. In addition I
thank Prof. Peter Sarnak for some interesting discussions.
Finally, I would like to thank Prof. Pierre Deligne for letting me
use his ideas about the geometrization of the Weil representation
which appeared in a letter he wrote to David Kazhdan in 1982.
\\
\\
In the course of my work I had the privilege to visit and to give
lectures in several institutions around the world. I would like to
thank them for their hospitality and for providing me with
excellent working conditions. In this list I count: MSRI, Berkeley
University, IHES, Gothenburg University, Max-Planck Institute at
Bonn, Bonn University, Courant Institute, Princeton University,
Yale University and Ohio state University.
\\
\\
I am happy to have the opportunity to thank my friends: S.
Artstein, D. Blanck, D. Gaitsgory, Y. Ostrover, Y. Kremnizer, D.
Mangobi, Y. Peterzil, N. Porat, E. Sayag, I. Tyomkin and D.
Yekutieli who supported me, each of them in his own special way.
\\
\\
Finally, I would like to thank my family: my parents Noga and Oded
and my brothers Navot, Hamutal and Ira.

\chapter*{}
\chapter*{Abstract}
\addcontentsline{toc}{chapter}{Abstract}
Consider the two dimensional symplectic torus $(\T,\ome)$ and an
hyperbolic automorphism $A$ of $\T$. The automorphism $A$ is known
to be ergodic. In 1980, using a non-trivial procedure called
quantization, the physicists J. Hannay and M.V. Berry attached to
this automorphism a quantum operator $\rhoh(A)$ acting on a
Hilbert space $\Hh$. One of the central questions of "Quantum
Chaos Theory", in this model, is whether the operator $\rhoh(A)$
is "quantum ergodic"?
\\
\\
We consider the following two distributions on the algebra $\A =
\F$ of smooth complex valued functions on $\T$. The first one is
given by the \textit{Haar} integral:
\begin{eqnarray*}
f \longmapsto \int_\T f \ome
\end{eqnarray*}
and the second one is given by the \textit{Wigner} distribution:
\begin{eqnarray*}
f\longmapsto \Wigner_\chi(f)
\end{eqnarray*}
defined as the expectation of the "quantum observable" $\Pih(f)$
in the Hecke state $v_\chi$, i.e. $\Wigner_\chi(f) := \left <
v_\chi | \Pih(f) v_\chi \right >$. Here the vector $v_\chi$ is a
common eigenvector, with eigencharacter $\chi$, of the
\textit{Hecke} group of symmetries of the quantum operator
$\rhoh(A)$.
\\
\\
The \textit{Kurlberg-Rudnick} rate conjecture is a quantitative
description of the behavior of the Wigner distribution attached to
the ergodic automorphism $A$. It states that for \textit{Planck}
constant of the form $\h = \frac{1}{p}$, where $p$ is a prime
number, one has:
\\
\\
\textbf{Rate Conjecture.} The following bound holds:
\begin{eqnarray*}
\left | \Wigner_\chi(f) - \int_\T f \ome \right | \leq
\frac{C_f}{\sqrt{p}}
\end{eqnarray*}
where $C_f$ is a constant that depends only on the function $f$.
\\
\\
In the current thesis we present a proof of the Kurlberg-Rudnick
conjecture. This is carried out using new representation theoretic
constructions and algebro-geometric sheaf realization of the Weil
metaplectic representation, which was proposed by P. Deligne in
1982.

\tableofcontents

\mainmatter
\chapter*{Introduction}
\markboth{INTRODUCTION}{INTRODUCTION}
\addcontentsline{toc}{chapter}{Introduction}
\section*{Hannay-Berry model} In the paper ``{\it Quantization of
linear maps on the torus - Fresnel diffraction by a periodic
grating}'', published in 1980 \cite{HB}, the physicists and J.
Hannay and M.V. Berry explore a
 model for quantum mechanics on the two dimensional symplectic torus
$(\T,\ome)$. Hannay and Berry suggested to \textit{quantize}
simultaneously the functions on the torus and the linear
symplectic group $\G= \GZ$.

\section*{Quantum chaos} One of their main motivations was to
study the phenomenon of quantum chaos \cite{R2, S2} in this model.
More precisely, they considered an ergodic discrete dynamical
system on the torus, which is generated by an hyperbolic
automorphism $A \in \GZ$. Quantizing the system, we replace: the
classical phase space $(\T,\ome)$  by a Hilbert space $\Hh$,
classical observables, i.e., functions $f \in \F$,  by operators
$\Pih(f) \in \End(\Hh)$ and classical symmetries by a unitary
representation  $\rhoh : \GZ \lto \mathrm{U}(\Hh)$. A fundamental
meta-question in the area of quantum chaos is to
\textit{understand} the ergodic properties of the quantum system
$\rhoh(A)$, at least in the semi-classical limit as $\h
\rightarrow 0$.

\section*{Hecke quantum unique ergodicity} This question was
addressed in a paper by Kurlberg and Rudnick  \cite{KR1}. In this
paper they formulated a rigorous definition of quantum ergodicity
for the case $\h = \frac{1}{p}$. The following is a brief
description of that work. The basic observation is that the
representation $\rhoh$ factors through the quotient group $\Gf
\iso \GG$. We denote by $\CA \subset \Gf$ the centralizer of the
element $A$, now considered as an element of the quotient group
$\Gf$. The group $\CA$ is called (cf. \cite{KR1}) the
\textit{Hecke torus} corresponding to the element $A$. The Hecke
torus acts semisimply on $\Hh$. Therefore we have a decomposition:
$$ \Hh = \bigoplus_{\chi : \CA \lto \C^*} \H_\chi $$
where $\H_\chi$ is the Hecke eigenspace corresponding to the
character $\chi$. Considering a unit vector $v \in \H_\chi$, one
defines the \textit{Wigner distribution} $\Wigner_\chi : \F   \lto
\C $ by the formula $\Wigner_\chi(f) := \left < v | \Pih(f) v
\right
>$. The main statement in \cite{KR1} asserts about an explicit
bound of the semi-classical asymptotic of $\Wigner_\chi(f)$:
$$ \left | \Wigner_\chi(f) - \int_\T f \ome \right | \leq \frac{C_f}{p^{1/4}}$$
where $C_f$ is a constant that depends only on the function $f$.
In Rudnick's lectures at MSRI, Berkeley 1999 \cite{R1} and ECM,
Barcelona 2000 \cite{R2} he conjectured that a stronger bound
should hold true, namely:
\\
\\
%
%
\textbf{Conjecture (Rate Conjecture).} The following bound holds:
\begin{eqnarray*}
\left | \Wigner_\chi(f) - \int_\T f \ome \right | \leq
\frac{C_f}{p^{1/2}}\,.
\end{eqnarray*}
\\
\\
The basic \textit{clues} suggesting the validity of this stronger
bound come from two main sources. The first source is
\textit{computer} simulations \cite{Ku} accomplished over the
years to give extremely precise bounds for considerably large
values of $p$. A more mathematical argument is based on the fact
that for special values of $p$, in which the Hecke torus
\textit{splits}, namely $\CA \simeq \Fpm$, one is able to compute
explicitly the eigenvector $v \in \H_\chi$ and as a consequence to
give an explicit \textit{formula} for the Wigner distribution
\cite{KR2, DGI}. More precisely, in case $\xi \in \Td$ , i.e., a
character, the distribution $\Wigner_\chi(\xi)$ turns out to be
equal to an exponential sum very much similar to the Kloosterman
sum:
\begin{equation*}
\frac{1}{p} \sum_{a\in \Fp^*} \psi\left(\frac{a+1}{a-1}\right)
\lgn(a) \chi(a)
\end{equation*}
where $\lgn$ denotes the Legendre character. In this case the
classical Weil bound \cite{W1} yields the result. In this thesis a
proof of the rate conjecture, for all tori (split or inert)
simultaneously, is presented. For this peropus we view things from
a more abstract perspective.
\section*{Geometric approach (Deligne sheaf)} The basic observation to be made is
that the theory of quantum mechanics on the torus, in case $\h =
\frac{1}{p}$, can be equivalently recast in the language of
representation theory of finite groups in characteristic $p$. We
will endeavor to give a more precise explanation of this matter.
Consider the quotient $\Fp$-vector space $\V = \Td / p \Td$, where
$\Td$ is the lattice of characters on $\T$. We denote by $\heiz =
\heiz(\V)$ the Heisenberg group. The group $\Gf \iso \GG$ is
naturally identified with the group of linear symplectomorphisms
of $\V$. We have an action of $\Gf$ on $\heiz$. The Stone-von
Neumann theorem states that there exists a unique irreducible
representation $\pi : \heiz \lto \GL(\H)$, with the non-trivial
central character $\psi$, for which its isomorphism class is fixed
by $\Gf$. This is equivalent to saying that $\H$ is equipped with
a compatible projective representation $\rho : \Gf \lto \PGL(\H)$.
Noting that $\heiz$ and $\Gf$ are the sets of rational points of
corresponding algebraic groups, it is natural to \textit{ask}
whether there exists an algebro-geometric object that underlies
the pair $(\pi,\rho)$?. The answer to this question is
\textit{positive}. The construction is proposed in an unpublished
letter that was sent in 1982 from Pierre Deligne to David Kazhdan
\cite{D1}. Parts of this letter will be published for the first
time in this thesis. In one sentence, the content of this letter
is a construction of \textit{Representation Sheaves} $\SK_\pi$ and
$\SK_\rho$ on the algebraic varieties $\Aheiz$ and $\AGG$
respectively. One obtains, as a consequence, the following general
principle:
\\
\\
\textbf{(*) Motivic principle}. All quantum mechanical quantities
in the Hannay-Berry model are motivic in nature.\label{principle}
\\
\\
By this we mean that every quantum-mechanical quantity
$\mathcal{Q}$, is associated with a vector space $\V_\mathcal{Q}$
endowed with a Frobenius action $\Fr : \V_\mathcal{Q} \lto
\V_\mathcal{Q}$ s.t.:
$$ \mathcal{Q} = \Tr (\Fr_{|_{\V_\mathcal{Q}}}).$$
The \textit{main contribution} of this paper is to implement this
principle. In particular we show that there exists  a two
dimensional vector space $\V_\chi$, endowed with an action $\Fr :
\V_\chi \lto \V_\chi$ s.t.:
$$ \Wigner_\chi(\xi) = \Tr (\Fr_{|_{\V_\chi}}).$$
This, combined with a bound on the modulus of the eigenvalues of
Frobenius, i.e., $$\left|\ev(\Fr_{|_{\V_\chi}})\right| \leq
\frac{1}{p^{1/2}},$$ completes the proof of the rate conjecture.
\\
\\
\section*{Side remarks} There are two remarks we would like to
make at this point:
\\
\\
\textbf{Remark 1: Discreteness principle.} ``Every'' quantity
  $\mathcal{Q}$ that appears in the
  Hannay-Berry model admits discrete spectrum in the following
  arithmetic sense: the modulus $ | \mathcal{Q} | $ can take only
  values of the form $p^{i/2}$ for $i \in \Z$. This is a
  consequence of principle (*) and Deligne's weight theory \cite{D2}. We
  believe that this principle can be effectively used in various situations
  in order to derive strong bounds out of weaker
  bounds. A \textit{striking example} is an alternative trivial "proof" for
  the bound $|\Wigner_\chi(\xi)| \leq \frac{C_\xi}{p^{1/2}}$:

$$|\Wigner_\chi(\xi)| \leq \frac{C_\xi}{p^{1/4}} \r  \Rightarrow \r |\Wigner_\chi(\xi)| \leq \frac{C_\xi}{p^{1/2}}\,.$$
  Kurlberg and Rudnick proved in their paper \cite{KR1} the weak bound
  $|\Wigner_\chi(\xi)| \leq \frac{C_\xi}{p^{1/4}}$. This directly
  implies (under certain mild assumptions) that the stronger
  bound $ |\Wigner_\chi(\xi)| \leq \frac{C_\xi}{{p}^{1/2}}$ is valid.
\\
\\
\textbf{Remark 2: Higher dimensional exponential sums.} Proving
the bound $| \Wigner_\chi(f) | \leq \frac{C_f}{\sqrt{p}}$ can be
equivalently stated as bounding by $\frac{C_f}{\sqrt{p}}$ the
spectral radius of the operator $\Av_{\CA} (f) := \frac{1}{|\CA|}
\sum\limits_{B \in \CA} \rhoh(B) \Pih(f) \rhoh(B^{-1})$. This
implies a bound on the $L_N$ norms, for every $N \in \Z^+$:
\begin{equation} \label{Lnorm}
\left \| \Av_{\CA} (f) \right \|_{N} \leq \frac{C_f}{p^N}.
\end{equation}
In particular for $0 \neq f = \xi \in \Td$ one can compute
explicitly the left hand side of (\ref{Lnorm}) and obtain:
$$\left \| \Av_{\CA} (\xi) \right \|_{N} := \Tr ( |\Av_{\CA} (\xi)|^N )
=  \frac{1}{|\CA|^{2N}}\sum_{(x_1,\ldots,x_{2N}) \in
X}\psi(\sum_{i < j} \ome(x_i,x_j))$$
where $X := \{(x_1,\ldots,x_{2N})|\r x_i \in {\cal O}_\xi, \r\sum
x_i = 0 \}$ and ${\cal O}_\xi := \CA \cdot \xi \subset \V$ denotes
the orbit of $\xi$ under the action of $\CA$. Therefore referring
to (\ref{Lnorm}) we obtained a \textit{non-trivial} bound for a
higher dimensional exponential sum. It would be interesting to
know whether there exists an independent proof for this bound and
whether this representation theoretic approach can be used to
prove optimal bounds for other interesting higher dimensional
exponential sums.
\section*{Sato-Tate conjecture}
The next level of the theory is to understand the \textit{complete
statistics} of the Hecke-Wigner distributions for different Hecke
states. More precisely, let us fix a character $\xi \in \Td$. For
every character $\chi :\CA \lto \C^*$ we consider the normalized
value $ \nWigner_\chi(\xi) := \frac{1}{2 \sqrt{p}}
\Wigner_\chi(\xi)$, which lies in the interval $[-1,1]$. Now
running over all multiplicative characters we define the following
atomic measure on the interval $[-1,1]$:
%
%
$$ \mu_p := \frac{1}{|\CA|} \sum_{\chi} \delta_{\nWigner_\chi(\xi)}. $$
One would like to describe the limit measure (if it exists!). This
is the content of another conjecture of Kurlberg and Rudnick
\cite{KR2}:
\\
\\
\textbf{Conjecture (Sato-Tate Conjecture).} The following limit
exists:
$$ \lim_{p \rightarrow \infty} \mu_p = \mu_{ST} $$
where $\mu_{ST}$ is the projection of the Haar measure on $S^1$ to
the interval $[-1,1]$.
\\
\\
We hope that by using the methodology described in this paper one
will be able to gain some progress in proving this conjecture.
\\
\\
\textbf{Remark.} Note that the family $\{ \nWigner_\chi(\xi)
\}_{\chi \in \CAD}$ runs over a non-algebraic space of parameters.
Hence Deligne's equidistribution theory (cf. Weil II \cite{D2})
can not be applied directly in order to solve the Sato-Tate
Conjecture.
\section*{Results} \label{results}
\begin{enumerate}
\item \textbf{Kurlberg-Rudnick conjecture.} The main result of the current work is Theorem \ref{pr}, which
is the \textit{proof} of the Kurlberg-Rudnick rate conjecture on
the asymptotic behavior  of the Hecke-Wigner distributions.
\item \textbf{Weil representation and the Hannay-Berry model.} We introduce \textit{two new constructions}
  of the Weil representation over $\Z$ and over the finite fields $\Fq$ of characteristic $\neq 2$.
  As an application we obtained a construction of the Hannay-Berry model.
  \begin{enumerate}
  \item The \textit{first} construction is stated in Theorem \ref{GH2}, Corollary \ref{prq} and
  Corollary \ref{prqh}. It is based on the Rieffel \textit{quantum torus} $\Ad$, for $\h \in \Q$.
  This approach is essentially equivalent to the classical approach (cf. \cite{Kl, W2}) that
  uses the representation theory of the Heisenberg group in characteristic $p$.
  As an application we obtained (Chapters \ref{BHm}, \ref{2d} and Appendix \ref{hd})
  a construction of the \textit{Hannay-Berry model} of quantum
  mechanics for Tori in all dimensions. This is a
  \textit{new realization} of the Hannay-Berry model. This was an
  important achievement, since the original Hannay-Berry model was
  formulated in physical terms. In particular,
  using this new approach we were able to construct (Chapters \ref{BHm} and \ref{2d}) a slightly more general model
  which has a larger and algebraic group of symmetries, namely the \textit{whole symplectic group} $\G = \GZ$.
  This was then generalized (Appendix \ref{hd}) to the higher dimensional tori and the
  groups $\mathrm{Sp}(2n,\Z)$.
  \item \textbf{Canonical Hilbert space (Kazhdan's question).} The \textit{second} construction uses
  our new \textit{"Method of Canonical Hilbert Space"} (see Chapter
\ref{metaplectique}). This approach is based on
  the following statement:
\\
\\
\textbf{Proposition (Canonical Hilbert Space).} Let $(\V,\ome)$ be
a two dimensional symplectic vector space over the finite field
$\Fq$. There exists a canonical Hilbert space $\Hc$ attached to
$\V$.
\\
\\
An immediate consequence of this proposition is that all
symmetries of $(\V,\ome)$ automatically act on $\H_\V$. In
particular, we obtain a \textit{linear} representation of the
group $\Sp := \Sp(\V,\ome)$ on $\H_\V$. Probably this approach has
higher dimensional generalization, for the case where $\V$ is of
dimension $2n$. This will be a subject of a future publication.
\\
\\
\textbf{Remark.} Note the main difference of our construction from
the classical approach due to Weil (cf. \cite{W2}). The
\textit{classical construction} proceeds in two stages. Firstly,
one obtains a projective representation of $\Sp$ and secondly
using general arguments about the group $\Sp$, one proves the
existence of a linearization. A consequence of \textit{our
approach} is that there exists a \textit{distinguished} linear
representation and its existence is not related to any group
theoretic property of $\Sp$. We would like to mention that this
approach answers, in the case of the two dimensional Heisenberg
group, \textit{a question of David Kazhdan} \cite{Ka} dealing with
the existence of Canonical Hilbert Spaces for co-adjoint orbits of
general unipotent groups. The main motive behind our construction
is the notion of \textit{oriented Lagrangian subspace}. This idea
was suggested to us by Joseph Bernstein \cite{B}.
\end{enumerate}
\item \textbf{Deligne's Weil representation sheaf and applications.} In our work
we developed \textit{$\ell$-adic} geometric techniques for the
investigation of the Weil representation in general and the
Hannay-Berry model in particular. These techniques are based on
\textit{Deligne's letter to Kazhdan} \cite{D1}. We include for the
sake of completeness (see Chapter \ref{metaplectique} section
\ref{delignes_letter}) a formal presentation of Deligne's letter,
that places the Weil representation on a complete
algebro-geometric ground. These techniques play a \textit{central
rule} in the proof of the Kurlberg-Rudnick conjecture.
\item \textbf{The higher-dimensional Kurlberg-Rudnick conjecture.}
Lately we were able \cite{GH4} to \textit{extend} some of our
results to the higher-dimensional tori. This will a subject of a
future research and publication. However, we want to note here an
\textit{interesting} phenomenon (see Appendix \ref{counter}).
Namely, the two dimensional Kurlberg-Rudnick conjecture in its
original formulation, that claims that \textit{any} ergodic
automorphism of the torus is represented by a Hecke quantum
ergodic operator, is \textit{not true} in higher dimensions:
\\
\\
\textbf{Observation.} Let $\T$ be the $2n$ dimensional torus, $n >
1$. There exists an element $A \in \GZZ$ which acts ergodically on
$\T$ such that the corresponding quantum operator $\rhoh(A)$ is
not Hecke ergodic.
\\
\\
The discussion on the \textit{meaning} of this observation and the
search for a \textit{"correct formulation"} to the conjectures,
that will be valid in any dimension, will be a subject of a future
research in the field of quantum chaos.
\end{enumerate}
\section*{Structure of the thesis}
The paper is naturally separated into four parts:
\\
\\
\textbf{ Part I.} Chapter \textbf{\ref{BHm}}. In this chapter we
\textit{present} the \textit{Hannay-Berry model}. In section
\ref{classicaltorus} we discuss classical mechanics on the torus.
In section \ref{quantization} we discuss quantum mechanics
\'{a}-la Hannay and Berry, using the Rieffel quantum torus model.
This part of the paper is self-contained and consists of mainly
linear algebraic considerations.
\\
\\
\textbf{Part II.}  Chapter \textbf{\ref{Proof}}. This is the main
part of the paper, consisting of the formulation and the
\textit{proof} of the Kurlberg-Rudnick conjecture. In section
\ref{QHUE} we \textit{formulate} the Hecke quantum unique
ergodicity \textit{conjecture} of Kurlberg-Rudnick (Theorem
\ref{pr}). In section \ref{theproof} the proof is given in two
stages. The first stage consists of mainly linear algebra
manipulations to obtain a more transparent formulation of the
statement, resulting in Theorem \ref{GH4}. In the second stage we
venture into algebraic geometry. All linear algebraic
constructions are replaced by sheaf theoretic objects, concluding
with the \textit{Geometrization Theorem}, i.e., Theorem
\ref{deligne}. Next, the statement of Theorem \ref{GH4} is reduced
to a geometric statement, the \textit{Vanishing Lemma}, i.e.,
Lemma \ref{vanishing}. The remainder of the chapter is devoted to
the proof of Lemma \ref{vanishing}. For the convenience of the
reader we include a large body of intuitive explanations for all
the constructions involved. In particular, we devote some space
explaining the Grothendieck's {\it Sheaf to Function
Correspondence} procedure which is the basic bridge connecting
sections \ref{QHUE} and \ref{theproof}.
\\
\\
\textbf{Part III.} Chapter \textbf{\ref{metaplectique}}. In
section \ref{canonical_hilbert} we describe the \textit{method of
canonical Hilbert space}. In section \ref{weilrep} we describe the
Weil representation in this manifestation. In section
\ref{realization} we relate the invariant construction to the more
classical constructions, supplying explicit formulas that will be
used later. In section \ref{delignes_letter} we give a formal
presentation of \textit{Deligne's letter to Kazhdan} \cite{D1}.
The main statement of this section is Theorem \ref{main_thm}, in
which the \textit{Weil representation sheaf} $\SK$ is introduced.
We include in our presentation only the parts of that letter which
are most relevant to our needs. In particular, we consider only
the two dimensional case of this letter. In section
\ref{app_proofs} we supply proofs for all technical lemmas and
propositions appearing in the previous sections of the chapter.
\\
\\
\textbf{Part IV.} Chapter \textbf{\ref{2d}}. Here we present the
formal construction of the two dimensional Hannay-Berry model that
were used in the previous chapters.
\\
\\
\textbf{Part V.} Appendices \textbf{\ref{hd}} and
\textbf{\ref{counter}}. In Appendix \ref{hd} we give the
construction of the Hannay-Berry model for the higher-dimensional
tori. In Appendix \ref{counter} we give the \textit{example} of a
symplectic automorphism $A \in \GZZ$ which acts ergodically on the
higher dimensional torus $\T$, but is represented by a quantum
operator $\rhoh(A)$ which is not Hecke ergodic.
\\
\\
\textbf{Part VI.} Appendix \textbf{\ref{proofs}}. In this Appendix
we supply the proofs for all statements appearing in Part I and
Part II. In particular, we give the \textit{proof} of the
\textit{Geometrization Theorem} (Theorem \ref{deligne}) which
essentially consists of taking the \textit{Trace} of Deligne's
{\it Weil representation sheaf} $\SK$.

\numberwithin{equation}{section}
\begin{chapter}{The Hannay-Berry Model}\label{BHm}
\section{Classical Torus} \label{classicaltorus}
Let $(\T,\ome)$ be the two dimensional symplectic torus. Together
with its linear symplectomorphisms $\G \iso \GZ$ it serves as a
simple model of classical mechanics (a compact version of the
phase space of the harmonic oscillator). More precisely, let $\T=
\W/ \Lambda$ where $\W$ is a two dimensional real vector space,
i.e.,  $\W \simeq \R^2$ and $\Lambda$ is a rank two lattice in
$\W$, i.e., $\Lambda \simeq \Z^2$. We obtain the symplectic form
on $\T$ by taking a non-degenerate symplectic form on $\W$:
$$\ome: \W \times \W \lto \R.$$
We require $\ome$ to be integral, namely $\ome : \Lambda \times
\Lambda \lto
\Z$ and normalized, i.e., Vol$(\T) =1$.\\\\
Let ${\mathrm{Sp}}(\W, \ome)$ be the group of linear
symplectomorphisms, i.e., $\mathrm{Sp}(\W,\ome) \simeq
\mathrm{SL}_2(\R)$. Consider the subgroup $\G \subset
\mathrm{Sp}(\W,\ome)$ of elements that preserve the lattice
$\Lambda$, i.e., $\G (\Lambda) \subseteq \Lambda $. Then $\G
\simeq \GZ$. The
subgroup $\G$ is the group of linear symplectomorphisms of $\T$.\\
We denote by $\Lm \subseteq \W^*$ the dual lattice, $\Lm = \{ \xi
\in \W^* | \r\r \xi (\Lambda) \subset \Z \} $. The lattice $\Lm$
is identified with the lattice of characters of $\T$ by the
following map:
\begin{equation*}
  \xi \in \Lm \longmapsto e^{2 \pi i <\xi, \cdot>} \in \r \Td
\end{equation*}
where $\Td := \Hom(\T,\C^*)$.
\subsection{Classical mechanical system}
We consider a very simple discrete mechanical system. An
hyperbolic element $A \in \G$, i.e., $|\Tr(A)| > 2$, generates an
ergodic discrete dynamical system. The {\it Birkhoff's Ergodic
Theorem} states that:
\begin{equation*} 
  \lim_{N \rightarrow \infty} \frac{1}{N}\sum_{k=1}^{N} f(A^k
  x)= \int_{\T}f \ome
\end{equation*}
for every $f \in \FST$ and for almost every point $x \in \T$. Here
$\FST$ stands for a good class of functions, for example
trigonometric polynomials or smooth functions.\\\\
We fix an hyperbolic element $A \in \G$ for the remainder of the
paper.
\section{Quantization of the Torus} \label{quantization}
Quantization is one of the big mysteries of modern mathematics,
indeed it is not clear at all what is the precise structure which
underlies quantization in general. Although physicists have been
using quantization for almost a century, for mathematicians the
concept remains all-together unclear. Yet, in specific cases,
there are certain formal models for quantization that are well
justified mathematically. The case of the symplectic torus is one
of these cases. Before we employ the formal model, it is
worthwhile to discuss the general phenomenological
principles of quantization which are surely common for all models.\\\\
Let us start with a model of classical mechanics, namely a
symplectic manifold, serving as a classical phase space. In our
case this manifold is the symplectic torus $\T$. Principally,
quantization is a protocol by which one associates a quantum
"phase" space $\H$ to the classical phase space $\T$, where $\H$
is a Hilbert space. In addition, the protocol gives a rule by
which one associates to every classical observable, namely a
function $f \in \FST$, a quantum observable ${\mathrm{Op}}(f) : \H
\lto \H$, an operator on the Hilbert space. This rule should send
a
real function into a self adjoint operator.\\\\
To be more precise, quantization should be considered not as a
single protocol, but as a one parameter family of protocols,
parameterized by $\hb$, the Planck constant. For every fixed value
of the parameter $\hb$ there is a protocol which associates to
$\T$ a Hilbert space $\H_{\hb}$ and for every function $f \in
\FST$ an operator ${\mathrm{Op}_{\hb}}(f) : \H_{\hb} \lto \H_{\hb}
$. Again the association rule should send real
functions to self adjoint operators.\\\\
Accepting the general principles of quantization, one searches for
a formal model by which to quantize, that is a mathematical model
which will manufacture a family of Hilbert spaces $\H_{\hb}$ and
association rules $\FST \leadsto \End(\H_{\hb})$. In this work we
employ a model of quantization called the {\it Weyl Quantization
model}.
\subsection {The Weyl quantization model}
The Weyl quantization model works as follows. Let $\Ad$ be a one
parameter deformation of the algebra $\A$ of trigonometric
polynomials on the torus. This algebra is known in the literature
as the Rieffel torus \cite{Ri}. The algebra $\Ad $ is constructed
by taking the free algebra over $\C$ generated by the symbols
$\{s(\xi) \r | \r \xi \in \Lm \}$ and quotient out by the relation
$ s(\xi + \eta) = e^{\pih \ome(\xi,\eta)}s(\xi)s(\eta)$. We point
out two facts about the algebra $\Ad$. First, when substituting
$\hb = 0$ one gets the group algebra of $\Lm$, which is exactly
equal to the algebra of trigonometric polynomials on the torus.
Second, the algebra $\Ad$ contains as a standard basis the lattice
$\Lm$:
\begin{equation*}                
s: \Lm \lto \Ad.
\end{equation*}
Therefore one can identify the algebras $\Ad \simeq \A$ as vector
spaces. Therefore, every function $f \in \A$ can be viewed as an
element of
$\Ad$.\\\\
For a fixed $\hbar$ a representation $\Pih:\Ad \lto \End(\Hh)$
serves as a quantization protocol, namely for every function $f
\in \A$ one has:
\begin{equation*}\label{quantprotocol}
 f \in {\A}  \simeq {\Ad}  \longmapsto \Pih (f) \in \End(\Hh)\def\ornlag
 {\cal{L}ag^{\circ}}.
\end{equation*}
An equivalent way of saying this is:
\begin{equation*}\label{quantprot2}
  f \longmapsto \sum_{\xi \in \Lm} a_{\xi} \Pih (\xi)
\end{equation*}
where $f = \sum\limits_{\xi \in \Lm} a_{\xi} \cdot \xi $ is the Fourier expansion of $f$.\\\\
To summarize: every family of representations $\Pih : \Ad \lto
\End(\Hh)$ gives us a complete quantization protocol. Yet, a
serious question now arises, namely what representations to
choose? Is there a correct choice of representations, both
mathematically, but also perhaps physically? A possible
restriction on the choice is to choose an irreducible
representation. Yet, some ambiguity still remains because there
are several irreducible classes for specific values of $\hbar$.\\\\
We present here a partial solution to this problem in the case
where the parameter $\hbar$ is restricted to take only rational
values (see Chapter \ref{2d} for the construction in this
generality). Even more particularly, for our purpose we will take
$\hbar$ to be of the form $\hbar = \frac{1}{p}$ where $p$ is an
odd prime number. Before any formal discussion one should recall
that our classical object is the symplectic torus $\T$
\textit{together} with its linear symplectomorphisms $\G$. We
would like to quantize not only the observables $\A$, but also the
symmetries $\G$. Next, we are going to construct an equivariant
quantization of $\T$.
\subsection {Equivariant Weyl quantization of the torus} \label{weilquant}
Let $\hbar = \frac{1}{p}$ and consider a non-trivial additive
character $\psi :\Fp \lto \C^*$. We give here a slightly different
presentation of the algebra $\Ad$. Let $\Ad$ be the free
$\C$-algebra generated by the symbols $\{s(\xi) \;|\; \xi \in \Lm
\}$ and the relations $s(\xi + \eta) = \psi( \half \ome(\xi,\eta)
) s(\xi)s(\eta)$. Here we consider $\ome$ as a map $\ome : \Lm
\times \Lm \lto \Fp$. The lattice $\Lm$ serves as a standard basis
for $\Ad$:
\begin{equation*}                      
s: \Lm \lto \Ad.
\end{equation*}
The group $\G$ acts on the lattice $\Lm$, therefore it acts on
$\Ad$. It is easy to see that $\G$ acts on $\Ad$ by homomorphisms
of algebras. For an element $B \in \G$, we denote by $ f
\longmapsto f^B$ the action of $B$ on an element $f \in \Ad$.\\\\
An equivariant quantization of the torus is a pair:
\begin{eqnarray*}          
\Pih   :  {\Ad} & \lto & \End(\Hh),
\\
\rhoh  :  \G & \lto & \PGL(\Hh)
\end{eqnarray*}
where $\Pih$ is a representation of $\Ad$ and $\rhoh$ is a
projective representation of $\G$. These two should be compatible
in the following manner:
\begin{equation}\label{eqvquant3}
  \rhoh(B) \Pih(f) \rhoh(B)^{-1} = \Pih(f^B)
\end{equation}
for every $B \in \G$ and $f \in \Ad$. Equation (\ref{eqvquant3})  is called the {\it Egorov identity}.\\\\
Let us suggest now a construction of an equivariant quantization
of
the torus.\\\\
Given a representation $\pi: \Ad \lto \End(\H)$ and an element $B
\in \G$, we construct a new representation $\pi^B:\Ad \lto
\End(\H)$:
\begin{equation}\label{actcat}
  \pi^B(f) := \pi(f^B).
\end{equation}
This gives an action of $\G$ on the set $\Irr$ of classes of
irreducible representations. The set $\Irr$ has a very regular
structure as a principal homogeneous space over $\T$. Moreover,
every irreducible representation of $\Ad$ is finite dimensional
and of dimension $p$. The following theorem (see Chapter \ref{2d}
for the proof) plays a central role in the construction.
\begin{theorem}[Canonical invariant representation]\label{gh}
Let $\hbar = \frac{1}{p},$ where $p$ is a prime. There exists a
\textit{unique} $($up to isomorphism$)$ irreducible representation
$(\Pih,\Hh)$ of $\Ad$ for which its equivalence class is fixed by
$\G$.
\end{theorem}
Let $(\Pih,\Hh)$ be a representative of the fixed irreducible
equivalence class. Then for every $B \in \G$ we have:
\begin{equation}\label{iso}
\Pih^{B} \simeq \Pih.
\end{equation}
This means that for every element $B \in \G$ there exists an
operator $\rhoh(B)$ acting on $\Hh$ which realizes the isomorphism
(\ref{iso}). The collection $\{\rhoh(B) : B \in \G \}$ constitutes
a projective representation\footnote{This is the famous Weil
  representation (cf. \cite{W2}, Chapter \ref{2d} and Appendix \ref{hd}) of $\GZ$.}:
\begin{equation}\label{projrep}
  \rhoh:\G \lto \PGL(\Hh).
\end{equation}
Equations (\ref{actcat}) and (\ref{iso}) also imply the  Egorov
identity (\ref{eqvquant3}).\\\\
The group $\G \simeq \GZ$ is almost a free group and it is
finitely presented. A brief analysis (Chapter \ref{2d}) shows that
every projective representation of $\G$ can be lifted (linearized)
into a true representation. More precisely, it can be linearized
in 12 different ways, where 12 is the number of characters of
$\G$. In particular, the projective representation (\ref{projrep})
can be linearized (\textit{not} uniquely) into an honest
representation. The next theorem asserts the existence of a
canonical linearization. Let $\Gf \iso \GG$ denotes the quotient
group of $\G$ modulo $p$.
\begin{theorem}[Canonical linearization]\label{GH2} Let $\hbar =
\frac{1}{p}$, where $p$ is an odd prime. There exists a
\textit{unique} linearization:
\begin{equation*}
\rhoh:\G \lto \GL(\Hh)
\end{equation*}
characterized by the property that it factors through the quotient
group $\Gf$:
\[
\qtriangle[\G`\Gf`\GL(\Hh);`\rhoh`\bar{\rho}_{_\h}]
\]
\end{theorem}
From now on $\rhoh$ means the linearization of Theorem \ref{GH2}.
\\
\\
\textbf{Summary.} Theorem \ref{gh} confirms the existence of a
unique invariant representation of $\Ad$, for every $\hbar =
\frac{1}{p}$. This gives a canonical equivariant quantization
$(\Pih,\rhoh,\Hh)$. Moreover, for $p$ odd, by Theorem \ref{GH2},
the projective representation $\rhoh$ can be linearized in a
canonical way to give an honest representation of $\G$ which
factors through $\Gf$\footnote{This is the Weil representation of
$\GG$.}. Altogether this gives a pair:
\begin{eqnarray*}
\Pih  :  {\Ad} & \lto & \End(\Hh),
\\
\rhoh  :  \Gf & \lto & \GL(\Hh)
\end{eqnarray*}
satisfying the following compatibility condition (Egorov
identity):
\begin{equation*}
  \rhoh(B) \Pih(f) \rhoh(B)^{-1} = \Pih(f^B)
\end{equation*}
for every $B \in \Gf$, $f \in \Ad$. The notation $\Pih(f^B)$ means
that we take any pre-image $\bar{B} \in \G$ of $B \in \Gf$ and act
by it on $f$, but the operator $\Pih(f^{\bar{B}})$ does not depend
on the choice of $\bar{B}$. In the following, we denote the Weil
representation $\bar{\rho}_{_\h}$ by $\rhoh$ and consider $\Gf$ to
be the default domain.
\subsection{Quantum mechanical system}
Let $(\Pih,\rhoh,\Hh)$ be the canonical equivariant quantization.
Let $A$ be our fixed hyperbolic element, considered as an element
of $\Gf$ . The element $A$ generates a quantum dynamical system.
For every (pure) quantum state $v \in S(\Hh) = \{ v \in \Hh
:\|v\|=1\}$:
\begin{equation}\label{Qsystem}
  v \longmapsto v^A := \rhoh(A)v.
\end{equation}
\end{chapter}

\chapter{The Kurlberg-Rudnick Conjecture}\label{Proof}
\section{Hecke Quantum Unique Ergodicity}\label{QHUE}
The main silent question of the current work is whether the system
(\ref{Qsystem}) is quantum ergodic. Before discussing this
question, one is obliged to define a notion of quantum ergodicity.
As a first approximation follow the classical definition, but
replace each classical notion by its quantum counterpart. Namely,
for every  $f \in \Ad$ and almost every quantum state $v \in
S(\Hh)$, the following holds:
\begin{equation}\label{quantumergodicity}
 \lim_{N \rightarrow \infty} \frac{1}{N}\sum_{k=1}^{N}
  < v|\Pih(f^{A^k})v > \above{?}= \int_{\T}f \ome.
\end{equation}
Unfortunately (\ref{quantumergodicity}) is literally not true. The
limit is never exactly equal to the integral for a fixed $\hbar$.
Let us now give a true statement which is a slight modification of
(\ref{quantumergodicity}), called the {\it Hecke Quantum Unique
Ergodicity}. First, rewrite (\ref{quantumergodicity}) in an
equivalent form. We have:
\begin{equation}\label{rw1}
  <v|\Pih(f^{A^k})v> = <v|\rhoh(A^k) \Pih(f) \rhoh(A^k)^{-1} v>
\end{equation}
using the Egorov identity (\ref{eqvquant3}).\\\\
Now, note that the elements $A^k$ run inside the finite group
$\Gf$. Denote by $\GrA \subseteq \Gf$ the cyclic subgroup
generated by $A$. It is easy to see, using (\ref{rw1}), that:
\begin{equation*}
\lim_{N \rightarrow \infty} \frac{1}{N}\sum_{k=1}^{N}
<v|\Pih(f^{A^k})v>= \frac{1}{|\GrA|}\sum_{B \in \GrA} <v|\rhoh(B)
\Pih(f) \rhoh(B)^{-1} v>.
\end{equation*}
Altogether (\ref{quantumergodicity}) can be written in the form:
\begin{equation}\label{rw3}
\Av_{_{\GrA}} (<v|\Pih(f)v>) \above{?}= \int_{\T}f \ome
\end{equation}
where $\Av_{_{\GrA}}$ denotes the average of $ <v|\Pih(f)v>$ with
respect to the group $\GrA$.
\subsection{Formulation of the Kurlberg-Rudnick conjecture} Denote
by $\CA$ the centralizer of $A$ in $\Gf \iso \GG$. The finite
group $\CA$ is an algebraic group. More particulary, as an
algebraic group, it is a torus. We call $\CA$ the {\it Hecke
torus} (cf. \cite{KR1}). One has, $\GrA \subseteq \CA \subseteq
\Gf$. Now, in (\ref{rw3}) take the average with respect to the
group $\CA$ instead of the group $\GrA$. The precise statement of
the \textbf{Kurlberg-Rudnick rate conjecture} (cf. \cite{R1, R2})
is given in the following theorem:
\begin{theorem}[Hecke Quantum Unique Ergodicity]\label{pr} Let $\hbar =
\frac{1}{p}, \r p$ an odd prime. For every $f \in \Ad$ and $v \in
S(\Hh)$, we have:
\begin{equation}\label{qheckerg}
\left| \Av_{_{\CA}} ( <v| \Pih(f) v> ) - \int_{\T}f \ome \right|
\leq \frac{C_{f}}{\sqrt{p}},
\end{equation}
where $C_{f}$ is an explicit constant depending only on $f$.
\end{theorem}
The next section is devoted to proving Theorem \ref{pr}.
\section{Proof of the Kurlberg-Rudnick conjecture}\label{theproof} The proof is
given in two stages. The first stage is a preparation stage and
consists mainly of linear algebra considerations. We massage
statement (\ref{qheckerg}) in several steps into an equivalent
statement which will be better suited to our needs. In the second
stage we introduce the main part of the proof. Here we invoke
tools from algebraic geometry in the framework of $\ell$-adic
sheaves and $\ell$-adic cohomology (cf. \cite{M, BBD}).
\subsection{Preparation stage}
\textbf{Step 1.} It is enough to prove Theorem \ref{pr} for the
case when $f$ is a non-trivial character, $\xi \in \Lm$. Because
$\int_{\T}\xi \ome = 0$, statement (\ref{qheckerg}) becomes :
\begin{equation}\label{qheckerg2}
\left| \Av_{_{\CA}} ( <v|\Pih(\xi) v> )\right| \leq
\frac{C_\xi}{\sqrt{p}}.
\end{equation}
The statement for general $f \in \Ad$ follows directly from the
triangle inequality.
\\\\
\textbf{Step 2.} It is enough to prove (\ref{qheckerg2}) in case
$v \in S(\Hh)$ is a {\it Hecke} eigenvector. To be more precise,
the {\it Hecke} torus $\CA$ acts semisimply on $\Hh$ via the
representation $\rhoh$, thus  $\Hh$ decomposes to a direct sum of
character spaces:
\begin{equation}\label{decom}
\Hh = \bigoplus_{\chi:\CA \lto \C^*}\H_{\chi}.
\end{equation}
The sum in (\ref{decom}) is over multiplicative characters of the
torus $\CA$. For every $v \in \H_{\chi}$ and $B \in \CA$, we have:
\begin{equation*}
\rhoh(B)v = \chi(B)v.
\end{equation*}
Taking $v \in \H_{\chi}$, statement (\ref{qheckerg2}) becomes:
\begin{equation}\label{qheckerg3}
\left|<v|\Pih(\xi)v>\right| \leq \frac{C_\xi}{\sqrt{p}}.
\end{equation}
Here $C_\xi = 2$.\\\\
The  averaged operator:
\begin{equation*}
\Av_{_{\CA}} (\Pih(\xi)) := \frac{1}{|\CA|}\sum_{B \in \CA}
\rhoh(B) \Pih(\xi) \rhoh(B)^{-1}
\end{equation*}
is essentially\footnote{This follows from Remark \ref{lemma1}. If
$\CA$ does not split over $\Fp$ then $\Av_{_{\CA}}(\Pih(\xi))$ is
diagonal in the Hecke basis. In case $\CA$ splits then for the
Legendre character $\sigma$ we have that dim $\H_\sigma = 2$.
However, in the later case one can prove (\ref{qheckerg2}) for
$v\in \H_\sigma$ by a computation of explicit eigenvectors (cf.
\cite{KR2}).} diagonal in the Hecke base. Knowing this, statement
(\ref{qheckerg2}) follows from (\ref{qheckerg3}) by invoking the
triangle inequality.\\\\
\textbf{Step 3.} Let $P_{\chi}:\Hh \lto \Hh$ be the orthogonal
projector on the eigenspace $\H_{\chi}$.
\begin{remark}\label{lemma1}
For $\chi$ other then the quadratic character of $\;\CA$ we have
$\dim\;\H_{\chi} = 1.$\footnote{This fact, which is needed if we
want to stick with the matrix coefficient formulation of the
conjecture, can be proven by algebro-geometric techniques or
alternatively by a direct computation (cf. \cite{Ge}).}
\end{remark}
Using Remark \ref{lemma1} we can rewrite (\ref{qheckerg3}) in the
form:
\begin{equation*}
\left|\Tr(P_{\chi} \Pih(\xi)) \right| \leq \frac{2}{\sqrt{p}}.
\end{equation*}
The projector $P_{\chi}$ can be defined in terms of the
representation $\rhoh$:
\begin{equation*}               
P_{\chi} = \frac{1}{|\CA|}\sum_{B \in \CA} \chi(B) \rhoh(B).
\end{equation*}
Now write (\ref{qheckerg3}):
\begin{equation}\label{qheckerg4}
\frac{1}{|\CA|} \left| \sum_{B \in \CA} \Tr( \rhoh(B) \Pih(\xi))
\chi(B) \right| \leq \frac{2}{\sqrt{p}}.
\end{equation}
On noting that $|\CA| = p \pm1$ and multiplying both sides of
(\ref{qheckerg4}) by $|\CA|$ we obtain the following statement:\\
\begin{theorem}[Hecke Quantum Unique Ergodicity (Restated)]\label{GH4}
Let $\hbar = \frac{1}{p}$, where $p$ is an odd prime. For every
$\xi \in \Lm$ and every character $\chi$ the following holds:
\begin{equation*}
\left| \sum_{B \in \CA} \Tr( \rhoh(B) \Pih(\xi)) \chi(B) \right|
\leq 2 \sqrt{p}.
\end{equation*}
\end{theorem}
We prove the Hecke ergodicity theorem in the form of Theorem
\ref{GH4}.
\subsection{The trace function} We prove Theorem \ref{GH4} using
Sheaf theoretic techniques. Before diving into geometric
considerations, we investigate further the ingredients appearing
in Theorem \ref{GH4}. Denote by $F$ the function $F : \Gf \times
\Lm \lto \C$ defined by $F(B,\xi) = \Tr(\rho(B) \Pih(\xi))$. We
denote by $\V := \Lm / p \Lm$ the quotient vector space, i.e., $\V
\simeq \Fp^2$. The symplectic form $\ome$ specializes to give a
symplectic form on $\V$. The group $\Gf$ is the group of linear
symplectomorphisms of $\V$, i.e., $\Gf = \Sp(\V)$. Set $\Y := \Gf
\times \Lm$ and $\YY := \Gf \times \V$. One has the quotient map:
\begin{equation*}
\Y \lto \YY.
\end{equation*}
\begin{lemma}\label{factorization}
The function $F : \Y \lto \C$ factors through the quotient $\YY$.
\[
\qtriangle[\Y`\YY`\C;`F`\overline{F}]
\]
\end{lemma}
Denote the function $\overline{F}$ also by $F$ and from now on
$\YY$ will be considered as the default domain. The function
$F:\YY \lto \C$ is invariant under a certain group action of
$\Gf$. To be more precise let $S \in \Gf$. Then:
\begin{equation*}
\Tr(\rhoh(B) \Pih(\xi)) = \Tr(\rhoh(S) \rhoh(B) \rhoh(S)^{-1}
\rhoh(S) \Pih(\xi) \rhoh(S)^{-1}).
\end{equation*}
Applying the Egorov identity (\ref{eqvquant3}) and using the fact
that $\rhoh$ is a representation we get:
\begin{equation*}
\Tr( \rhoh(S) \rhoh(B) \rhoh(S)^{-1}  \rhoh(S) \Pih(\xi)
\rhoh(S)^{-1} )= \Tr(\Pih(S \xi) \rhoh(S B S^{-1})).
\end{equation*}
Altogether we have:
\begin{equation}\label{invar3}
F(B, \xi) = F(SBS^{-1} , S \xi).
\end{equation}
%
%
Putting (\ref{invar3}) in a more diagrammatic form: there is an
action of $\Gf$ on $\YY$ given by the following formula:
\begin{equation}\label{actionset}
  \begin{CD}
   \Gf \times \YY    @>\alpha>> \YY,\\
   (S,(B, \xi))      @>>>  (SBS^{-1} , S \xi).
   \end{CD}
\end{equation}
Consider the following diagram:
\begin{equation*}
  \begin{CD}
   \YY    @<pr<<    \Gf\times \YY    @>\alpha>> \YY \\
  \end{CD}
\end{equation*}
where $pr$ is the projection on the $\YY$ variable. Formula
(\ref{invar3}) can be stated equivalently as:
\begin{equation*}
  \alpha^{*}(F) = pr^{*}(F)
\end{equation*}
where $\alpha^{*}(F)$ and $pr^{*}(F)$ are the pullbacks of the
function $F$ on $\YY$ via the maps $\alpha$ and $pr$ respectively.
\subsection{Geometrization (Sheafification)} Next, we will phrase a
geometric statement that will imply Theorem \ref{GH4}. Moving into
the geometric setting, we replace the set $\YY$ by an algebraic
variety and the functions $F$ and $\CHI$ by sheaf theoretic objects,
also of a geometric flavor.\\\\
\textbf{Step 1.}  The set $\YY$ is not an arbitrary finite set but
it is the set of rational points of an  algebraic variety $\AYY$
defined over $\Fp$. To be more precise, $\AYY \simeq \ASp \times
\AV$. The variety $\AYY$ is equipped with an endomorphism:
\begin{equation*}   
  \Fr:\AYY \lto \AYY
\end{equation*}
called Frobenius. The set $\YY$ is identified with the set of
fixed points of Frobenius:
\begin{equation*}
  \YY = \AYY^{\Fr} = \{ y \in \AYY : \Fr(y) = y \}.
\end{equation*}
Note that the finite group $\Gf$ is the set of rational points of
the  algebraic group $\ASp$. The vector space $\V$ is the set of
rational points of the variety $\AV$, where $\AV$ is isomorphic to
the affine plane $\bA^2$. We denote by $\alpha$ the action of
$\ASp$ on the variety $\AYY$ (cf. (\ref{actionset})).
\\
\\
Having all finite sets replaced by corresponding algebraic
varieties, we replace functions by sheaf theoretic
objects as shown.\\\\
%
%
%
%
%
\textbf{Step 2.} The following theorem proposes an appropriate
sheaf theoretic object standing in place of the function $F : \YY
\lto \C$. Denote by $\Db (\AYY)$  the bounded derived category of
constructible $\ell$-adic Weil sheaves on $\AYY$ (cf. \cite{M,
BBD}, in addition see \cite{BL} for equivariant sheaves theory).
\begin{theorem}[Geometrization Theorem]\label{deligne}
There exists an object $\SF \in \Db(\AYY) $ satisfying the
following properties:
\end{theorem}
%
%
\begin{enumerate}
\item \label{prop_del1}$($Function$)$ It  is associated, via the \textit{sheaf-to-function correspondence}, to the
function $F : \YY \lto \C$:
\begin{equation*}
  f^{\SF} = F.
\end{equation*}
\item \label{prop_del2} $($Weight$)$ It is of weight: $$w(\SF) \leq
0.$$
\item \label{prop_del3} $($Equivariance$)$  For every element $S \in
  \ASp$ there exists an isomorphism $$\alpha_S ^* \SF \simeq
  \SF.$$
\item \label{prop_del4} $($Formula$)$ On introducing coordinates $\AV
  \simeq \bA^2$ we identify $\ASp \simeq \AGG$. Then there exists an
  isomorphism:
\begin{equation*}
\SF_{|_{\AT \times \AV}} \simeq \SL_{\psi(\half
\lambda\mu\frac{a+1}{a-1})} \otimes \SL_{\lgn(a)}.\footnote{By
this we mean that $\SF_{|_{\AT \times \AV}}$ is isomorphic to the
extension of the sheaf defined by the formula in the right-hand
side.}
\end{equation*}
Here $\AT := \{ \left( \begin{smallmatrix} a & 0 \\ 0 & a^{-1}
\end{smallmatrix} \right) \}$ stands for the standard torus and
$(\lambda,\mu)$ are the coordinates on $\AV$.
\end{enumerate}
%
%
We give here an \textit{intuitive} explanation of Theorem
{\ref{deligne}, part by part, as it was stated. An object $\SF \in
\Db(\AYY)$ can be considered as a vector bundle $\SF$ over $\AYY$:
\begin{equation*}
  \begin{CD}
  \SF \\
  @VVV \\
  \AYY
\end{CD}
\end{equation*}
The letter  ``w''  in the notation $\Db$ means that $\SF$ is a
\textit{Weil sheaf}, i.e., it is equipped with a lifting of the
Frobenius:
\begin{equation*}
  \begin{CD}
   \SF  @>\Fr>>  \SF \\
   @VVV        @VVV \\
   \AYY @>\Fr>>  \AYY
  \end{CD}
\end{equation*}
To be even more precise, think of $\SF$ not as a single vector
bundle, but as a complex $\SF = \SF^{\bullet}$ of vector bundles
over $\AYY$:
\begin{equation*}
  \begin{CD}
   ...  @>d>> \SF^{-1}  @>d>> \SF^{0}  @>d>> \SF^{1}  @>d>> ...
   \end{CD}
\end{equation*}
The complex $\SF^{\bullet}$ is equipped with a lifting of
Frobenius:
\begin{equation*}
  \begin{CD}
   ...  @>d>> \SF^{-1}  @>d>> \SF^{0}  @>d>> \SF^{1}  @>d>> ... \\
    &    &      @V\Fr VV         @V\Fr VV           @V\Fr VV            \\
   ...  @>d>> \SF^{-1}  @>d>> \SF^{0}  @>d>> \SF^{1}  @>d>> ...
   \end{CD}
\end{equation*}
Here the Frobenius commutes with the differentials.\\\\
Next, we explain the meaning of property \ref{prop_del2}, i.e.,
the \textit{statement} $w(\SF) \leq 0$. Let $y \in \AYY^{\Fr} =
\YY$ be a fixed point of Frobenius. Denote by $\SF_{y}$ the fiber
of $\SF$ at the point $y$. Thinking of $\SF$ as a complex of
vector bundles, it is clear what one means by taking the fiber at
a point. The fiber $\SF_{y}$ is just a complex of vector spaces.
Because the point $y$ is fixed by the Frobenius, it induces an
endomorphism of $\SF_{y}$:
\begin{equation}\label{liftingcompy}
  \begin{CD}
   ...  @>d>> \SF^{-1}_{y}  @>d>> \SF^{0}_{y}  @>d>> \SF^{1}_{y}  @>d>> ... \\
    &    &      @V\Fr VV         @V\Fr VV           @V\Fr VV            \\
   ...  @>d>> \SF^{-1}_{y}  @>d>> \SF^{0}_{y}  @>d>> \SF^{1}_{y}  @>d>> ...
   \end{CD}
\end{equation}
The Frobenius acting as in (\ref{liftingcompy}) commutes with the
differentials. Hence, it induces an action on cohomologies. For
every $i \in \Z $ we have an endomorphism:
\begin{equation}\label{actcohom}
  \Fr:\coH^{i}(\SF_{y}) \lto \coH^{i}(\SF_{y}).
\end{equation}
Saying that an object $\SF$ has $w(\SF) \leq w$ means that for
every point $y \in \AYY^{\Fr}$ and for every $i \in \Z$ the
absolute value of the eigenvalues of Frobenius acting on the
$i$'th cohomology (\ref{actcohom}) satisfy:
\begin{equation*}
 \left |\ev(\Fr \big{|}_{\coH^{i}(\SF_{y})}) \right | \leq
 \sqrt{p}^{w+i}.
\end{equation*}
In our case $w = 0$ and therefore:
\begin{equation}\label{weightdefinition}
 \left |\ev(\Fr \big{|}_{\coH^{i}(\SF_{y})}) \right | \leq
 \sqrt{p}^{\; i}.
\end{equation}
Property \ref{prop_del1} of Theorem \ref{deligne} concerns a
function $f^{\SF}: \YY \lto \C$ associated to the sheaf $\SF$. To
define $f^{\SF}$, we only have to describe its value at every
point $y \in \YY$. Let $y \in \YY = \AYY^{\Fr}$. Frobenius acts on
the cohomologies of the fiber $\SF_{y}$  (cf. (\ref{actcohom}) ).
Now put:
\begin{equation*}
  f^{\SF}(y) := \sum_{i \in \Z} (-1)^i \Tr(\Fr
  \big{|}_{\coH^i(\SF_y)}).
\end{equation*}
In words: $f^{\SF}(y)$ is the alternating sum of traces of
Frobenius acting on the cohomologies of the fiber $\SF_y$. This
alternating sum is called the {\it Euler characteristic} of
Frobenius and is denoted by:
\begin{equation*}
  f^{\SF}(y) = \chi_{_{\Fr}}(\SF_y).
\end{equation*}
Theorem \ref{deligne} confirms that $f^{\SF}$ is the function $F$
defined earlier. Associating the function $f^{\SF}$ on the set
$\AYY^{\Fr}$ to the sheaf $\SF$ on $\AYY$ is a particular case of
a general procedure called {\it Sheaf-to-Function Correspondence}
\cite{G}. As this procedure will be used later, next we spend some
space explaining it in greater details (cf.
\cite{Ga}).\\\\
\underline{\textbf{Grothendieck's Sheaf-to-Function
Correspondence}}
\\
\\
Let $\AX$ be an algebraic variety defined over $\Fq$. This means
that there exists a Frobenius endomorphism:
\begin{equation*}
  \Fr : \AX \lto \AX.
\end{equation*}
The set $ X = \AX^\Fr$ is called the set of rational points of
$\AX$. Let ${\cal L} \in \Db(\AX)$ be a Weil sheaf. One can
associate to ${\cal L}$ a function $f^{\cal L}$ on the set $X$ by
the following formula:
\begin{equation*}
  f^{\cal L}(x) := \sum_{i \in \Z} (-1)^i \Tr(\Fr \big{|}_{\coH^i({\cal
  L}_x)}).
\end{equation*}
This procedure is called {\it Sheaf-To-Function correspondence}.
Next, we describe some important functorial
properties of this procedure.\\\\
Let $\AX_1$, $\AX_2$ be algebraic varieties defined over $\Fq$.
Let $X_1 = \AX^{\Fr}_1$ and $X_2 = \AX^{\Fr}_2$ be the
corresponding sets of rational points. Let $\pi : \AX_1 \lto
\AX_2$ be a morphism of algebraic varieties. Denote also by
$ \pi : X_1 \lto X_2$ the induced map on the level of sets.\\\\
%
%
%
\textbf{First statement}. Suppose we have a sheaf ${\cal L} \in
\Db(\AX_2)$. The following holds:
\begin{equation}\label{sfc1}
 f^{\pi^{*}({\cal L})} = \pi^{*}(f^{\cal L})
\end{equation}
where on the function level $\pi^{*}$ is just the pull  back of
functions. On the sheaf theoretic level $\pi^{*}$ is the pull-back
functor of sheaves (think of pulling back a vector bundle).
Equation (\ref{sfc1}) states that the
{\it Sheaf-to-Function Correspondence} commutes with the operation of pull back.\\\\
%
%
\textbf{Second statement}. Suppose we have a sheaf ${\cal L} \in
\Db(\AX_1)$.  The following holds:
\begin{equation}\label{sfc2}
 f^{\pi_{!}({\cal L})} = \pi_{!}(f^{\cal L})
\end{equation}
where on the function level $\pi_{!}$ means to sum up the values
of the function along the fibers of the map $\pi$. On the sheaf
theoretic level $\pi_{!}$ is a compact integration of sheaves
(here we have no analogue under the vector bundle interpretation).
Equation (\ref{sfc2}) states that the {\it Sheaf-to-Function
Correspondence} commutes with integration.\\\\
%
%
\textbf{Third statement}. Suppose we have two sheaves ${\cal
L}_{1}, {\cal L}_{2} \in \Db(\AX_1)$. The following holds:
\begin{equation}\label{sfc3}
f^{{\cal L}_{1} \otimes {\cal L}_{2}} = f^{{\cal L}_{1}} \cdot
f^{{\cal L}_{2}}.
\end{equation}
In words: {\it Sheaf-to-Function Correspondence} takes tensor
product of sheaves to multiplication of the corresponding
functions.
\subsection{Geometric statement}
Fix an element $\xi \in \Lm$ with $\xi \neq 0$. We denote by
$\i_XI$ the inclusion map $\i_XI : \CA \times \xi \lto \YY $.
%
%
Going back to Theorem \ref{GH4} and putting its content in a
functorial notation, we write the following inequality:
\begin{equation*}
  \left| pr_! ( \i_XI ^* (F) \cdot \CHI) \right| \leq 2\sqrt {p}.
\end{equation*}
In words, taking the function $F : \YY \lto \C$ and:
\begin{itemize}
  \item Restrict $F$ to $\CA \times \xi $ and get $\i_XI^ * (F)$.
  \item Multiply $\i_XI^ * F $ by the character $\CHI$ to get
  $\i_XI^ * (F) \cdot \CHI$.
  \item Integrate $\i_XI^ * (F)  \cdot \CHI$ to the point, this means to sum up all
  its values, and get a scalar $a_{\CHI} := pr_! ( \i_XI^ * (F)  \cdot
\CHI)$. Here $pr$ stands for
  the projection $pr : \CA \times \xi \lto pt$.
\end{itemize}
Then Theorem \ref{GH4} asserts that the scalar $a_{\CHI}$ is of  an absolute value less than $2\sqrt{p}$.\\\\
%
%
Repeat the same steps in the geometric setting. We denote again by
$\i_XI$ the closed imbedding $\i_XI: \ACA \times \xi \lto \AYY $.
Take the sheaf $\SF$ on $\AYY$ and apply the following sequence of
operations:

\begin{itemize}
  \item Pull-back $\SF$ to the closed subvariety $ \ACA \times \xi $ and  get the sheaf $\i_XI^ * (\SF)$.

  \item Take the tensor product of $\i_XI^ * (\SF)$ with the Kummer
  sheaf ${\SL}_{\CHI}$  and get $\i_XI^ * (\SF) \otimes
  {\SL}_\CHI$.

  \item Integrate $\i_XI^ * (\SF) \otimes {\SL}_\CHI$ to the point
    and get the sheaf $pr_! (\i_XI^ * (\SF) \otimes {\SL}_\CHI)$ on the point.
\end{itemize}
The Kummer sheaf $\SL_\CHI$ is the {\it character sheaf} (cf.
\cite{Ga}) associated via {\it
Sheaf-to-Function Correspondence} to the character $\CHI$.\\\\
The operation of {\it Sheaf-to-Function Correspondence} commutes
both with pullback (\ref{sfc1}), with integration (\ref{sfc2}) and
takes the tensor product of sheaves to the multiplication of
functions (\ref{sfc3}). This means that it intertwines the
operations carried out on the level of sheaves with those carried
out on the level of functions. The following diagram describes
pictorially what has been said so far:
\begin{equation*}
\begin{CD}
   \SF                                        @>\chiFr >>              & F               \\
   @A\i_XIAA            &                                            @A \i_XI AA            \\
  \i_XI^{*}(\SF) \otimes {\SL}_{\CHI}      @>\chiFr >>    & \i_XI^{*}(F) \cdot \CHI    \\
   @V pr  VV           &                                             @V pr VV        \\
 {pr}_{!}(\i_XI^{*}(\SF) \otimes {\SL}_{\CHI})  @>\chiFr >>     &  {pr}_{!}(\i_XI^{*}(F)\cdot \CHI)
\end{CD}
\end{equation*}
%
%
Recall $w(\SF) \leq 0$. Now, the effect of functors $\i_XI^
*$, $pr_!$ and tensor product $\otimes$ on the property of weight should be examined.\\\\
%
%
The functor $\i_XI^{*}$ does not increase weight. Observing the
definition of weight this claim is immediate. Therefore we get:
\begin{equation*}
  w(\i_XI^ * (\SF)) \leq 0.
\end{equation*}
%
%
Assume we have two sheaves  ${\cal L}_1$ and  ${\cal L}_2$ with
weights $w({\cal L}_1) \leq w_{_1}$ and $w({\cal L}_{2}) \leq
w_{_2}$. The weight of the tensor product satisfies $w({\cal
L}_{1}\otimes {\cal L}_{2}) \leq w_{_1} + w_{_2}$. This is again immediate from the definition of weight.\\\\
%
%
Knowing that the Kummer sheaf has weight $w(\SL_{\CHI}) \leq 0$ we
deduce:
\begin{equation*}
  w(\i_XI^*(\SF) \otimes \SL_{\CHI}) \leq 0.
\end{equation*}
%
%
Finally, one has to understand the affect of the functor $pr_!$.
The following theorem, proposed by Deligne \cite{D2}, is a very
deep and important result in the theory of weights. Briefly
speaking, the theorem states that compact integration of sheaves
does not increase weight. Here is the precise statement:
\begin{theorem}[Deligne, Weil II \cite{D2}]\label{deligne2}
Let $\pi :\AX_1 \lto \AX_2$ be a morphism of algebraic varieties.
Let ${\cal L} \in \Db(\AX_1)$ be a sheaf of weight $w({\cal L})
\leq w$ then $w(\pi_! ({\cal L})) \leq w$.
\end{theorem}
Using Theorem \ref{deligne2} we get:
\begin{equation*}
 w(pr_! (\i_XI^*(\SF) \otimes \SL_{\CHI})) \leq 0.
\end{equation*}
Consider the sheaf $\SG := pr_! (\i_XI^*(\SF) \otimes
\SL_{\CHI})$. It is an object in $\Db(pt)$. This means it is
merely a complex of vector spaces, $\SG = \SG^{\bullet}$, together
with an action of Frobenius:
\begin{equation*}
  \begin{CD}
   ...  @>d>> \SG^{-1}  @>d>> \SG^{0}  @>d>> \SG^{1}  @>d>> ... \\
    &    &      @V\Fr VV         @V\Fr VV           @V\Fr VV            \\
   ...  @>d>> \SG^{-1}  @>d>> \SG^{0}  @>d>> \SG^{1}  @>d>> ...
   \end{CD}
\end{equation*}
The complex $\SG^{\bullet}$ is associated by {\it
Sheaf-To-Function correspondence} to the scalar $a_{\CHI}$:
\begin{equation}\label{geulerchar}
  a_{\CHI} = \sum_{i \in \Z} (-1)^i \Tr(\Fr \big{|}_{\coH^i(\SG)}).
\end{equation}
Finally, we can give the geometric statement  about $\SG$,  which
will imply Theorem \ref{GH4}.
\begin{lemma}[Vanishing Lemma]\label{vanishing}
Let $\SG = pr_! (\i_XI^*(\SF) \otimes \SL_{\CHI})$. All
cohomologies $\coH^{i}(\SG)$ vanish except for $i=1$. Moreover,
$\coH^1(\SG)$ is a two dimensional vector space.
\end{lemma}
Theorem \ref{GH4} now follows easily. By Lemma \ref{vanishing}
only the first cohomology $\coH^1(\SG)$ does not vanish and it is
two dimensional. Having $w(\SG) \leq 0$ implies (cf.
\ref{weightdefinition}) that the eigenvalues of Frobenius acting
on $\coH^1(\SG)$ are of absolute value $ \leq \sqrt{p}$. Hence,
using formula (\ref{geulerchar}) we get:
%
\begin{equation*}
  | a_{\CHI} | \leq 2 \sqrt{p}.
\end{equation*}
The remainder of the chapter is devoted to the proof of Lemma
\ref{vanishing}.
\subsection{Proof of the Vanishing Lemma}
The proof will be given in several steps.\\\\
\textbf{Step 1.} All tori in $\ASp$ are conjugated. On introducing
coordinates, i.e., $\AV \simeq \bA^2$, we make the identification
$\ASp \simeq \AGG$.  In these terms there exists an element $\S0
\in \AGG$ conjugating the {\it Hecke} torus $\ACA \subset \AGG$
with the standard torus
$\AT = \left \{ ( \begin{smallmatrix} a & 0 \\
0 & a^{-1} \end{smallmatrix} \right ) \} \subset \AGG $, namely:
\begin{equation*}
\S0 \ACA \S0^{-1} = \AT.
\end{equation*}
%
%
The situation is displayed in the following diagram:
\begin{equation*}
  \begin{CD}
    \AGG \times \bA^2                        @>\alpha_{_{\S0}}>>                          \AGG \times \bA^2 \\
      @A\i_XIAA                                   @                                             A \iXII AA \\
    \ACA \times \xi                        @>\alpha_{_{\S0}}>>                          \AT \times \XII \\
     @V pr VV                                                                              @V pr VV     \\
      pt                                  @=                                                   pt
 \end{CD}
\end{equation*}
where $\XII = \S0 \cdot \XI$ and  $\alpha_{_{\S0}}$ is the
restriction of the action $\alpha$ to the element $\S0$.
\\
\\
%
%
\textbf{Step 2.} Using the equivariance property of the sheaf
$\SF$ (see Theorem \ref{deligne}, property  \ref{prop_del3}) we
will show that it is \textit{sufficient} to prove the Vanishing
Lemma for the sheaf $\SG_{st} := pr_! (\iXII ^* \SF \otimes
{\alpha_{_\S0}}_! \Skummer )$.
\\
\\
We have:
\begin{equation}\label{seq1}
  pr_! ( \i_XI^* \SF  \otimes \Skummer ) \iso pr_! {\alpha_{_\S0}}_{_{!}} ( \i_XI^* \SF  \otimes
  \Skummer).
\end{equation}
The morphism $\alpha_{_\S0}$ is an isomorphism, therefore
${\alpha_{_\S0}}_!$ commutes with taking $\otimes$, hence we
obtain:
\begin{equation}\label{seq2}
 pr_! {\alpha_{_{\S0}}}_{_{!}} (\i_XI^*(\SF) \otimes \Skummer ) \iso pr_! ({\alpha_{_{\S0}}}_! (\i_XI^* \SF) \otimes
{\alpha_{_{\S0}}}_! (\Skummer)).
\end{equation}
Applying base change we obtain:
\begin{equation}\label{seq3}
 {\alpha_{_{\S0}}}_! \i_XI^* \SF  \iso \iXII^* {\alpha_{_{\S0}}}_!
 \SF.
\end{equation}
Now using the equivariance property of the sheaf $\SF$ we have the
isomorphism:
\begin{equation}\label{seq4}
   {\alpha_{_{\S0}}}_! \SF  \simeq \SF.
\end{equation}
Combining (\ref{seq1}), (\ref{seq2}), (\ref{seq3}) and
(\ref{seq4})  we get:
\begin{equation}\label{seq5}
   pr_!( \i_XI^* \SF  \otimes \Skummer ) \iso  pr_! (\iXII^* \SF  \otimes {\alpha_{_{\S0}}}_! \Skummer
   ).
\end{equation}
Therefore we see from (\ref{seq5}) that it is sufficient to prove
vanishing of cohomologies for:
\begin{equation}\label{seq6}
  \SG_{st} := pr_! (\iXII^* \SF  \otimes {\alpha_{_{\S0}}}_!
  \Skummer).
\end{equation}
But this is  a situation over the standard torus and we can
compute explicitly all the sheaves involved!
\\
\\
\textbf{Step 3.} The Vanishing Lemma holds for the sheaf $\SG_{st}$.\\\\
We are left to compute (\ref{seq6}). We write $ \XII =
(\lambda,\mu)$. By Theorem \ref{deligne} Property \ref{prop_del4}
we have $ \iXII^* \SF \simeq \SL_{\psi( \half \lambda\mu
\frac{a+1}{a-1})} \otimes \SL_{\lgn(a)} $, where $a$ is the
coordinate of the standard torus $\AT$ and $\lambda \cdot \mu \neq
0$\footnote{This is a direct
  consequence of the fact that $A \in \GZ$ is an hyperbolic element and
  does not have eigenvectors in $\Lm$.} . The sheaf $ {\alpha_{_{\S0}}}_!  \Skummer $
  is a character sheaf on the torus $\AT$. Hence we get that
(\ref{seq6}) is a kind of a Kloosterman-sum sheaf. A direct
computation (Appendix \ref{proofs} section \ref{compvanishing})
proves the Vanishing Lemma for this sheaf. This completes the
proof of the Hecke quantum unique ergodicity conjecture. $\EProof$

\begin{chapter}{Metaplectique}\label{metaplectique}
In the first part of this chapter we give new construction of the
{\it Weil} ({\it metaplectic}) {\it representation}
$(\rho,\mathrm{Sp}(\V),\Hc)$, attached to a two dimensional
symplectic vector space $(\V,\ome)$ over $\Fq$, which appears in
the body of the thesis. The difference is that here the
construction is slightly more general. But even more importantly,
it is obtained in completely natural geometric terms. The focal
step in our approach is the introduction of a \textit{canonical
Hilbert space} on which the Weil representation is naturally
manifested. The motivation to look for this space was initiated by
a question of David Kazhdan \cite{Ka}. The key idea behind this
construction was suggested to us by Joseph Bernstein \cite{B}. The
upshot is to replace the notion of a Lagrangian subspace by a more
refined notion of an {\it oriented Lagrangian subspace}
\footnote{We thank A. Polishchuk for pointing out to us that this
is an $\Fq$-analogue of well known considerations with usual
oriented Lagrangians giving explicitly the metaplectic covering of
$\mathrm{Sp}(2n,\R)$ (cf. \cite{LV}).}.
\\
\\
In the second part of this chapter we apply a
\textit{geometrization} procedure to the construction given in the
first part, meaning that all sets are replaced by algebraic
varieties and all functions are replaced by $\ell$-adic sheaves.
This part is based on a letter of Deligne to Kazhdan from 1982
\cite{D1}. We extract from that work only the part that is most
relevant to this thesis. Although all basic ideas appear already
in the letter, we tried to give here a slightly more general and
detailed account of the construction. As far as we know, the
contents of this mathematical work has never been published. This
might be a good enough reason for writing this part.
\\
\\
The following is a description of the chapter. In section
\ref{canonical_hilbert} we introduce the notion of oriented
Lagrangian subspace and the construction of the canonical Hilbert
space. In section \ref{weilrep} we obtain a natural realization of
the Weil representation. In section \ref{realization} we give the
standard Schr\"{o}dinger realization (cf. \cite{W2}). We also
include several formulas for the kernels of basic operators. These
formulas will be used in section \ref{delignes_letter} where the
geometrization procedure is described. In section \ref{app_proofs}
we give proofs of all lemmas and propositions which appear in
previous sections.
\\
\\
For the remainder of this chapter we fix the following notations.
Let $\Fq$ denote the finite field of characteristic $p \neq 2$ and
$q$ elements. Fix $\psi : \Fq \lto \C^*$ a non-trivial additive
character. Denote by $\lgn : \Fqm \lto \C^*$ the {\it Legendre}
multiplicative quadratic character.
\section{Canonical Hilbert space}\label{canonical_hilbert}
\subsection{Oriented Lagrangian subspace}
Let $(\V,\ome)$ be a $2$-dimensional symplectic vector space over
$\Fq$.
%
%
\begin{definition} \label{oriented_lagrangian_def}
An oriented Lagrangian subspace is a pair $(\L,\orn_{_\L})$, where
$\L$ is a Lagrangian subspace of $\V$ and $\orn_{_\L}: \L
\smallsetminus \{0\} \to \{\pm 1\}$ is a function which satisfies
the following equivariant property:
%
%
\begin{equation*}
\orn_{_\L}(t \cdot l) = \lgn(t) \orn_{_\L}(l)
\end{equation*}
where $t \in \Fqm$ and $\lgn$ the Legendre character of $\Fqm$.
\end{definition}
We denote by $\ornlag$ the space of oriented Lagrangians
subspaces. There is a forgetful map $\ornlag \lto \lag$, where
$\lag$ is the space of Lagrangian subspaces, $\lag \simeq
\projI(\Fq)$. In the sequel we use the notation $\L^{\circ}$ to
specify that $\L$ is equipped with an orientation.
\subsection{The Heisenberg group}
Let $\heiz$ be the Heisenberg group. As a set we have $\heiz = \V
\times \Fq$. The multiplication is defined by the following
formula:
%
%
\begin{equation*}
(v,\lambda) \cdot (v',\lambda ') := (v+v',\lambda + \lambda ' +
\half \ome(v,v')).
\end{equation*}
We have a projection $\pi: \heiz \lto \V$. We fix a section of
this projection:
\begin{equation}\label{section}
 s:\V \dashrightarrow \heiz, \rev s(v) := (v,0).
\end{equation}
\subsection{Models of irreducible representation}
Given $\ornL = (\L,\orn_{_\L}) \in \ornlag$, we construct the
Hilbert space $\H_{\ornL} = {\mathrm{Ind}}_{\tilde{\L}}^{\heiz} \;
\C_{\tilde{\psi}}$, where $\tilde{\L} = \pi^{-1}(\L)$ and
$\tilde{\psi}$ is the extension of the additive character $\psi$
to $\tilde{\L}$ using the section $s$, i.e., $\tilde{\psi} :
\tilde{\L} = \L  \times \Fq   \lto \C^*$ is given by the formula:
%
%
\begin{equation*}
\tilde{\psi}(l,\lambda) := \psi(\lambda).
\end{equation*}
More concretely: $\H_{\ornL} = \{ f:\heiz \lto \C \; | \;
f(\lambda l e) = \psi(\lambda) f(e) \}$. The group $\heiz$ acts on
$\H_{\ornL}$ by multiplication from the right. It is well known
(and easy to prove) that the representations $\H_{\ornL}$ of
$\heiz$ are irreducible and for different $\ornL$'s they are all
isomorphic. These are different models of the same irreducible
representation. This is  stated in the following theorem:
%
%
\begin{theorem}[Stone-von Neumann]\label{isomorphic_models}
For an oriented Lagrangian subspace $\ornL$, the representation
$\H_{\ornL}$ of $\heiz$ is irreducible. Moreover, for any two
oriented Lagrangians $\ornLI,\ornLII \in \ornlag$ one has
$\H_{\ornLI} \simeq \H_{\ornLII}$ as representations of $\heiz$.
$\EProof$
\end{theorem}
\subsection{Canonical intertwining operators}
Let $\ornLI,\ornLII \in \ornlag$ be two oriented Lagrangians. Let
$\H_\ornLI, \H_\ornLII$ be the corresponding representations of
$\heiz$. We denote by $\interLILII := \Hom_{_\heiz}
(\H_\ornLI,\H_\ornLII)$ the space of intertwining operators
between the two representations. Because all representations are
irreducible and isomorphic to each other we have, $\dim \;
\interLILII = 1$. Next, we construct a canonical element in
$\interLILII$.
\\
\\
Let $\ornLI = (\LI,\orn_{_\LI})$, $\ornLII = (\LII, \orn_{_\LII})$
be two oriented Lagrangian subspaces. Assume they are in general
position, i.e., $\LI \neq \LII$. We define the following specific
element  $\thetaLILII \in \interLILII, \;\; \thetaLILII :
\H_{\ornLI} \lto \H_{\ornLII}$. It is defined by the following
formula:
\begin{equation}\label{different_presentation}
\thetaLILII = \aLILII \cdot \thnLILII
\end{equation}
where $\thnLILII : \H_\ornLI \lto \H_\ornLII$ denotes the standard
averaging operator and $\aLILII$ denotes the normalization factor.
The formulas are:
%
%
\begin{equation*}
\thnLILII(f)(e) = \sum_{l_2 \in \LII} f(l_2 e)
\end{equation*}
where $f \in \H_\ornLI$.
%
%
\begin{equation*}
\aLILII = \frac{1}{q} \sum_{l_1 \in \LI} \psi( \half \ome(l_1,
\xiLII)) \orn_{_\LI}(l_1) \orn_{_\LII}(\xiLII)
\end{equation*}
where $\xiLII$ is a fixed non-zero vector in $\LII$. Note that
$\aLILII$ does not depend on $\xiLII$.
\\
\\
Now we extend the definition of $\thetaLILII$ to the case where
$\LI =\LII$. Define:
%
%
\begin{equation*}
\thetaLILII = \left \{
\begin{matrix}
\;\;\mathrm{I}, & & \orn_{_\LI} & = & \orn_{_\LII} \\
    \mathrm{-I},& & \orn_{_\LI} & = &  -\orn_{_\LII} \\
\end{matrix} \right .
\end{equation*}
The \textit{main claim} is that the collection $\{ \thetaLILII
\}_{_{\ornLI,\ornLII \in\ornlag} }$ is associative. This is
formulated in the following theorem:
%
%
\begin{theorem}[Associativity]\label{associativity_thm}
Let $\ornLI,\ornLII,\ornLIII \in \ornlag$ be a triple of oriented
Lagrangian subspaces. The following associativity condition holds:
\begin{equation*}
\thetaLIILIII \circ \thetaLILII = \thetaLILIII.
\end{equation*}
\end{theorem}
\subsection{Canonical Hilbert space}
Define the {\it canonical Hilbert space} $\Hc \subset
\bigoplus\limits_{\ornL \in \ornlag} \H_{\ornL}$ as the subspace
of compatible systems of vectors, namely:
$$\Hc := \{(f_{_\ornL})_{_{\ornL \in \ornlag}} ;\rev
\thetaLILII(f_{_\ornLI}) = f_{_\ornLII}\}.$$
%
%
%
\section{The Weil representation} \label{weilrep} In this section
we construct the Weil representation using the Hilbert space
$\Hc$. We denote by $\Sp := \mathrm{Sp}(\V,\ome)$ the group of
linear symplectomorphisms of $\V$. Before giving any formulas,
note that the space $\Hc$ was constructed out of the symplectic
space $(\V,\ome)$ in a complete canonical way. This immediately
implies that all the symmetries of $(\V,\ome)$ automatically act
on $\Hc$. In particular, we obtain a \textit{linear}
representation of the group $\Sp$ in the space $\Hc$. This is the
famous Weil representation of $\Sp$ and we denote it by $\rho :
\Sp \lto \GL (\Hc)$. It is given by the following formula:
\\
\begin{equation}\label{Wrf} 
\rho(g)[(f_{_\ornL})] := (f_{_\ornL}^g).
\end{equation}
\\
Let us elaborate on this formula. The group $\Sp$ acts on the
space $\ornlag$. Any element $g \in \Sp$ induces an automorphism
$g : \ornlag \lto \ornlag$ defined by:
%
%
\begin{equation*}
(\L,\ornofL) \longmapsto (g \L,\ornofL^g)
\end{equation*}
where $\ornofL^g(l) = \ornofL(g^{-1} l)$. Moreover, $g$ induces an
isomorphism of vector spaces $g:\H_{\ornL} \lto \H_{g\ornL}$
defined by the following formula:
%
%
\begin{equation}\label{actionfiber}
f_{_\ornL}  \longmapsto f_{_\ornL}^g, \;\; f_{_\ornL}^g(e) :=
f_{_\ornL}(g^{-1} e)
\end{equation}
where the action of $ g \in \Sp$ on $e = (v,\lambda) \in \heiz$ is
given by $g(v,\lambda) = (gv,\lambda)$. It is easy to verify that
the action (\ref{actionfiber}) of $\Sp$ commutes with the
canonical intertwining operators, i.e., for any two
$\ornLI,\ornLII \in \ornlag$ and any element $g \in \Sp$ the
following diagram is commutative:
\begin{equation*}  
\begin{CD}
\H_{\ornLI}   @>\thetaLILII>>     \H_{\ornLII} \\
  @VgVV                             @VgVV       \\
\H_{g\ornLI}   @>\thetagLIgLII>>   \H_{g\ornLII}
\end{CD}
\end{equation*}
From this we deduce that formula (\ref{Wrf}) indeed gives the
action of $\Sp$ on $\Hc$.
\section{Realization and formulas}\label{realization} In this
section we give the standard Schr\"{o}dinger realization of the
Weil representation. Several formulas for the kernels of basic
operators are also included.
\subsection{Schr\"{o}dinger realization}
Fix  $\V = \VI \oplus \VII$ to be a Lagrangian decomposition of
$\V$. Fix $\orn_{_{\VII}}$ to be an orientation on $\VII$. Denote
by $\ornVII = (\VII,\orn_{_{\VII}})$ the oriented space. Using the
system of canonical intertwining operators we identify $\Hc$ with
a specific representative $\H_{\ornVII}$. Using the section $s:\V
\dashrightarrow \heiz$ (cf. \ref{section}) we further make the
identification $s: \H_{\ornVII} \simeq \FSVI$, where $\FSVI$ is
the space of complex valued functions on $\VI$. We denote $\H :=
\FSVI$. In this realization the Weil representation, $\rho : \Sp
\lto \GL (\H)$, is given by the following formula:
\begin{equation*}  
\rho(g)(f) = \thetagVIIVII(f^g)
\end{equation*}
where $f \in \H \simeq \H_{\ornVII}$ and $ g \in \Sp$.
\subsection{Formulas for the Weil representation}
First we introduce a basis $e \in \VI$ and the dual basis $e^* \in
\VII$ normalized so that $\ome(e,e^*) = 1$. In terms of this basis
we have the following identifications: $\V \simeq \Fq^2$,
$\VI,\VII \simeq \Fq$, $\Sp \simeq \GG$ and $\heiz \simeq \Fq^2
\times \Fq$ (as sets). We also have $\H \simeq \FSFq$.
\\
\\
For every element $g \in \Sp$ the operator $\rho(g) : \H \lto \H$
is represented by a kernel $K_g:\Fq^2 \lto \C$. The multiplication
of operators becomes convolution of kernels.  The collection $\{
K_g \}_{g \in \Sp}$ gives a single function of ``kernels'' which
we denote by $K_\rho : \Sp \times \Fq^2 \lto \C$. For every
element $g \in \Sp$ the kernel $K_\rho(g)$ is of the form:
%
%
\begin{equation*}  
K_\rho(g,x,y) = a_g \cdot \psi(R_g(x,y))
\end{equation*}
where $a_g$ is a certain normalizing coefficient and $R_g : \Fq^2
\lto \Fq$ is a quadratic function supported on some linear
subspace of $\Fq^2$. Next, we give an explicit description of the
kernels $K_\rho(g)$.
\\
\\
Consider the (opposite) Bruhat decomposition $\Sp = \oB w \oB \cup
\oB$ where:
%
%
\begin{equation*}
\oB := \left (
\begin{matrix}
*  &   \\
*  &  * \\
\end{matrix}
\right )
\end{equation*}
and $w := \left(\begin{smallmatrix} 0 & 1 \\ -1 & 0
\end{smallmatrix} \right)$ is the standard Weyl element.
\\
\\
If $g \in \oB w \oB$ then:
%
%
\begin{equation*}  
g = \left (
\begin{matrix}
a & b \\
c & d
\end{matrix}
\right )
\end{equation*}
where $ b \neq 0$. In this case we have:
%
%
\begin{eqnarray*}  
a_g & = & \frac{1}{q} \sum_{t \in \Fq} \psi(\frac{b}{2}t)
\lgn(t),\\
%
%
R_g(x,y) & = & \frac{-b^{-1}d}{2} x^2 + \frac{b^{-1} -c +
ab^{-1}d}{2} xy - \frac{ab^{-1}}{2} y^2.
\end{eqnarray*}
Altogether we have:
%
%
\begin{equation*}  
K_\rho (g,x,y) = a_g \cdot \psi(\frac{-b^{-1}d}{2} x^2 +
\frac{b^{-1} -c + ab^{-1}d}{2} xy - \frac{ab^{-1}}{2} y^2).
\end{equation*}
If $g \in \oB$ then:
%
%
\begin{equation*}  
g = \left (
\begin{matrix}
a & 0 \\
r & a^{-1}
\end{matrix}
\right ).
\end{equation*}
In this case we have:
%
%
\begin{eqnarray*}
a_g & = & \lgn(a),\\  
%
%
R_g(x,y) & = & \frac{-r a^{-1}}{2} x^2 \cdot \delta_{y=a^{-1}x}.  
\end{eqnarray*}
Altogether we have:
%
%
\begin{equation} \label{K_borel}
K_\rho (g,x,y) = a_g \cdot \psi(\frac{-r a^{-1}}{2} x^2)
\delta_{y=a^{-1}x}.
\end{equation}
\subsection{Formulas for the Heisenberg representation}
On $\H$ we also have a representation of the Heisenberg group
$\heiz$. We denote it by $\pi : \heiz \lto \GL(\H)$. For every
element $ e \in \heiz$ we have a kernel $K_e :\Fq^2 \lto \C$. We
denote by $K_\pi : \heiz \times \Fq^2 \lto \C$ the function of
kernels. For an element $e \in \heiz$ the kernel $K_\pi (e)$ has
the form $\psi(R_e(x,y))$ where $R_e$ is an affine function which
is supported on a certain one dimensional subspace of $\Fq^2$.
Here are the exact formulas:
\\
\\
For an element $e=(\cq,\cp,\lambda)$ we have:
%
%
$$ R_e(x,y) =  (\frac{\cp \cq}{2} + \cp x  + \lambda) \delta_{y=x+\cq},$$
%
%
\begin{eqnarray}\label{K_heizenberg}
K_\pi(e,x,y) & = & \psi(\frac{\cp \cq}{2} + \cp x  + \lambda)
\delta_{y=x+\cq}.
\end{eqnarray}
\subsection{Formulas for the representation of the semi-direct product}
The representations $\rho : \Sp \lto \GL(\H)$ and $\pi : \heiz
\lto \GL(\H)$ combine together to give a representation of the
semi-direct product $\semi = \Sp \ltimes \heiz$. We denote the
total representation by $ \rho \ltimes \pi : \semi \lto \GL(\H)$,
$\rho \ltimes \pi (g,e) = \rho(g) \cdot \pi(e)$. The
representation $ \rho \ltimes \pi$ is given by a kernel $K_{\rho
\ltimes \pi} : \semi  \times \Fq^2 \lto \C$. We denote this kernel
simply by $K$.
\\
\\
We give an explicit formula for the kernel $K$ only in the case
$(g,e) \in \oB w \oB \times \heiz$, i.e.,
%
%
\begin{equation*}  
g = \left (
\begin{matrix}
a & b \\
c & d
\end{matrix}
\right )
\end{equation*}
where $ b \neq 0$ and $ e = (\cq,\cp,\lambda)$. In this case:
%
%
\begin{eqnarray}
\mRSwE(g,e,x,y) & = & R_g(x,y-\cq) + R_e(y-\cq,y) \label{R_map}, \\
%
%
K(g,e,x,y) & = & a_g \cdot \psi(R_g(x,y-\cq) + R_e(y-\cq,y)).
\label{K_Delta}
\end{eqnarray}
\section{Deligne's letter} \label{delignes_letter} In this
section we \textit{geometrize} (Theorem \ref{main_thm}) the total
representation $\rho \ltimes \pi : \semi \lto \H$. First, we
realize all finite \textit{sets} as rational points of certain
algebraic \textit{varieties}. Vector spaces: $\V = \AV(\Fq)$, $\VI
= \AVI(\Fq)$. Groups:  $\heiz = \Aheiz(\Fq)$, where $\Aheiz = \AV
\times \Ga$, $\Sp = \ASp(\Fq)$ and finally $\semi = \Asemi(\Fq)$,
where $\Asemi = \ASp \times \Aheiz$. The second step is to replace
the \textit{kernel} $K := K_{\rho \ltimes \pi} : \semi \times
\Fq^2 \lto \C$ (see (\ref{K_Delta})) by \textit{sheaf} theoretic
object. Recall that the kernel $K$ is a {\it representation
kernel}, namely it is a kernel of a representation and hence
satisfies the convolution property:
\begin{equation}\label{mm}
m^*K = K*K
\end{equation}
where $m:\semi \times \semi \lto \semi$ denotes the multiplication
map and $*$ means convolution of kernels, i.e., matrix
multiplication. We replace the kernel $K$ by Deligne's
\textit{Weil representation sheaf} \cite{D1}. This is an object
$\SK \in \Db(\Asemi \times \bA^2)$ that
satisfies\footnote{However, in this work we will prove a weaker
property (see Theorem \ref{main_thm}) which is sufficient for our
purposes.} the analogue (to (\ref{mm})) convolution property:
\begin{equation*}
m^*\SK \iso \SK*\SK
\end{equation*}
and its function is:
$$
f^{\SK} = K.
$$
Here $m:\Asemi \times \Asemi \lto \Asemi$ denotes the
multiplication morphism and $*$ means convolution of sheaves.
\subsection{Uniqueness and existence of the Weil representation
sheaf}
\subsubsection{The strategy} The method of constructing the
Weil representation sheaf $\SK$ is close in spirit to the
construction of an analytic function via analytic continuation. In
the realm of perverse sheaves one uses the operation of
\textit{perverse extension} (cf. \cite{BBD}). The main idea is to
construct, using formulas, an explicit irreducible perverse sheaf
$\SKO$ on a "good" open subvariety $\O \subset \Asemi \times
\bA^2$ and then to obtain the sheaf $\SK$ by perverse extension of
$\SKO$ to the whole variety $\Asemi \times \bA^2$.
\subsubsection{Explicit construction on open subvariety} On
introducing coordinates we get: $\AV \simeq \bA^2,\; \AV_1 \iso
\bA^1$ and $\ASp \simeq \AGG$. We denote by $\O$ the open
subvariety:
$$ \mathbb{O} := \ASw \times \Aheiz \times \bA^2 $$
where $\ASw$ denotes the (opposite) big Bruhat cell $\AoB w \AoB
\subset \AGG$.
\\
\\
Let us fix some standard notations from the theory of $\ell$-adic
sheaves (see \cite{BBD} for the notions of $\ell$-adic sheaves).
We denote by $\Sartin$ the Artin-Schreier sheaf on the group $\Ga$
that corresponds to the character $\psi$. We denote by
$\SL_{\lgn}$ the Kummer sheaf on the multiplicative group $\Gm$
that corresponds to the Legendre character $\lgn$. In the sequel
we will frequently make use of the \textit{character} property
(cf. \cite{Ga}) of the sheaves $\Sartin$ and $\SL_{\lgn}$, namely:
%
%
\begin{eqnarray}
s^*\Sartin & \simeq & \Sartin \boxtimes \Sartin, \label{cspSa}
\\
m^*\SL_{\lgn} & \simeq & \SL_{\lgn} \boxtimes \SL_{\lgn}
\label{cspSl}
\end{eqnarray}
where $s: \Ga \times \Ga \to \Ga$ and $m: \Gm \times \Gm \to \Gm$
denote the addition and multiplication morphisms correspondingly,
and $\boxtimes$ means exterior tensor product of sheaves.
\\
\\
At last, given a sheaf ${\cal L}$ we use the notation ${\cal L}
[i]$ for the translation functors and the notation ${\cal L} (i)$
for the i'th Tate twist.
\\
\\
\textbf{\underline{The construction of the explicit sheaf $\SKO$}}
\\
\\
We sheafify the kernel $K_{\rho \ltimes \pi}$ of the total
representation, when restricted to the set $O := \Sw \times \heiz
\times  \Fq^2$, using the formula (\ref{K_Delta}). We obtain a
sheaf on the open subvariety $\ASw \times \Aheiz \times \bA^2$,
which we denote by $\SKO$. We define:
\begin{equation*}
\SKO := \SASwE \otimes \SKnSwE
\end{equation*}
where $\SKnSwE$ is the sheaf of the non-normalized kernels and
$\SASwE$ is the sheaf of the normalization coefficients. The
sheaves $\SKnSwE$ and $\SASwE$ are constructed as follows: define
the morphism $\mRSwE : \ASw \times \Aheiz \times \bA^2 \lto \bA^1$
by formula (\ref{R_map}) and let $pr: \ASw \times \Aheiz \times
\bA^2 \lto \ASw$ be the projection morphism. Now take:
\begin{eqnarray*}
\SKnSwE & := & {\mRSwE}^* \Sartin, \\
\SASwE & := & pr ^* \SASw
\end{eqnarray*}
where:
$$\SASw := {pr_1}_{!} (\nu^* \Sartin \otimes pr_2^*\SL_{\lgn})[2](1).$$
Here $\nu: \ASw \times \bA^1 \lto \bA^1$ is the morphism defined
by $\nu(g,t) = \half b t$, where $g = \left(
\begin{smallmatrix} a & b
\\ c & d \end{smallmatrix}\right )$, and $pr_1: \ASw \times \bA^1 \lto
\ASw$, $pr_2: \ASw \times \bA^1 \lto \bA^1$ denote the natural
projections on the first and second coordinates correspondingly.
From the construction we deduce:
\begin{corollary}\label{propofSKO}
The sheaf $\SKO$  is perverse irreducible of pure weight zero and
its function agrees with K on $O$, i.e., $f^{\SKO} = K_{|_O}$.
\end{corollary}
\subsubsection{Main theorem} We are now ready to state and prove
the main theorem of this section:
%
%
\begin{theorem}[Weil representation sheaf \cite{D1}]\label{main_thm}
There exists a unique $($up to isomorphism$)$ object $\SK \in
\Db(\Asemi \times \bA^2)$ which satisfies the following
properties:
\end{theorem}
\begin{enumerate}
\item \label{prop1}$($Restriction$)$ There exists an isomorphism $\SK_{|_{\O}} \simeq \SKO$.
\item \label{prop2}$($Convolution property$)$ For every element $g \in \Asemi$ there exists
isomorphisms:
$$\SK_{|_g} * \SK \simeq L_g^*( \SK )$$
and:
$$\SK * \SK_{|_g} \simeq R_g^*( \SK ).$$
Here $\SK_{|_g}$ denotes the restriction of $\SK$ to the
subvariety $g \times \bA^2,\r *$ means convolution of sheaves and
$R_g, L_g :\Asemi \lto \Asemi$ are the morphisms of right and left
multiplication by $g$ respectively.
\end{enumerate}
\begin{corollary}\label{propofSK}
The Weil representation sheaf $\SK$ has the following two
properties:
\end{corollary}
\begin{enumerate}
\item[a.] (Function) $f^{\SK} = K$.
\item[b.] (Weight) $\ome(\SK) = 0$.
\end{enumerate}
$\it{Proof}$ (of Theorem \ref{main_thm}). \textbf{Uniqueness.} Let
$\SK, \SK'$ be two sheaves satisfying properties \ref{prop1} and
\ref{prop2}. By property \ref{prop1} there exists isomorphisms:
\begin{eqnarray*}
\SK_{|_{\O}} & \simeq & \SKO,\\
\SK'_{|_{\O}} & \simeq & \SKO.
\end{eqnarray*}
This implies that $\SK_{|_{\O}}$ and  $\SK'_{|_{\O}}$ are two
isomorphic irreducible perverse sheaves. Moreover, $\dim
\;\Hom(\SK_{|_{\O}}, \SK'_{|_{\O}})=1$. Applying property
\ref{prop2} for the Weyl element $w \in \AGG$, we obtain the
following isomorphisms:
\begin{eqnarray*}
\SK_{|_{w \O}} & \simeq & \SKw * L_{w^{-1}}^* \SK_{|_{\O}}, \\
\SK'_{|_{w \O}} &  \simeq & \SKw * L_{w^{-1}}^* \SK'_{|_{\O}}
\end{eqnarray*}
where $\SKw := {\SKO}_{|_w}$. Convolving  with $\SKw$ is basically
applying the Fourier transform, therefore it takes irreducible
perverse sheaves into irreducible perverse sheaves (cf.
\cite{KL}). Hence, we get that $ \SK_{|_{w \O}}$ and $\SK'_{|_{w
\O}}$ are two isomorphic irreducible perverse sheaves and in
particular $\dim \; \Hom(\SK_{|_{w \O}}, \SK'_{|_{ w \O}}) =1$.
Having that $\Asemi \times \bA^2 =\O \cup  w \O$ we are left to
show that one can choose $\theta_{|_{\O}} : \SK_{|_{\O}} \lto
\SK'_{|_{\O}}$ and $\theta_{|_{w\O}} : \SK_{|_{w \O}} \lto
\SK'_{|_{ w \O}}$ that agree on the intersection $\O \cap w \O$.
This can be done since the restrictions $\SK_{|_{\O \cap w \O}}$
and $\SK'_{|_{\O \cap w \O}}$ are again two isomorphic irreducible
perverse sheaves. Hence we obtained the required isomorphism.
\\
\\
\textbf{Existence.} The construction is immediate. We take our
sheaf to be the \textit{perverse extension}:
$$\SK := j_{!*} ( \SKO)$$
where $j$ stands for the open imbedding $j: \O \lto \Asemi \times
\bA^2$.
\begin{claim}\label{SK-prop}
The sheaf $\SK$ satisfies properties $\ref{prop1}$ and
$\ref{prop2}$.
\end{claim}
This completes the proof of Theorem \ref{main_thm}. $\EProof$
\\
\\
From the construction we learn:
\begin{corollary}
The sheaf $\; \SK$ is irreducible perverse. $\EProof$
\end{corollary}
\section{Proofs} \label{app_proofs}
In this section we give the proofs for all technical facts that
appeared in Chapter \ref{metaplectique}.
\\
\\
\textbf{Proof of Theorem \ref{associativity_thm}}. Before giving
the proof, we introduce a structure which is inherent to
configurations of triple Lagrangian subspaces. Let $\LI,\LII,\LIII
\subset \V$ be three Lagrangian subspaces which are in a general
position. In our case, these are just three different lines in the
plane. Then the space $\mathrm{L}_j$ induces an isomorphism $\ri :
\mathrm{L}_{k} \lto \mathrm{L}_{i}$, $i\neq j \neq k$, which is
given by the rule $\ri(l_{k}) = l_{i}$ where $l_{k} + l_{i} \in
\mathrm{L}_j$.
\\
\\
The actual proof of the theorem will be given in two parts. In the
first part we deal with the case where the three lines
$\LI,\LII,\LII \in \lag$ are in a general position. In the second
part we deal with the case when two of the  three lines are equal
to each other.
\\
\\
\textbf{Part 1.} (General position) Let $\ornLI,\ornLII,\ornLIII
\in \ornlag$ be three oriented lines in a general position. Using
the presentation (\ref{different_presentation}) we can write:
%
%
\begin{eqnarray*}
\thetaLIILIII \circ \thetaLILII & = & \aLIILIII \cdot  \aLILII \cdot \thnLIILIII \circ \thnLILII, \\
\thetaLILIII & = & \aLILIII \cdot \thnLILIII.
\end{eqnarray*}
The result for Part 1 is a consequence of the following three
simple lemmas:
%
%
\begin{lemma} \label{ass_cocycle_lemma}
The following equality holds:
\begin{equation*}
\thnLIILIII \circ \thnLILII = \Cconst \cdot \thnLILIII
\end{equation*}
where $\Cconst = \sum\limits_{l_2 \in \LII} \psi(\half
\ome(l_2,\rLIILIII(l_2)))$
\end{lemma}
%
%
\begin{lemma} \label{ass_factors_lemma}
The following equality holds:
\begin{equation*}
\aLIILIII \cdot \aLILII = \Dconst \cdot \aLILIII
\end{equation*}
where $\Dconst = \frac{1}{q} \sum\limits_{l_2 \in \LII}
\psi(-\half \ome(l_2, \rLIILIII(\xiLII))) \orn_{_\LII}(l_2)
\orn_{_\LII}(\xiLII)$
\end{lemma}
%
%
\begin{lemma}\label{DCequal_lm}
The following equality holds:
\begin{equation*}
\Dconst \cdot \Cconst = 1.
\end{equation*}
\end{lemma}
\textbf{Part 2.} (Non-general position) It is enough to check the
following equalities:
%
%
\begin{eqnarray}
\thetaLIILI \circ \thetaLILII & = & \;\;\mathrm{I} \label{equall1_eq}, \\
%
%
\thetaLIILIm \circ \thetaLILII & = & -\mathrm{I}
\label{equall2_eq}
\end{eqnarray}
where $\ornLIm$ has the opposite orientation to $\ornLI$. We
verify equation (\ref{equall1_eq}). The verification of
(\ref{equall2_eq}) is done in the same way, therefore we omit it.
\\
\\
Write:
\begin{equation} \label{part2_eq1}
\thnLIILI(\thnLILII(f))(e) = \sum_{l_1 \in \LI} \sum_{l_2 \in
\LII} f(l_2 l_1 e)
\end{equation}
where $ f \in \H_\LI$ and $e \in \heiz$. Both sides of
(\ref{equall1_eq}) are self intertwining operators of $\H_\LI$
therefore they are proportional. Hence it is sufficient to compute
(\ref{part2_eq1}) for a specific function $f$ and specific element
$ e\in \heiz$. We take $e=0$ and $ f = \delta_0$, where
$\delta_0(\lambda l e) := \psi(\lambda)$ if $e=0$  and equals 0
otherwise. We get:
\begin{equation*}
 \sum_{l_1 \in \LI} \sum_{l_2 \in \LII} f(l_2 l_1 e) = q.
\end{equation*}
Now write:
\begin{equation*}
\aLIILI \cdot \aLILII = \frac{1}{q^2} \sum_{l_1 \in \LI,\; l_2 \in
\LII} \psi(\half \ome (l_2,\xiLI) + \half \ome(l_1,\xiLII))
\orn_{_\LII}(l_2) \orn_{_\LI}(\xiLI) \orn_{_\LI}(l_1)
\orn_{_\LII}(\xiLII).
\end{equation*}
We identify $\LII$ and $\LI$  with the field $\Fq$ by the rules $
s \cdot 1 \longmapsto s \cdot \xiLII$ and $t \cdot 1 \longmapsto t
\cdot \xiLI$ correspondingly. In terms of these identifications we
get:
%
%
\begin{equation} \label{part2_eq4}
\aLIILI \cdot \aLILII = \frac{1}{q^2} \sum_{ t,\; s \in \Fq}
\psi(\half s \ome(\xiLII,\xiLI) + \half t \ome(\xiLI,\xiLII))
\lgn(t) \lgn(s).
\end{equation}
Denote by $a = \ome(\xiLII,\xiLI)$. The right-hand side of
(\ref{part2_eq4}) is equal to:
\begin{equation*}
\frac{1}{q^2} \sum_{s \in \Fq} \psi( \half a s) \lgn(s) \cdot
\sum_{ t \in \Fq} \psi( - \half a t)\lgn(t) =\frac{q}{q^2} =
\frac{1}{q}.
\end{equation*}
All together we get:
\begin{equation*}
\thetaLIILI \circ \thetaLILII = \mathrm{I}.
\end{equation*}
\\
\\
\textbf{Proof of Lemma \ref{ass_cocycle_lemma}.} The proof is by
direct computation. Write:
%
%
\begin{eqnarray}
\thnLILIII(f)(e) & = & \sum_{l_3 \in \LIII} f(l_3 e),  \label{L1_to_L3_eq} \\
\thnLIILIII( \; \thnLILII(f) \;)(e) & = & \sum_{l_3 \in \LIII}
\sum_{l_2 \in \LII} f(l_2 l_3 e) \label{L1_to_L2_to_L3_eq}
\end{eqnarray}
where $f \in \H_\LI$ and $e \in \heiz$. Both (\ref{L1_to_L3_eq})
and (\ref{L1_to_L2_to_L3_eq}) are intertwining operators from
$\H_\LI$ to $\H_\LIII$, therefore they are proportional. In order
to compute the proportionality coefficient $\Cconst$ it is enough
to compute (\ref{L1_to_L3_eq}) and (\ref{L1_to_L2_to_L3_eq}) for
specific $f$ and specific $e$. We take $e = 0$ and $f = \delta_0$
where  $\delta_0(\cq,\cp,\lambda) := \psi(\lambda)$. We get:
%
%
\begin{equation*}
\thnLILIII(\delta_0) = 1,
\end{equation*}
\begin{equation} \label{specific_L1_to_L2_to_L3_eq}
\thnLIILIII( \; \thnLILII(\delta_0) \; )(0) = \sum_{l_2 + l_3 \in
\LI} \psi( \half \ome(l_2,l_3)).
\end{equation}
But the right-hand side of (\ref{specific_L1_to_L2_to_L3_eq}) is
equal to:
\begin{equation*}  
\sum_{l_2 \in \LII} \psi( \; \half \ome(l_2,\rLIILIII(l_2))).
\end{equation*}
$\EProof$
\\
\\
%
%
%
%
%
\textbf{Proof of Lemma \ref{ass_factors_lemma}.} The proof is by
direct computation. Write:
%
%
\begin{equation*}
\aLILIII = \frac{1}{q} \sum_{l_1 \in \LI} \psi( \half \ome(l_1,
\xiLIII)) \orn_{_\LI}(l_1) \orn_{_\LII}(\xiLIII),
\end{equation*}
%
%
\begin{equation} \label{ass_factor_eq2}
\aLIILIII \cdot \aLILII = \frac{1}{q^2} \sum_{ l_1 \in \LI,\; l_2
\in \LII} \psi( \half \ome(l_1,\xiLII) + \half \ome(l_2,\xiLIII))
\orn_{_\LI}(l_1) \orn_{_\LII}(\xiLII) \orn_{_\LII}(l_2)
\orn_{_\LIII}(\xiLIII).
\end{equation}
The term $\psi( \half \ome(l_1,\xiLII) + \half \ome(l_2,\xiLIII))$
is equal to:
%
%
\begin{equation} \label{ass_factor_eq3}
\psi(\half \ome(l_1,\xiLII - \xiLIII) + \half \ome(l_2,\xiLIII))
\cdot \psi( \half \ome(l_1,\xiLIII)).
\end{equation}
We are free to choose $\xiLIII$ such that $\xiLII-\xiLIII \in
\LI$. Therefore using (\ref{ass_factor_eq3}) we get that the
right-hand side of (\ref{ass_factor_eq2}) is equal to:
%
%
\begin{equation} \label{ass_factor_eq4}
\frac{1}{q} \sum_{l_2 \in \LII}\psi(\half \ome(l_2,\xiLIII))
\orn_{_\LII}(\xiLII) \orn_{_\LII}(l_2) \cdot \aLILIII.
\end{equation}
Now, substituting $\xiLIII = -\rLIILIII(\xiLII)$ in
(\ref{ass_factor_eq4}) we obtain:
%
%
\begin{equation*} 
\frac{1}{q} \sum_{l_2 \in \LII}  \psi(-\half
\ome(l_2,\rLIILIII(\xiLII))) \orn_{_\LII}(l_2)
\orn_{_\LII}(\xiLII) \cdot \aLILIII.
\end{equation*}
$\EProof$
\\
\\
%
%
%
%
%
%
\textbf{Proof of Lemma \ref{DCequal_lm}.} Identify $\LII$ with
$\Fq$ by the rule $ t \cdot 1 \longmapsto  t \cdot \xiLII$. In
terms of this identification we get:
%
%
\begin{eqnarray*}
\Dconst & = & \frac{1}{q} \sum_{t \in \Fq} \psi(- \half \ome(t
\xiLII , \rLIILIII(\xiLII))) \orn(t),
\\
\Cconst & = & \sum_{t \in \Fq} \psi(\half \ome(t \xiLII ,
\rLIILIII(t\xiLII))).
\end{eqnarray*}
Denote by $a := \ome(\xiLII , \rLIILIII(\xiLII))$. Then:
\begin{eqnarray*}
\Dconst & = & \frac{1}{q} \sum_{t \in \Fq} \psi(-\half a t)
\orn(t),
\\
%
%
\Cconst & = & \sum_{ t \in \Fq} \psi (\half a t^2).
\end{eqnarray*}
Now, we have the following remarkable equality:
%
%
\begin{equation*}  
\sum_{t \in \Fq} \psi(\half a t) \orn(t) = \sum_{ t \in \Fq} \psi
(\half a t^2).
\end{equation*}
This, combined with $\Cconst \cdot \overline{\Cconst} = q$, gives
the result. $\EProof$
\\
\\
This completes the proof of Part 2 and of Theorem
\ref{associativity_thm} $\EProof$
\\
\\
\textbf{Proof of Claim \ref{SK-prop}.} Property \ref{prop1}
follows immediately from the construction. We give the proof of
property \ref{prop2} with respect to left multiplication (the
proof of the equivariance property with respect to the right
multiplication is the same). In the course of the proof we are
going to use the following auxiliary sheaves:
\\
\\
$\bullet$ We sheafify the kernel $K_\pi$ using the formula
(\ref{K_heizenberg}) and obtain a sheaf on $\Aheiz \times \bA^2$,
which we denote by $\SK_\pi$. Define the morphisms ${\mathrm R}:
\Aheiz \times \bA^1 \lto \bA^1$ and $ i: \Aheiz \times \bA^1 \lto
\Aheiz \times \bA^2$ by the formulas ${\mathrm
R}((\cq,\cp,\lambda),x) = \half \cp \cq + \cp x + \lambda$ and
$i((\cq,\cp,\lambda),x) = ( (\cq,\cp,\lambda),(x,x+\cq))$. Now
take:
$$\SKpi := i_! {\mathrm R}^* \Sartin.$$
\\
\\
$\bullet$ We sheafify the kernel $K_\rho$ when restricted to the
set $\oB \times \Fq^2$ using the formula (\ref{K_borel}) and
obtain a sheaf on the variety $\AoB \times \bA^2$, which we denote
by $\SKoB$. We define:
$$\SKoB := \SAoB \otimes \SKnoB$$
where the sheaf $\SKnoB$ stands for the non-normalized kernels and
the sheaf $\SAoB$ stands for the normalization coefficients. The
sheaves $\SKnoB$ and $\SAoB$ are constructed as follows: define
the morphisms ${\mathrm R}: \AoB \times \bA^1 \lto \bA^1$, $\nu:
\AoB \lto \bA^1$ and $i: \AoB \times \bA^1 \lto \AoB \times \bA^2$
 by the formulas ${\mathrm R}(\ob,x) = \half r a^{-1} x^2$,
$\nu(\ob) = a$ and $i(\ob,x) = (\ob,x,a^{-1} x)$ correspondingly,
where $\ob = \left (
\begin{smallmatrix} a & 0 \\ r & a^{-1} \end{smallmatrix} \right
)$. Now take:
$$\SKnoB := i_! {\mathrm R} ^* \Sartin$$
and:
$$\SAoB :=\nu^* \SL_{\lgn}.$$
\\
\\
$\bullet$ We will frequently make use of several other sheaves
obtained by restrictions from $\SKO, \r {\mathcal A}_{\O}, \r
\SKoB$ and $\SAoB$. Suppose ${\mathbb X} \subset \O_w \times
\Aheiz$ is a subvariety. Then we define ${\SK}_{\mathbb X} :=
{\SKO}_{|_{\mathbb X \times \bA^2}}$ and ${\mathcal A}_{\mathbb X}
:= {{\mathcal A}_{\O}}_{|_{\mathbb X \times \bA^2}}$. The same
when ${\mathbb X} \subset \AoB$. Finally, we denote by $\delta_0$
the {\it sky-scraper} sheaf on $\bA^1$ which corresponds to the
delta function at zero.
\\
\\
\textbf{Proof of property \ref{prop2}.} The proof will be given in
several steps.
\\
\\
\textbf{Step 1.} It is sufficient to prove the equivariance
property separately for the Weyl element $w$, an element $\ob \in
\AoB$ and an element $e \in \Aheiz$. Indeed, Step 1 follows from
the Bruhat decomposition, Corollary \ref{corr1} below and the
following decomposition lemma:
%
%
\begin{lemma}\label{DL}
There exist isomorphisms:
\begin{equation*}
\SKO \simeq \SKoB \ast \SKw \ast  \SKoU \ast \SKpi
\end{equation*}
where $\AoU$ denote the unipotent radical of $\AoB$ and $w$ is the
Weyl element.
\end{lemma}
%
%
\textbf{Step 2.} We prove property \ref{prop2} for the Weyl
element, i.e., $ g = w$. We want to construct an isomorphism:
%
%
\begin{equation} \label{existence_eq1}
\SK_{|_w} * \SK \simeq L_w ^* \SK.
\end{equation}
Noting that both sides of (\ref{existence_eq1}) are irreducible
(shifted) perverse sheaves, it is sufficient to construct an
isomorphism on the open set ${\calU} := \O \cap w\O$. The
advantage of working with this subvariety is that over $\calU$ we
have formulas for $\SK$, and moreover, $L_w$ maps $\calU$ into
itself. On $\calU$ we have two coordinate systems:
\begin{eqnarray}
\calU & \simeq & {\AUn}^\times \times \AoB \times \Aheiz, \label{cs1}\\
\calU & \simeq  & {\AoU^\times} w \times \AoB \times \Aheiz
\nonumber
\end{eqnarray}
where ${\AoU}^{\times} :=  {\AoU} \smallsetminus \{{\mathrm I}\}$
and $\AUn$ denotes the standard unipotent radical. With respect to
these coordinate systems we have the following decompositions:
\begin{claim}\label{ics}
There exists isomorphisms:
\end{claim}
\begin{enumerate}
\item $\SKU (\u \ob e) \simeq \SKUx (\u) \ast \SKoB (\ob) \ast \SKpi(e) \label{ics1}$.
\item $ \SKU (\ou w \ob e)  \simeq   \SKoUxw(\ou w) \ast \SKoB (\ob) \ast \SKpi(e)$. \label{ics2}
\end{enumerate}
Now, restricting to $\calU$ and using the coordinate system
(\ref{cs1}) our \textit{ main statement} is the existence of an
isomorphism:
\begin{equation}\label{main}
\SK_w \ast \SKU(\u \ob e)  \simeq  \SKU(w \u \ob e).
\end{equation}
Indeed, on developing the right-hand side of (\ref{main}) we
obtain:
\begin{eqnarray*}
\SKU(w \u \ob e) & = & \SKU({\u}^w w \ob e) \\
& \simeq & \SKoUxw({\u}^w w) \ast \SKoB (\ob) \ast \SKpi(e) \\
& \simeq & \SK_w \ast (\SKUx (\u) \ast \SKoB (\ob) \ast \SKpi(e))\\
& \simeq & \SK_w \ast \SKU (\u \ob e)
\end{eqnarray*}
where ${\u}^w := w \u w^{-1}$. The first and third isomorphisms
are applications of Claim \ref{ics} parts \ref{ics2} and
\ref{ics1} respectively. The second isomorphism is a result of
associativity of convolution and the following \textit{central
lemma}:
%
%
\begin{lemma} \label{centralemma}
There exists an isomorphism:
\begin{equation} \label{main_iso}
\SKoUxw (\u^w w) \simeq  \SKw * \SKUx (\u).
\end{equation}
\end{lemma}
The following is a consequence of (\ref{existence_eq1}).
%
%
\begin{corollary}\label{corr1}
There exists an isomorphism:
\begin{equation} \label{corr1_eq}
\SK_{|_{\AoB \times \Aheiz}} \simeq \SKoB \ast \SKpi.
\end{equation}
\end{corollary}
$\BProof$ On developing the left hand side of (\ref{corr1_eq}) we
obtain:
\begin{eqnarray*}
\SK_{|_{\AoB \times \Aheiz}} & \simeq & \SKw * \SK_{|{w^{-1} \AoB
\times \Aheiz}}
\\
& \simeq & \SKw * (\SKwI * \SKoB \ast \SKpi)
\\
& \simeq & (\SKw * \SKwI ) * \SKoB \ast \SKpi
\\
& \simeq & \SKoB \ast \SKpi.
\end{eqnarray*}
The first isomorphism is a consequence  of (\ref{existence_eq1}).
The second isomorphism is a consequence of Lemma \ref{DL}. The
third isomorphism is the associativity property of convolution.
The last isomorphism is a property of the Fourier transform
\cite{KL}, namely $\SKw * \SKwI \simeq \SI$, where $\SI$ is the
kernel of the identity operator. $\EProof$
\\
\\
\textbf{Step 3.} We prove property \ref{prop2} for element $\ob
\in \AoB$. Using corollary \ref{corr1} we have $\SK_{|_{\ob}}
\simeq {\SKoB}_{|_{\ob}} := \SKob$. We want to construct an
isomorphism:
\begin{equation}\label{existence_eq3}
\SKob * \SK \simeq L_{\ob} ^* \SK.
\end{equation}
Since both sides of (\ref{existence_eq3}) are irreducible
(shifted) perverse sheaves, it is enough to construct an
isomorphism on the open set $\O := \ASw \times \Aheiz \times
\bA^2$. Write
$$ \SKob * \SK_{|_{\O}} \simeq \SKob * \SKO \simeq \SKob * (\SKoB * \SKw *
\SKoU \ast \SKpi ) \simeq (\SKob * \SKoB) * (\SKw * \SKoU \ast
\SKpi).$$
The first isomorphism is by construction. The second isomorphism
is an application of Lemma \ref{DL}. The third isomorphism is the
associativity property of the convolution operation between
sheaves. From the last isomorphism we see that it is enough to
construct an isomorphism $\SKob * \SKoB \simeq L_{\ob}^*(\SKoB)$,
where $\L_{\ob} : \AoB \lto \AoB$. The construction is an easy
consequence of formula (\ref{K_borel}) and the character sheaf
property (\ref{cspSa}) of $\Sartin$.
\\
\\
\textbf{Step 4.} We prove property \ref{prop2} for an element $ e
\in \Aheiz$. We want to construct an isomorphism:
\begin{equation} \label{existence_eq4}
\SK_{|_e} * \SK \simeq L_e ^* \SK.
\end{equation}
Both sides of (\ref{existence_eq4}) are irreducible (shifted)
perverse sheaves, therefore it is sufficient to construct an
isomorphism on the open set $\O$. This is done by a direct
computation, similar to what has been done before, hence we omit
it. This completes the proof of Claim \ref{SK-prop}. $\EProof$
\\
\\
%
%
%
%
%
%
%
%
%
%
%
%
%
%
%
%
\textbf{Proof of Lemma \ref{DL}}. We will prove the Lemma in two
steps.
\\
\\
%
%
%
%
\textbf{Step 1.} We prove that $\SKO  \simeq \SK_{\O_w} \ast
\SKpi$. In a more explicit form we want to show:
%
%
\begin{equation} \label{lm3_explicit}
\SA_\O \otimes \SKnSwE \simeq \SASw \otimes \SKnSw * \SKnE.
\end{equation}
It is sufficient to show the existence of an isomorphism $\SKnSwE
\simeq \SKnSw * \SKnE$. On developing the left-hand side of
(\ref{lm3_explicit}) we obtain:
$$ \SKnSwE(g,e,x,y) := \SL_{\psi(R(g,e,x,y))}.$$
On developing the right-hand side we obtain:
\begin{eqnarray*}
\SKnSw * \SKnE (\; (g,e) \; ,x,y) & := & \sint_{z \in \bA^1} \SKnSw(g,x,z) \otimes \SKnE(e,z,y) \\
& := & \sint_{\bA^1} \SL_{\psi (R_g(x,z))} \otimes \SL_{\psi(R_e(z,y))} \otimes \delta_{y=z-\cq} \\
&  \simeq & \SL_{\psi(R_g(x,y-\cq))} \otimes \SL_{\psi(R_e(y-\cq,y))} \\
&  \simeq & \SL_{\psi(R_g(x,y-\cq) + R_e(y-\cq,y))} \\
& = & \SL_{\psi(R(g,e,x,y))}.
\end{eqnarray*}
The only non-trivial isomorphism is the last one and  it is a
consequence of the Artin-Schreier sheaf being a character sheaf on
the additive group $\Ga$.
\\
\\
\textbf{Step 2.} We prove that $\SK_{\O_w} \simeq  \SKoB \ast \SKw
\ast \SKoU$. In a more explicit form we want to show:
%
%
\begin{equation*} 
\SASw \otimes \SKnSw \simeq  \SAoB \otimes \SAw \otimes  \SAoU
\otimes  \SKnoB * \SKnw * \SKnoU.
\end{equation*}
We will separately show the existence of two isomorphisms:
%
%
\begin{eqnarray}
\SKnSw & \simeq & \SKnoB  * \SKnw *  \SKnoU, \label{lm1_iso1}\\
%
%
 \SASw & \simeq & \SAoB  \otimes \SAw \otimes \SAoU. \label{lm1_iso2}
\end{eqnarray}
\textbf{Isomorphism (\ref{lm1_iso1}).} Introduce the coordinate
system $\ASw \iso \AoB \times w \times \AoU$. Let $ \ob = \left (
\begin{smallmatrix} a & 0
  \\ r & a^{-1} \end{smallmatrix} \right )$ and  $\ou = \left (
\begin{smallmatrix} 1 & 0 \\ s & 1 \end{smallmatrix} \right )$  be
general elements in the groups $\AoU$ and $\AoB$ respectively. In
terms of the coordinates $ (a,r,s) $ a general element in $\ASw$
is of the form $ g = \left ( \begin{smallmatrix} as & a \\ rs -
a^{-1} & r
\end{smallmatrix} \right )$. Developing the left-hand side of
(\ref{lm1_iso1}) in terms of the coordinates $ (a,r,s)$ we obtain:
$$ \SKnSw(\ob w \ou ,x,y) := \SL_{\psi(-\half a^{-1} r x^2 + a^{-1}xy
-\half s y^2)}.$$
On developing the right-hand side of (\ref{lm1_iso1}) we obtain:
\begin{eqnarray*}
\SKnoB * \SKnw * \SKnoU (\ob w \ou,x,y) & := &  \sint\limits_{z,z'
\in \bA^1} \SKnoB(\ob,x,z) \otimes \SKnw(w,z,z') \otimes
\SKnoU(\ou,z',y)
\\
& := & \sint\limits_{z,z' \in \bA^1} \SL_{\psi(-\half r a^{-1}
x^2)} \otimes \delta_{x = a z} \otimes \SL_{\psi(zz')} \otimes
\SL_{\psi(-\half s z'^2)} \otimes \delta_{y=z'}
\\
& \simeq & \SL_{\psi(-\half r a^{-1} x^2)} \otimes
\SL_{\psi(a^{-1} x y)} \otimes \SL_{\psi(-\half s y^2)}
\\
& \simeq & \SL_{\psi(-\half ra^{-1} x^2 + a^{-1}xy -\half s y^2)}.
\end{eqnarray*}
The last isomorphism is a consequence of the fact that the
Artin-Schreier sheaf is a character sheaf (\ref{cspSa}).
Altogether we obtained isomorphism (\ref{lm1_iso1}).
\\
\\
\textbf{Isomorphism (\ref{lm1_iso2}).} On developing the left-hand
side of (\ref{lm1_iso2}) in terms of the coordinates $(a,r,s)$ we
obtain:
\begin{eqnarray*}
\SASw( \ob w \ou ) & := & \SG(\psi_a,\lgn)[2](-1)\\
& \simeq & \SG(\psi,\lgn_{a^{-1}}) [2](-1)\\
& \simeq & \SL_{\lgn(a^{-1})} \otimes \SG(\psi,\lgn) [2](-1)\\
%
& \simeq & \SL_{\lgn(a)} \otimes  \SG(\psi,\lgn) [2](-1) \\
& =: & \SAoB \otimes \SAw \otimes \SAoU( \ob w \ou )
\end{eqnarray*}
where $\SG(\psi_s,\lgn_a) := \int_{\bA^1} \SL_{\psi(\half sz)}
\otimes \SL_{\lgn(az)}$ denotes the quadratic {\it Gauss-sum}
sheaf. The second isomorphism is a change of coordinates $z
\mapsto az$ under the integration. The third isomorphism is a
consequence of the Kummer sheaf $\SL_\lgn$ being a character sheaf
on the multiplicative group $\Gm$ (\ref{cspSl}). The fourth
isomorphism is a specific property of the Kummer sheaf which is
associated to the quadratic character $\lgn$. This completes the
construction of isomorphism (\ref{lm1_iso2}). $\EProof$
\\
\\
%
%
%
\textbf{Proof of claim \ref{ics}.} Carried out in  exactly the
same way as the proof of the decomposition Lemma \ref{DL}. Namely,
using the explicit formulas of the sheaves $\SKO, \SKoB, \SKpi$
and the character sheaf property of the sheaves $\Sartin$ and
$\Slegendre$. $\EProof$
\\
\\
\textbf{Proof of Lemma \ref{centralemma}.} First, we write
isomorphism (\ref{main_iso}) in a more explicit form:
%
%
\begin{equation} \label{main_explicit}
\SAoUxw \otimes \SKnoUxw (\u^w w) \simeq  \SAw \otimes \SAUx
\otimes \SKnw * \SKnUx (\u).
\end{equation}
Let $ \u = \left ( \begin{smallmatrix} 1 & s \\ 0  & 1
\end{smallmatrix} \right ) \in \AUnx$ be a non-trivial unipotent,.
Then $\u^w w = w \u = \left (\begin{smallmatrix} 0 & 1 \\ -1 & -s
\end{smallmatrix} \right)$. On developing the left-hand side of
(\ref{main_explicit}) we obtain:
%
%
\begin{eqnarray*}
\SKnoUxw (\u^w w,x,y) & := & \SL_{\psi(\half s x^2 +xy)}, \\
\SAoUxw (\u^w w) & := & \SG(\psi,\lgn) [2](1)
\end{eqnarray*}
where $\SG(\psi_s,\lgn_a) := \int_{\bA^1} \SL_{\psi(\half s z)}
\otimes \SL_{\lgn(az)}$.
\\
\\
On developing the right-hand side of (\ref{main_explicit}) we
obtain:
\begin{eqnarray*}
\SKnw * \SKnUx (\u,x,y) & := &  \sint_{z \in \bA^1} \SKnw(x,z) \otimes \SKnUx(u,z,y) \\
& := & \sint_{\bA^1}\SL_{\psi(xz)} \otimes \SL_{\psi(-\half s^{-1} z^2 + s^{-1}zy -\half s^{-1} y^2)} \\
& \simeq & \sint_{\bA^1} \SL_{\psi(xz -\half s^{-1} z^2 + s^{-1} zy -\half s^{-1} y^2)} \\
& \simeq & \sint_{\bA^1} \SL_{\psi(-\half s^{-1} ( z-sx-y)^2)} \otimes \SL_{\psi(\half s x^2 +xy)} \\
& \simeq &  \sint_{\bA^1} \SL_{\psi(-\half s^{-1} z^2)} \otimes
\SL_{\psi(\half s x^2 +xy)}.
\end{eqnarray*}
By applying change of coordinates $z \mapsto sz $ under the last
integration we obtain:
\begin{equation}\label{main_eq1}
\sint_{\bA^1} \SL_{\psi(-\half s^{-1} z^2)} \otimes
\SL_{\psi(\half s x^2 +xy)} \simeq  \sint_{z \in \bA^1}
\SL_{\psi(-\half s z^2)} \otimes \SL_{\psi( \half s x^2 +xy)}.
\end{equation}
Now write:
\begin{equation} \label{main_eq2}
\SAw \otimes \SAUx (\u) := \SG(\psi,\lgn) [2](1) \otimes
\SG(\psi_s,\lgn)[2](1).
\end{equation}
Combining (\ref{main_eq1}) and (\ref{main_eq2}) we obtain that the
right-hand side of (\ref{main_explicit}) is isomorphic to:
\begin{equation*} 
\left (\SG(\psi_s,\lgn) [2](1) \otimes \sint_{\bA^1}
\SL_{\psi(-\half s z^2)} \right ) \otimes \left ( \SG(\psi,\lgn)
[2](1) \otimes \SL_{\psi( \half s x^2 +xy)} \right).
\end{equation*}
The main argument is the existence of the following isomorphism:
%
%
\begin{equation*} 
\SG(\psi_s,\lgn) [2](1) \otimes  \sint_{\bA^1} \SL_{\psi(- \half s
z^2)} \simeq \Qlb.
\end{equation*}
It is a direct consequence of the following lemma:
\begin{lemma}[Main lemma] \label{mainlemma}
There exists a canonical isomorphism of sheaves on $\Gm$:
\begin{equation*} 
\sint_{\bA^1} \SL_{\psi(\half s z)} \otimes \SL_{\lgn(z)} \simeq
\sint_{\bA^1} \SL_{\psi(\half s z^2)}
\end{equation*}
where $s \in \Gm$.
\end{lemma}
$\BProof$  The parameter $s$ does not play any essential role in
the argument, therefore it is sufficient to prove:
%
%
\begin{equation} \label{mainlemma_eq1}
\sint_{\bA^1} \SL_{\psi(z)} \otimes \SL_{\lgn(z)} \simeq
\sint_{\bA^1} \SL_{\psi(z^2)}.
\end{equation}
Define the morphism $ p:\Gm \lto \Gm$, $p(x) = x^2 $. The morphism
$p$ is an \`{e}tale double cover. We have $ p_* \Qlb \simeq
\SL_{\lgn} \oplus \Qlb$. Now on developing the left-hand-side of
(\ref{mainlemma_eq1}) we obtain:
$$ \sint_{\bA^1} \SL_{\psi(z)} \otimes \SL_{\lgn(z)}  := \pi_! (\Sartin
\otimes \SL_{\lgn}) \simeq \pi_!(\Sartin \otimes ( \SL_\lgn \oplus
\Qlb)).$$
The first step is just a translation to conventional notations,
where $\pi$ stands for the projection $\pi :\Gm \lto pt$. The
second isomorphism uses the fact that  $\pi_! \Sartin \simeq 0 $.
Next:
$$ \pi_!(\Sartin \otimes ( \SL_\lgn \oplus
\Qlb)) \simeq \pi_! (\Sartin \otimes p_* \Qlb) \simeq \pi_! p^*
\Sartin =: \sint_{\bA^1} \SL_{\psi(z^2)}. \rev \EProof$$
This completes the proof of proposition \ref{centralemma}.
$\EProof$
\\
\\
\textbf{Proof of Corollary \ref{propofSK}.} We have
$f^{\SK_{|_{\O}}} = K_{|_{O}}$. Applying the convolution property
for the Weyl element $w \in \AGG$ we obtain the following
isomorphism:
\begin{equation}\label{Fourier} \SK_{|_{w \O}} \simeq \SKw *
L_{w^{-1}}^* \SK_{|_{\O}}.
\end{equation}
Hence $f^{\SK_{|_{w\O}}} = K_{|_{wO}}$. Having that $\Asemi \times
\bA^2 =\O \cup w \O$ we conclude that $f^\SK = K$. In addition we
want to show that the sheaf $\SK$ has weight zero. Since weight is
a local property it is enough to prove the weight property for the
sheaves $\SK_{|_{\O}}$ and $\SK_{|_{w\O}}$. By corollary
\ref{propofSKO} we have $\ome(\SK_{|_{\O}}) = 0$. Moreover, these
two sheaves are related essentially via the Fourier transform
(\ref{Fourier}) which preserves weight \cite{KL}. Hence
$\ome(\SK_{|_{w\O}}) = 0$ . This completes the proof of the
corollary. $\EProof$
\end{chapter}

\begin{chapter}{The Two Dimensional Hannay-Berry Model}\label{2d}
The main goal of this chapter is to construct the Hannay-Berry
model of quantum mechanics, on a two dimensional symplectic torus,
that were used in the previous chapters. However, for the
convenience of the reader we give a self contained presentation of
this chapter which is independent from the rest of the paper. We
construct a simultaneous quantization of the algebra of functions
and the linear symplectic group $\G =$ SL$_2 (\Z)$. We obtain the
quantization via an action of $\G$ on the set of equivalence
classes of irreducible representations of Rieffel`s quantum torus
$\Ad$. For $\h \in \Q$ this action has a unique fixed point. This
gives a canonical projective equivariant quantization. There
exists a Hilbert space on which both $\G$ and $\Ad$ act
equivariantly. Combined with the fact that every projective
representation of $\G$  can be lifted to a linear representation,
we also obtain linear equivariant quantization.
\section{Introduction}
\subsection{Motivation}
In the paper ``{\it Quantization of linear maps on the torus -
Fresnel diffraction by a periodic grating}'', published in 1980
(cf. \cite{HB}), the physicists J. Hannay and M.V. Berry explore a
model for quantum mechanics on the 2-dimensional torus. Hannay and
Berry suggested to quantize simultaneously the functions on the
torus and the linear symplectic group $\G = $ SL$_2(\Z)$. They
found (cf. \cite{HB},\cite{Me}) that the theta subgroup $\G
_\Theta \sub \G$ is the largest that one can  quantize and asked
(cf. \cite{HB},\cite{Me}) whether the quantization of $\G$ satisfy
a multiplicativity property (i.e., is a linear representation of
the group). In this chapter we want to \textit{construct} the
Hannay-Berry's model for the bigger group of symmetries, i.e., the
whole symplectic group $\G$. The central \textit{question} is
whether there exists a Hilbert space on which a deformation of the
algebra of functions and the linear symplectic group $\G$ both act
in a compatible way.
\subsection{Results}
In this chapter we give an \textit{affirmative} answer to the
existence of the quantization procedure. We show a construction
(Theorem \ref{Cer}, Corollary \ref{prq} and Theorem \ref{lin}) of
the canonical equivariant quantization procedure for rational
Planck constants. It is \textit{unique} as a projective
quantization (see definitions below).  We show that the projective
representation of $\G$ can be lifted in exactly 12 different ways
to a linear representation (to obey the multiplicativity
property). These are the first examples of such equivariant
quantization for the whole symplectic group $\G$. Our construction
slightly \textit{improves} the known constructions \cite{HB, Me,
KR1} for which the group of quantizable elements is $\G_\Theta
\subset \G$ and gives a \textit{positive} answer to the
Hannay-Berry question on the linearization of the projective
representation of the group of quantizable elements. (cf.
\cite{HB}, \cite{Me}). Previously it was shown by Mezzadri and
Kurlberg-Rudnick (cf. \cite{Me}, \cite{KR1}) that one can
construct an equivariant quantization for the theta subgroup, in
case when the Planck constant is of the form $\h = \frac{1}{N},\;
N\in \N $.
\subsubsection{Classical torus}
Let $(\T,\ome)$ be the two dimensional symplectic torus. Together
with its linear symplectomorphisms $\G \iso \GZ$ it serves as a
simple model of classical mechanics (a compact version of the
phase space of the harmonic oscillator). More precisely, let $\T=
\W/ \Lambda$ where $\W$ is a two dimensional real vector space,
i.e.,  $\W \simeq \R^2$ and $\Lambda$ is a rank two lattice in
$\W$, i.e., $\Lambda \simeq \Z^2$. We obtain the symplectic form
on $\T$ by taking a non-degenerate symplectic form on $\W$:
$$\ome: \W \times \W \lto \R.$$
We require $\ome$ to be integral, namely $\ome : \Lambda \times
\Lambda \lto
\Z$ and normalized, i.e., Vol$(\T) =1$.\\\\
Let ${\mathrm{Sp}}(\W, \ome)$ be the group of linear
symplectomorphisms, i.e., $\mathrm{Sp}(\W,\ome) \simeq
\mathrm{SL}_2(\R)$. Consider the subgroup $\G \subset
\mathrm{Sp}(\W,\ome)$ of elements that preserve the lattice
$\Lambda$, i.e., $\G (\Lambda) \subseteq \Lambda $. Then $\G
\simeq \GZ$. The
subgroup $\G$ is the group of linear symplectomorphisms of $\T$.\\
We denote by $\Lm \subseteq \W^*$ the dual lattice, $\Lm = \{ \xi
\in \W^* | \r\r \xi (\Lambda) \subset \Z \} $. The lattice $\Lm$
is identified with the lattice $\Td := \Hom(\T,\C^*)$ of
characters of $\T$ by the following map:
\begin{equation*}
  \xi \in \Lm \longmapsto e^{2 \pi i <\xi, \cdot>} \in \r \Td.
\end{equation*}
The form $\ome$ allows us to identify the vector spaces $\W$ and
$\W^*$. For simplicity we will denote the induced form on $\W^*$
also by $\ome$.
\subsubsection{Equivariant quantization of the torus}
We will construct a particular type of quantization procedure for
the functions. Moreover this quantization will be equivariant with
respect to the action of the ``classical symmetries'' $\G$:
\begin{definition} By \textit{Weyl quantization} of ${\cal A}$ we mean a family of
$\C$-linear, $*-$ morphisms $\Pih:{\cal A} \lto $ End$(\H _\h), \r
\h \in \R$, where $\H _\h$ is a Hilbert space,  s.t. the following
property holds:
\begin{equation*}
    \pi _{_\h} (\xi+\eta) = e^{\pi i \h w(\xi,\eta)} \pi _{_\h}(\xi) \pi
_{_\h}(\eta)
\end{equation*}
for all $\xi,\eta \in \Lam ^*$ and $\h \in \R$.
\end{definition}
This type of quantization procedure  will obey the ``usual''
properties (cf. \cite{D4}):
\begin{eqnarray*}
|| \pi _{_\h}(fg) - \pi _{_\h}(f) \pi _{_\h}(g)||_{_{\H _\h}} &
\lto &  0, \rev   as \r \h  \to  0,
\\
||\frac{i}{\h}[\pi _{_\h}(f),\pi _{_\h}(g)]- \pi
_{_\h}(\{f,g\})||_{_{\H _\h}} & \lto &  0, \rev   as \r \h \to 0.
\end{eqnarray*}
where $\{,\}$ is the Poisson brackets on functions.
\begin{definition} By \textit{equivariant quantization} of $\T$ we mean a quantization of $\cal A$ with additional maps
$\rho _{_\h} : \G \lto  \mathrm{U} (\H _\h)$ s.t. the following
equivariant property $($called {\it Egorov's identity}$)$ holds:
\begin{equation}\label{eqp}
{\rho _{_\h}  (B)}^{-1} \pi _{_\h}(f) \rho _{_\h} (B) = \pi
_{_\h}(f\circ B)
\end{equation}
for all $ \h \in \R , \r f \in {\cal A}$ and $B \in \G$. Here
$\mathrm{U} (\H _\h)$ is the group of unitary operators on $\H
_\h$. If  $(\rho _{_\h} , \H _\h) $ is a projective
$($respectively linear$)$ representation of the group $\G$ then we
call the quantization \textit{projective} $($respectively
\textit{linear}$)$.
\end{definition}
The idea of the construction is as follows: We use a "deformation"
of the algebra $\cal A$ of functions on $\T$. We define an algebra
$\Ad$, usually called the two dimensional non-commutative torus
(cf. \cite{Ri}). If $\h = \frac{M}{N} \in \Q$, then we will see
that all irreducible representations of $\Ad$ have dimension $N$.
We denote by $\Irr$ the set of equivalence classes of irreducible
algebraic representations of the quantized algebra. We will see
that $\Irr$ is a set "equivalent" to a torus.
\\
\\
The group $\G$ naturally acts on a quantized algebra $\Ad$ and
hence on the set $\Irr$. Let $\h = \frac{M}{N}$ with
$\mathrm{gcd}(M,N) =1$. The following holds:
\begin{theorem}[Canonical equivariant representation]\label{Cer}
There exists a \textit{unique} $($up to isomorphism$)$
N-dimensional irreducible representation $(\Pih,\Hh)$ of $\Ad$ for
which its equivalence class is fixed by $\G$.
\end{theorem}
This means that:
\[
     \Pih  \iso  \Pih^B
\]
for all $B \in \G$.
\\
\\
Since the canonical representation $(\Pih,\Hh)$ is irreducible, by
Schur's lemma we get the canonical projective representation of
$\G$ compatible with $\Pih$:
\begin{corollary}[Canonical projective representation]\label{prq}
There exists a unique projective representation $\rhop: \G \lto
\mathrm{PGL}(\Hh)$ s.t:
\begin{equation*}
{\rhop (B)}^{-1} \Pih (f) \rhop (B) = \Pih (f\circ B)
\end{equation*}
for all $f \in {\cal A}$ and $B \in \G$.
\end{corollary}
\textbf{Remark.} Corollary \ref{prq} is an improvement to the
known constructions (cf. \cite{HB,Me,KR1}) which has the group $\G
_\Theta := \{\left(
\begin{smallmatrix}
 a & b \\
 c & d
\end{smallmatrix}\right)  | \r ab = cd = 0 \r (2)\}$ as the group of quantizable
elements.
\\
\\
Using a result of Coxeter-Moser \cite{CM} about the structure of
the group $\G$ we get:
\begin{theorem}[Linearization]\label{lin}
The projective representation $\rhop$ can be lifted to a linear
representation in exactly $12$ different ways.
\end{theorem}
\textbf{Remark.} The existence of the linear representation
$\rhoh$ in Theorem \ref{lin} answers Hannay-Berry's question (cf.
\cite{HB, Me}) on the multiplicativity of the map $\rhoh$.
\\
\\
\textbf{Summary.} For $\h \in \Q$ let $(\rhoh,\Pih,\Hh)$ be the
canonical (projective) equivariant quantization of $\T$. We can
endow the space $\Hh$ with a canonical unitary structure s.t
$\Pih$ is a $*$-representation and $\rhoh$ is unitary. This
``family'' of $*-$representations of $\Ad$ is by definition a Weyl
quantization of the functions on the torus. The above results show
the existence of a canonical projective equivariant quantization
of the torus, and the existence of a linear equivariant
quantization of the torus.
\section{Construction}
We consider the algebra ${\cal A} := \F$ of smooth complex valued
function on the torus and the dual lattice $\Lam ^* := \{ \xi \in
V^* | \; \xi(\Lam) \subset \Z \} $. Let $<,>$ be the pairing
between $\W$ and $\W^*$. The map $\xi \mapsto s(\xi)$ where
$s(\xi)(x) := e^{2\pi i <x,\xi>},\r x\in \T$ and $\xi\in \Lam^*$
defines a canonical isomorphism between $\Lam^*$ and the group
$\Td  := \mathrm{Hom} (\T,\C^\ast)$ of characters of $\T$.

\subsection{The quantum tori}
Fix $\h \in \R.$ The Rieffel's quantum torus (cf. \cite{Ri}) is
the non-commutative algebra $\Ad$ defined over $\C$ by generators
$\{s(\xi),\r \xi \in \Lam ^* \}$, and relations:
\begin{equation*}
s(\xi+\eta) = e^{\pi i \h \ome(\xi,\eta)} s(\xi)s(\eta)
\end{equation*}
for all $\xi,\eta \in \Lam^*$.
\\
\\
Note that the lattice $\Lam^*$ serves, using the map $\xi\mapsto
s(\xi)$, as a basis for the algebra $\Ad$. This induces an
identification of vector spaces $\Ad\iso \A$ for every $\h$. We
will use this identification in order to view elements of the
(commutative) space $\A$ as members of the (non-commutative) space
$\Ad$.
\subsection{Weyl quantization}
To get a Weyl quantization of $\cal A$ we use a specific
one-parameter family of representations (see subsection \ref{Ceq}
below) of the quantum tori. This defines an operator $\Pih(\xi)$
for every $\xi \in \Lam^*$. We extend the construction to every
function $f \in {\cal A}$ using the Fourier theory. Suppose:
\begin{equation*}
f = \sum\limits_{\xi \in \Lam^*} a_{_\xi}\cdot\xi
\end{equation*}
is its Fourier expansion. Then we define its {\it Weyl
quantization} by:
\begin{equation*}
\Pih(f) := \sum\limits_{\xi \in \Lam^*} a_{_\xi} \Pih(\xi).
\end{equation*}
The convergence of the last series is due to the rapid decay of
the Fourier coefficients of the function $f$.

\subsection{Projective equivariant  quantization}
The group $\G = \mathrm{SL}_2(\Z)$ acts on $\Lam$ preserving
$\ome$. Hence $\G$ acts on $\Ad$ and the formula of this action is
$s^B(\xi) := s(B\xi)$. Given a representation $(\Pih,\Hh)$ of
$\Ad$ and an element $B \in \G$, define $\Pih^B(s(\xi)) :=
\Pih(s^{B^{-1}}(\xi))$. This formula induces an action of $\G$ on
the set Irr$(\Ad)$ of equivalence classes of irreducible algebraic
representations of $\Ad$.
\begin{lemma}\label{Fdim} All irreducible representations of $\Ad$
are $N$-dimensional.
\end{lemma}
Now, suppose $(\Pih,\Ad,\Hh)$  is an irreducible  representation
for which its equivalence class is fixed by the action of $\G$.
This means that for any $B\in \G$ we have $\Pih \iso \Pih^B$, so
by definition there exists an operator $\rhoh(B) \in \GL(\Hh)$
such that:
\begin{equation*}
{\rhoh(B)}^{-1} \Pih(\xi) \rhoh(B) = \Pih(B\xi)
\end{equation*}
for all $\xi\in \Lam ^*$. This implies the Egorov identity
(\ref{eqp}) for any function . Now, since $(\Pih,\Hh)$ is an
irreducible representation then by Schur's lemma for every $B\in
\G$ the operator $\rhoh(B)$ is uniquely defined up to a scalar.
This implies that $(\rhoh,\Hh)$ is a projective representation of
$\G$.
\subsection{The canonical equivariant quantization}\label{Ceq}
In what follows we consider only the case  $\h \in \Q$. We write
$\h$ in the form $\h = \frac{M}{N}$ with $\mathrm{gcd}(M,N) =1$.
\begin{proposition}\label{uf}
There exists a unique $\Pih \in \Irr$ which is a fixed point for
the action of $\G$.
\end{proposition}
\subsection{Unitary structure}
Note that $\Ad$ becomes a $\r *-$ algebra using the formula
$s(\xi)^* := s(-\xi)$. Let $(\Pih,\Hh)$ be the canonical
representation of $\Ad$.
\begin{remark}
There exists a canonical $($unique up to scalar$)$ unitary
structure on $\Hh$  for which $\Pih$ is a $*-$representation.
\end{remark}
\section{Proofs}
\subsection{Proof of Lemma \ref{Fdim}} Suppose $(\Pih,\Hh)$ is an irreducible representation of $\Ad$.
\\
\\
\textbf{Step 1.} First we show that $\Hh$ is finite dimensional.
$\Ad$ is a finite module over $\cent (\Ad) = \{s(N\xi),\r \xi \in
\Lam ^* \}$ which is contained in the center of $\Ad$. Because
$\Hh$ has at most countable dimension (as a quotient space of
$\Ad$) and $\C$ is uncountable then by Kaplansky's trick (cf.
\cite{MR}) $\cent (\Ad)$ acts on $\Hh$ by scalars. Hence dim $\Hh
< \infty $.
\\
\\
\textbf{ Step 2.} We show that $\Hh$ is N-dimensional. Choose a
basis $(e_1,e_2)$ of $\Lam ^*$ s.t. $\ome(e_1,e_2)=1$. Suppose
$\lam \neq 0$ is an eigenvalue of $\Pih (e_1)$ and denote by $\H
_\lam$ the corresponding eigenspace. We have the following
commutation relation $\Pih (e_1) \Pih (e_2) = \gamma  \Pih (e_2)
\Pih (e_1)$  where $\gamma := e^{-2 \Pi i \frac{M}{N}}$. Hence
$\Pih (e_2): \H _{\gamma ^j \lam} \lto \H _{\gamma ^{j+1} \lam},$
and because $\mathrm{gcd}(M,N) = 1$ then $\r \H _{\gamma ^i \lam}
\neq \H _{\gamma ^j \lam} $ for $ 0\leq i\neq j \leq N-1$. Now,
let $v \in \H _\lam$ and recall that $\Pih (e_2) ^N = \Pih(Ne_2)$
is a scalar operator. Then the space $\mathrm{span}\{ v,\Pih
(e_2)v,\ldots,\Pih (e_2) ^{N-1} v\}$ is N-dimensional $\Ad
-$invariant subspace hence it equals $\r \Hh$. $\EProof$
\subsection{Proof of Proposition \ref{uf}}
Let us show the existence of a unique fixed point for the action
of $\G$ on Irr$(\Ad)$.
\\
\\
Suppose $(\Pih,\Hh)$  is an irreducible representation of $\Ad$.
By Schur's lemma for every $\xi \in \Lam^*$ the operator
$\Pih(N\xi)$ is a scalar operator, i.e., $\Pih (N\xi) = q
_{_{\Pih}}(\xi) \cdot \mathrm{I}$. We have $\Pih(0) = \mathrm{I}$
and hence $q _{_{\Pih}}(\xi)\neq 0$ for all $\xi \in \Lam^*$. Thus
to any irreducible representation we have attached a scalar
function $q _{_{\Pih}}:\Lam ^* \lto \C^*$. Consider the set $Q
_{_\h}$ of \textit{twisted characters} of $\Lam ^*$:
\begin{equation*}
Q _{_\h} := \{ q:\Lam ^* \lto \C ^* , \r q(\xi+\eta) = (-1)^{M N
w(\xi,\eta)} q(\xi)q(\eta) \}.
\end{equation*}
The group $\G$ acts naturally on this space by
$q^B(\xi):=q(B^{-1}\xi)$. It is easy to see that we have defined a
map $\q:\Irr \lto Q _{_\h}$ given by $\Pih \mapsto q _{_{\Pih}}$
and it is obvious that this map is compatible with the action of
$\G$. We use the space of twisted characters in order to give a
description for the set Irr$(\Ad)$:
\begin{lemma}\label{realizati}
The map $\Pih  \mapsto q _{_{\Pih}}$ is a  $\G$-equivariant
bijection:
\[
\begin{CD}
\q : \mathrm{Irr}(\Ad)    @>  >>  Q _{_\h}.
\end{CD}
\]
\end{lemma}
Now, Proposition \ref{uf} follows from the following claim:
\begin{claim}\label{ufpoint}
There exists a unique $q _{_o} \in Q _{_\h}$ which is a fixed
point for the action of $\G$.
\end{claim}
\textbf{Proof of Lemma \ref{realizati}.} \textbf{Step 1.} The map
$\q$ is surjective. Denote by $\AT:= \mathrm{Hom} (\Lam ^*,\C ^*)
\r $ the group of complex characters of $\Lam ^*$. We define an
action of $\AT$ on $\Irr$ and on $Q _{_\h}$ by $\Pih \mapsto
\chi\Pih$ and $q \mapsto \chi^N q$, where $\chi \in \AT$, $\Pih
\in \Irr$ and $q \in Q _{_\h}$. The map $\q$ is clearly a
$\AT$-equivariant map with respect to these actions. Since $\q$ is
$\AT$-equivariant, it is enough to show that the action of $\AT$
on $Q _{_\h}$ is transitive.
 Suppose $q _{_1} , q _{_2} \in Q _{_\h}$. By definition there exists a character $\chi _{_1} \in \AT$
 for which $\chi _{_1} q _{_1}  = q _{_2}$. Let
$\chi$ be one of the $N$'s roots of  $\chi _{_1}$ then $\chi ^N q
_{_1} = q _{_2}$.
\\
\\
\textbf{Step 2.} The map $\q$ is one to one. Suppose $(\Pih,\Hh)$
is an irreducible representation of $\Ad$. It is easy to deduce
from the proof of Lemma \ref{Fdim} (Step 2) that for $\xi \notin
N\Lam ^*$ we have $\mathrm{tr}(\Pih(\xi)) = 0$. But we know from
character theory that an isomorphism class of a finite dimensional
irreducible representation of an algebra is recovered from its
character. This completes the proof of Lemma \ref{realizati}.
$\EProof$
\\
\\
\textbf{Proof of Claim \ref{ufpoint}.} {\it Uniqueness}. Fix $q
\in Q _{_\h}$. The map $\chi \mapsto \chi q$ give a bijection of
$\AT$ with $Q _{_\h}$. But the trivial character $\textbf{1} \in
\AT$ is the unique fixed point for the action of $\G$ on $\AT$.
\\
{\it Existence.} Choose a basis $(e_1,e_2)$  of $\Lam ^*$ s.t
$\ome(e_1,e_2) = 1$. This allows to identify $\Lam ^*$ with $\Z
\oplus \Z$. It is easy to see that the function:
\begin{equation*}
q _{_o}(m,n) =(-1)^{MN(mn+m+n)}
\end{equation*}
is a twisted character which is fixed by $\G$. This completes the
proof of Claim \ref{ufpoint} and of Proposition \ref{uf}.
$\EProof$
\\
\\
\subsection{Proof of Theorem \ref{lin}} The theorem follows from
the following proposition:
\begin{proposition}
Fix a projective representation $\rhop : \G \lto
\mathrm{GL}(\Hh)$. Then it can be lifted to a linear
representation in exactly $12$ ways.
\end{proposition}
\BProof Existence. We want to find constants $c(B)$ for every $B
\in \G$ s.t. $\rhoh := c(\cdot) \rhop$ is a linear representation
of $\G$. This is possible to carry out due to the following fact:
\begin{lemma}[\cite{CM}]\label{present}
The group $\G$ is isomorphic to the group generated by three
letters $S,\; B$ and $Z$ subjected to the relations: $Z^2 = 1$ and
$\r S^2 = B^3 = Z$.
\end{lemma}
Lemma \ref{present} $\Rightarrow$ Existence. We need to find
constants $\r c_{_Z}, c_{_B}, c_{_S} \r $ so that the operators
$\rhoh (Z) := c_{_Z} \rhop (Z),\r \rhoh (B) := c_{_B} \rhop (B),\r
\rhoh (S) := c_{_S} \rhop (S)$ will satisfy the identities:
\begin{equation*}
\rhoh (Z) ^2 = I, \r \rhoh (B) ^3 = \rhoh (Z),\r \rhoh (S) ^2  =
\rhoh (Z).
\end{equation*}
This can be done by taking appropriate scalars.
\\
\\
Now, fix one lifting $\rho_{_0}$. Then for the collection of
operators $\rhoh(B)$ which lifts $\rhop$ define a function
$\chi(B)$ by $\rhoh(B) = \chi(B) \rho_{_0}(B)$. It is obvious that
$\rhoh$ is a representation if and only if $\chi$ is a character.
Thus liftings corresponds to characters. By Lemma \ref{present}
the group of characters $\G^\vee := \mathrm{Hom}(\G,\C^*)$ is
isomorphic to $\Z/12\Z$.$\EProof$.
\end{chapter}

\appendix
\addcontentsline{toc}{chapter}{Appendices}
\begin{chapter} {The Higher Dimensional Hannay-Berry
Model}\label{hd} The aim of this chapter is to construct the
Hannay-Berry model of quantum mechanics on a $2n$-dimensional
symplectic torus. We construct a simultaneous quantization of the
algebra $\cal A$ of functions on the torus and the linear
symplectic group $\G = \GZZ$. In the construction we use the
quantum torus $\Aeh$, which is a deformation of $\cal A$, together
with a $\G$-action on it. We obtain the quantization via the
action of $\G$ on the set of equivalence classes of irreducible
representations of $\Aeh$. For $\h \in \Q$ this action has a
unique fixed point. This gives an equivariant quantization. There
exists a Hilbert space on which both $\G$ and $\Aeh$ act in a
compatible way.
\section{Introduction}
\subsection{Motivation}
In this chapter we want to extend our construction (see Chapter
\ref{2d}), of the two-dimensional Hannay-Berry model, to the
higher dimensional tori. The central question is whether there
exists a vector space on which a deformation of the algebra of
functions and the linear symplectic group $\GZZ$, both act in a
compatible way. This construction is the first step toward the
investigation of quantum chaos questions in the higher dimensional
model. This study will be a subject of future publication.
\\
\\
Previously it was shown by  Bouzouina  and De Bievre \cite{BDB}
that one can quantize simultaneously the functions on the torus
and one ergodic element $A \in \GZZ$ in case where the Planck
constant is of the form $\h = \frac{1}{N},\; N\in \N $.
\subsection{Definitions}
\subsubsection{Classical torus}
Let $(\T,\ome)$ be the $2n$-dimensional symplectic torus. Together
with its linear symplectomorphisms $\G \iso \GZZ$ it serves as a
simple model of classical mechanics (a compact version of the
phase space of the $n$-dimensional harmonic oscillator). More
precisely, let $\T := \W/ \Lambda$ where $\W$ is a
$2n$-dimensional real vector space, i.e.,  $\W \simeq \R^{2n}$ and
$\Lambda$ is a rank $2n$ lattice in $\W$, i.e., $\Lambda \simeq
\Z^{2n}$. We obtain the symplectic form on $\T$ by taking a
non-degenerate symplectic form on $\W$:
$$\ome: \W \times \W \lto \R.$$
We require $\ome$ to be integral, namely $\ome : \Lambda \times
\Lambda \lto
\Z$ and normalized, i.e., Vol$(\T) =1$.\\\\
Let ${\mathrm{Sp}}(\W, \ome)$ be the group of linear
symplectomorphisms, i.e., $\mathrm{Sp}(\W,\ome) \simeq
\mathrm{Sp}(2n,\R)$. Consider the subgroup $\G \subset
\mathrm{Sp}(\W,\ome)$ of elements that preserve the lattice
$\Lambda$, i.e., $\G (\Lambda) \subseteq \Lambda $. Then $\G
\simeq \GZ$. The
subgroup $\G$ is the group of linear symplectomorphisms of $\T$.\\
We denote by $\Lm \subseteq \W^*$ the dual lattice, $\Lm := \{ \xi
\in \W^* | \r\r \xi (\Lambda) \subset \Z \} $. The lattice $\Lm$
is identified with the lattice $\Td := \Hom(\T,\C^*)$ of
characters of $\T$ by the following map:
\begin{equation*}
  \xi \in \Lm \longmapsto e^{2 \pi i <\xi, \cdot>} \in \r \Td.
\end{equation*}
The form $\ome$ allows us to identify the vector spaces $\W$ and
$\W^*$. For simplicity we will denote the induced form on $\W^*$
also by $\ome$.
\\
\\
Consider the algebra ${\cal A} := \F$ of smooth complex valued
functions on $\T$. By Fourier theory the lattice $\Lam^*$ serves
as a basis of $\A$.
\subsubsection{Equivariant quantization of the torus}\label{DefQ}
We will construct a particular type of quantization procedure for
the functions. Moreover this quantization will be equivariant with
respect to the action of the group of ``classical symmetries''
$\G$:
\begin{enumerate}
\item By \textit{Weyl quantization} of ${\cal A}$ we mean a family of
$\C$-linear morphisms $\pi _{_\h}:{\cal A} \lto
\mathrm{End}(\Hh),$ where $\h \in \R$ and $\Hh$ is a Hilbert space
s.t. the following property hold:
\begin{equation*}
\pi _{_\h} (\xi)\pi _{_\h} (\eta) = e^{2 \pi i \h \ome(\eta,\xi)}
\pi _{_\h}(\eta) \pi _{_\h}(\xi)
\end{equation*}
for all $\xi,\eta \in \Lam ^*$ and $\h \in \R$.
\item By \textit{equivariant quantization} of $\T$ we mean a quantization
of $\cal A$ with additional maps $\rhoh:\G \lto  \mathrm{GL}
(\Hh)$ s.t. the following equivariant property $($called ''Egorov
identity''$)$ holds:
\begin{equation}\label{Egorov}
{\rho _{_\h}  (B)}^{-1} \pi _{_\h}(f) \rho _{_\h} (B) = \pi
_{_\h}(f\circ B)
\end{equation}
for all $\h \in \R , \r f \in {\cal A} \r$ and $\r B \in \G$.
\end{enumerate}
\subsection{Results}
In this chapter we give an affirmative answer to the existence of
the quantization procedure. We show a construction (Theorem
\ref{GH3} and Corollary \ref{prqh}) of the quantization procedure
for rational Planck constants. As far as we know this is the first
construction of equivariant quantization for higher dimensional
tori, together with the whole linear symplectic group $\GZZ$.
\\
\\
The idea of the construction is as follows: We use a "deformation"
of the algebra $\cal A$ of functions on $\T$. We define (see
\ref{qs}) two algebras $\Aeh, \r \veps = 0,1$. The algebra ${\cal
A} _{0,\hbar}$ is the usual Rieffel's quantum torus (see
\cite{Ri}) and ${\cal A} _{1,\hbar}$ is some twisted version of
it. If $\h = \frac{M}{N} \in \Q$, then we will see that all
irreducible representations of $\Aeh$ have dimension $\Nd$. We
denote by $\Irreh$ the set of equivalence classes of irreducible
algebraic representations of the quantized algebra. We will see
that $\Irreh$ is a set "equivalent" to a torus.
\\
\\
The group $\G$ naturally acts on a quantized algebra $\Aeh$ and
hence on the set $\Irreh$. Let $\h = \frac{M}{N}$ with
$\mathrm{gcd}(M,N) =1$. Set $\varepsilon := MN$ (mod 2). Then:
\begin{theorem}[Equivariant representation]\label{GH3}  There exists a \textit{unique} $($up to
    isomorphism$)$ irreducible representation
    $(\Pih,\Hh)$ of $\Aeh$ for which its equivalence class is fixed by $\G$.
\end{theorem}
This means that:
$$
\Pih  \iso  \Pih^B
$$
for every $B \in \G$.
\\
\\
Since the canonical representation $(\Pih,\Hh)$ is irreducible
then by Schur's lemma we get the canonical projective
representation of $\G$ compatible with $\Pih$:
\begin{corollary}[Canonical projective representation]\label{prqh}
For every $B \in \G$ there exists an operator $\rhoh(B)$ on $\Hh$
s.t.:
\begin{equation*}
{\rhoh (B)}^{-1} \Pih (f) \rhoh (B) = \Pih (f\circ B)
\end{equation*}
for all $f \in {\cal A}$. Moreover the correspondence $B \mapsto
\rhoh(B)$ constitutes a projective representation of $\G$.
\end{corollary}
\begin{remark} The family $(\rhoh,\Pih,\Hh), \r \h \in \Q$ presented in Corollary
$\ref{prqh}$ gives an equivariant Weyl quantization of the torus.
We can endow $($see $\ref{us})$ the space $\Hh$ with a canonical
unitary structure s.t. $\Pih$ is a $*$-representation and $\rhoh$
unitary. This answers the question whether this quantization is
also unitarizable and hence fits to the idea that quantum
mechanics should be realized on a Hilbert space.
\end{remark}
\section{Construction}\label{Con}
Consider the algebra ${\cal A} := \F$ of smooth complex valued
functions on the torus and the dual lattice $\Lam ^* := \{ \xi \in
\W^* | \; \xi(\Lam) \subset \Z \} $. Let $<,>$ be the pairing
between $\W$ and $\W^*$. The map $\xi \mapsto s(\xi)$ where
$s(\xi)(x) := e^{2\pi i <x,\xi>},\r x\in \T,\r \xi\in \Lam^*$
defines a canonical isomorphism between $\Lam^*$ and the group
$\Td  := \mathrm{Hom} (\T,\C^\ast)$ of characters of $\T$.
\subsection{The quantum tori}\label{qs}
Fix $\h \in \R.$ Define two algebras (see also \cite{Ri} and
\cite{GH1})) $\Aeh,\r \veps = 0,1$ as follows. The algebra $\Aeh$
is defined over $\C$ by generators $\{s(\xi),\r \xi \in \Lam ^*
\}$, and relations:
\begin{equation*}
s(\xi+\eta) = \veps(\xi,\eta)e^{\pi i \h \ome(\xi,\eta)}
s(\xi)s(\eta)
\end{equation*}
where $\veps(\xi,\eta) := (-1)^{\veps\ome(\xi,\eta)}$ and
$\xi,\eta \in \Lam^*$.
\\
\\
Note that the lattice $\Lam^*$ serves, using the map $\xi\mapsto
s(\xi)$, as a basis for the algebra $\Ad$. This induces an
identification of vector spaces $\Ad\iso \A$ for every $\h$.
\subsection{Weyl quantization}
To get a Weyl quantization of $\cal A$ we use a specific
one-parameter family of representations (see subsection \ref{ceq}
below) of the quantum tori. This defines an operator $\Pih(\xi)$
for every $\xi \in \Lam^*$. We extend the construction to every
function $f \in {\cal A}$ using Fourier theory. Suppose:
\begin{equation*}
f = \sum\limits_{\xi \in \Lam^*} a_{_\xi}\cdot\xi
\end{equation*}
is its Fourier expansion. Then we define its {\it Weyl
quantization} by:
\begin{equation*}
\Pih(f) = \sum\limits_{\xi \in \Lam^*} a_{_\xi} \Pih(\xi).
\end{equation*}
The convergence of the last series is due to the rapid decay of the fourier
coefficients of $f$.

\subsection{Equivariant quantization}
We describe a strategy how to get an equivariant quantization of
$\T$. The group $\G$ acts on $\Lam$ preserving $\ome$. Hence $\G$
acts on $\Aeh$ by automorphisms of algebras. Suppose $(\Pih,\Hh)$
is a representation of $\Aeh$. For an element $B \in \G$, define
$\Pih^B(\xi) := \Pih(B^{-1}\xi)$. This formula defines an action
of $\G$ on the set $\Irreh$ of equivalence classes of irreducible
algebraic representations of $\Aeh$.
\begin{lemma}\label{dim} All irreducible representations of $\Aeh$
are $\Nd$-dimensional.
\end{lemma}
Now, suppose $(\Pih,\Hh)$  is a representation for which its
equivalence class is fixed by the action of $\G$. This means that
for any $B\in \G$ we have $\Pih \iso \Pih^B$ and hence there exist
an operator $\rhoh(B)$ on $\Hh$ s.t.:
\begin{equation*}
{\rhoh(B)}^{-1} \Pih(\xi) \rhoh(B) = \Pih(B\xi)
\end{equation*}
for all $\xi\in \Lam ^*$. This implies the Egorov identity
(\ref{Egorov}) for any function. Suppose in addition that
$(\Pih,\Hh)$ is an irreducible representation. Then by Schur's
lemma for every $B\in \G$ the operator $\rhoh(B)$ is uniquely
defined up to a scalar. This implies that $(\rhoh,\Hh)$ is a
projective representation of $\G$.
\subsection{The equivariant quantization}\label{ceq}
In what follows we consider only the case  $\h \in \Q$. We write
$\h$ in the form $\h = \frac{M}{N}$ with $\mathrm{gcd}(M,N) =1$.
Set $\veps = MN$ (mod 2).
\begin{proposition}\label{ufp}
There exists a unique $\Pih \in \Irreh$ which is a fixed point for
the action of $\G$.
\end{proposition}
\subsection{Unitary structure}\label{us}
Note that $\Aeh$ becomes a $*-$ algebra by the formula $s(\xi)^*
:= s(-\xi)$. Let $(\Pih,\Hh)$ be the canonical representation of
$\Aeh$.
\begin{remark}\label{us}
There exists a canonical $($unique up to scalar$)$ unitary
structure on $\Hh$  for which $\Pih$ is a $*-$representation.
\end{remark}
\section{Proofs}\label{P}
\subsection{Proof of Lemma \ref{dim}} Suppose $(\Pih,\Hh)$ is an
irreducible representation of $\Aeh$.
\\
\\
\textbf{Step 1.} First we show that $\Hh$ is finite dimensional.
The algebra $\Aeh$ is a finite module over $\cent(\Aeh) =
\{s(N\xi),\r \xi \in \Lam ^* \}$ which is contained in the center
of $\Aeh$. Because $\Hh$ has at most countable dimension (as a
quotient space of $\Aeh$) and $\C$ is uncountable then by
Kaplansky's trick (See \cite{MR}) $\cent (\Aeh)$ acts on $\Hh$ by
scalars. Hence dim $\Hh < \infty $.
\\
\\
\textbf{Step 2.} We show that $\Hh$ is $\Nd-$dimensional. Choose a
basis $(e_1,\ldots,e_n,e_1',\ldots,e_n')$ of $\Lam^*$ s.t.
$\ome(e_i,e_j) = \ome(e_i',e_j')= 0$ and $\ome(e_i,e _j') =
\delta_{ij}$ the Kronecker's delta. Denote by $E$ the commutative
subalgebra of $\Aeh$ generated by $\{s(e_i)\}_1^d$. Suppose $\lam
\in E^*$ is an eigencharacter of $E$ and denote by $\H _\lam :=
\H_{(\lam_1,\ldots,\lam_n)}$ the corresponding eigenspace, $\lam_i
:= \lam(e_i)$. We have the following commutation relation $\Pih
(e_i) \Pih (e_j') = \gamma^{\delta_{ij}} \Pih (e_j') \Pih (e_i)$
where $\gamma := e^{-2 \pi i \frac{M}{N}}$. Hence $\Pih (e_j'): \H
_{(\lam_1,\ldots,\gamma ^k\lam_j,\ldots,\lam_n)} \lto \H
_{(\lam_1,\ldots,\gamma ^{k+1}\lam_j,\ldots,\lam_n)}$. Since
$\mathrm{gcd}(M,N) = 1$ the eigencharacters
$(\gamma^{k_1}\lam_1,\ldots,\ldots,\gamma^{k_n}\lam_n), \r 0 \leq
k_j \leq N-1$, are all different. Let $0 \neq v \in \H _\lam$ and
recall that $\Pih (e_j') ^N = \Pih(N e_j')$ is a scalar operator.
Then the space $\mathrm{span}\{\Pih(e_j')^{k}v\}$ is
$\Nd$-dimensional $\Aeh -$invariant subspace hence it equals
$\Hh$. $\EProof$
\subsection{Proof of Proposition \ref{ufp}} Suppose $(\Pih,\Hh)$  is an
irreducible representation of $\Aeh$.  By Schur's lemma for every
$\xi\in \Lam^*$ the operator $\Pih(N\xi)$ is a scalar operator
$\Pih (N\xi) = \chi_{_{\Pih}}(\xi) \cdot \mathrm{I}$. We have
$\Pih(0) = \mathrm{I}$, hence $\chi_{_{\Pih}}(\xi) \neq 0$ for all
$\xi \in \Lam^*$. Thus to any irreducible representation we have
attached a scalar function $\chi _{\Pih}:\Lam ^* \lto \C^\ast$. It
is easy to see that $\chi_{\Pih}(\xi+\eta) =
\chi_{\Pih}(\xi)\chi_{\Pih}(\eta)$. Consider the group $\AT :=
\mathrm{Hom} (\Lam ^*,\C^\ast)$ of complex characters of $\Lam
^*$. We have defined a map $\Irreh \lto \AT$ given by $\Pih
\mapsto \chi _{\Pih}$. This map is obviously compatible with the
action of $\G$, where the group $\G$ acts on characters by
$\chi^B(\xi):=\chi(B^{-1}\xi)$.
\begin{lemma}\label{realiz}
The map $\Pih  \mapsto \chi _{\Pih}$ gives a $\G$-equivariant
bijection:
\begin{equation}\label{chi}
\begin{CD}
\mathrm{Irr}(\Aeh)    @>  >>  \AT.
\end{CD}
\end{equation}
\end{lemma}
From Lemma \ref{realiz} we easily deduce that there exists a
unique equivalence class $\Pih \in \Irreh$ which is fixed by the
action of $\G$. This is the one that corresponds to the trivial
character $\textbf{1} \in \AT$ which is the unique fixed point for
the action of $\G$ on $\AT$.
\\
\\
\textbf{Proof of Lemma \ref{realiz}.} \textbf{Step 1.} The map
$\Pih \mapsto \chi _{\Pih}$ is onto. We define an action of $\AT$
on $\Irreh$ and on itself by $\Pih \mapsto \chi\cdot\Pih$ and
$\psi\mapsto \chi^N \cdot\psi$, where $\chi \in \AT$, $\Pih \in
\Irreh$ and $\psi \in \AT$. The map (\ref{chi}) is clearly a
$\AT$-equivariant map with respect to these actions. The claim
follows since the above action of $\AT$ on itself is transitive.
$\EProof$
\\
\\
\textbf{Step 2.} We show that the map $\Pih \mapsto \chi _{\Pih}$
is one to one. Suppose $(\Pih,\Hh)$ is an irreducible
representation of $\Aeh$. It is easy to deduce from the proof of
Lemma \ref{dim} (Step 2) that for $\xi \notin N\Lam ^*$ we have
$\mathrm{tr}(\Pih(\xi)) = 0$. But we know from character theory
that an isomorphism class of a finite dimensional irreducible
representation of an algebra is recovered from its character.
$\EProof$
\end{chapter}

\begin{chapter}{Two higher-dimensional (counter)
examples}\label{counter} In this section we consider two examples
that show the need for new ideas, already at the level of
formulation, for the quantum chaos statement.
\\
\\
\textbf{ Example 1.} The following is an example of an ergodic
automorphism of the 4-dimensional torus $\T := \W/\Lambda$ (this
example works in every dimension $2n,\r n > 1$ under appropriate
modifications) which is not Hecke quantum ergodic.
\\
\\
We fix an element $B \in \GL_2(\Z)\smallsetminus \GZ$ with
eigenvalues which are not roots of unity. Take:
\begin{equation*}
 A := \left( \begin{smallmatrix} B & 0 \\ 0 & {}^tB^{-1} \end{smallmatrix} \right)
\end{equation*}
where $t$ denotes the transpose operation. It is well known that
$A \in \mathrm{Sp}(4,\Z)$ is an ergodic automorphism of $\T$. Now,
we can choose\footnote{This property does not hold in the
2-dimensional case.} a character  $0 \neq \xi \in \Lm \iso \Td$
that belongs to an $A$-invariant Lagrangian sublattice $\L^*
\subset \Lm$. Fix $\h= \frac{1}{p}$ and denote by $\V :=
\L^*/p\L^* \iso \Fp^2$ the quotient lattice. Quantizing the system
we attach to $A$ the quantum operator:
\begin{equation*}
\rhoh(A): \Hh \lto \Hh
\end{equation*}
where $\Hh := {\cal S}(\V)$ is the space of functions on $\V$.
\\
\\
In this realization elements of the form $\left( \begin{smallmatrix} B & 0 \\
0 & {}^tB^{-1} \end{smallmatrix} \right)$ acts on functions by:
\begin{equation*}
\left[\left( \begin{smallmatrix} B & 0 \\ 0 & {}^tB^{-1}
\end{smallmatrix} \right)\varphi\right](x) = \lgn(\mathrm{det}\;B)\varphi(B^{-1}x)
\end{equation*}
where $\lgn$ is the Legendre character. Hence the
\textit{function} $\varphi\equiv \textbf{1}$ is a common
eigenfunction for the Hecke \textit{torus} $\CA \subset
\mathrm{Sp}(4,\Fp)$. Recall that we have chosen $\xi \in \V$ and
hence the operator $\Pih(\xi)$ acts on functions via translation
by $\xi$, so we obtain:
\begin{equation*}
\lang \varphi| \Pih(\xi)\varphi \rang = {\bf 1 \neq 0} = \int_\T
\xi \ome.
\end{equation*}
\\
\textbf{Example 2.} There exists ergodic automorphisms $A \in
\mathrm{Sp}(2n,\Z)$ of $\T$ for which the \textit{centralizer}
${\mathrm C}_A \subset \GGG$ is \textit{not a torus} (or even a
commutative group). If in Example 1 above we take $B \in \GZ$
hyperbolic then we have ${\mathrm C}_A \iso \GL(2,\Fp)$. Moreover,
the element $A \in \GGG$ might belongs to several non-isomorphic
maximal commutative subgroups of $\GGG$. We see that in this case
it is \textit{not} \textit{clear} what should be the statement of
Hecke quantum ergodicity.

\chapter{Proofs for Chapters \ref{BHm} and \ref{Proof}}\label{proofs} For the
remainder of this chapter we fix the following notations. Let $\hb
= \frac{1}{p}$, where $p \geq 5$ is a fixed prime. Consider the
lattice $\Lm$ of characters of the torus $\T$ and the quotient
vector space $ \V := \Lm / p\Lm $. The integral symplectic form on
$\Lm$ is specialized to give a symplectic form on $\V$, i.e.,
$\ome : \V \times \V \longrightarrow \Fp$. Fix $\psi: \Fp
\longrightarrow \C^*$ a non-trivial additive character. Let $\Ad$
be the algebra of functions on the quantum torus
and $\G \simeq \GZ$ its group of symmetries.\\
\section{Proof of Theorem \ref{GH2}}\label{proofGH} {\it Basic
set-up}: let $ (\Pih,\Hh)$ be a representation of $\Ad$, which is
a representative of the unique irreducible class which is fixed by
$\G$ (cf. Theorem \ref{gh}). Let $\rhoh : \G \longrightarrow
\PGL(\Hh)$ be the associated projective representation. Here we
give a proof that $\rhoh$ can be linearized in a unique way which
factors through the quotient group $\Gf \simeq \GG$:
\[
\qtriangle[\G`\Gf`\GL(\Hh);`\rhoh`\bar{\rho}_{_\h}]
\]
The proof will be given in several steps.\\\\
\textbf{Step 1.} {\it Uniqueness}. The uniqueness of the
linearization follows directly  from the fact that the group
$\GG$, $p \geq 5$, has no characters.\\\\
\textbf{Step 2.} {\it Existence}. The main technical tool in the
proof of existence is a construction of a finite dimensional
quotient of the algebra $\Ad$. Let $\Adf$ be the algebra generated
over $\C$ by the symbols $\{ s(u) \r | \r u \in \Lmmf \}$ and
quotient out by the relations:
\begin{equation} \label{relations}
 s(u + v) = \psi(\half \omega(u,v)) s(u)s(v).
\end{equation}
The algebra $\Adf$ is non-trivial and the vector space $\Lmmf$
gives on it a standard basis. These facts will be proven in the
sequel. We have the following map:
\begin{equation*} 
 s : \Lmmf \longrightarrow \Adf, \r v \mapsto s(v).
\end{equation*}
The group $\Gf$ acts on $\Adf$ by automorphisms through its action
on $\Lmmf$. We have a homomorphism of algebras:
\begin{equation} \label{AdtoAdf}
q : \Ad \longrightarrow \Adf.
\end{equation}
The homomorphism (\ref{AdtoAdf}) respects the actions of the group
of symmetries $\G$ and $\Gf$ respectively. This is summarized in
the following commutative diagram:
\begin{equation}\label{respect}
\begin{CD}
   \G \times \Ad   @>>>   \Ad \\
        @V(p,q)VV       @VqVV \\
   \Gf \times \Adf @>>>   \Adf
\end{CD}
\end{equation}
where $p : \G \longrightarrow \Gf$ is the canonical quotient map.\\\\
\textbf{Step 3.} Next, we construct an explicit representation of
$\Adf$:
$$\Pif : \Adf \longrightarrow \End(\H).$$
Let $\Lmmf = \VI \bigoplus \VII$ be a Lagrangian decomposition of
$\Lmmf$. In our case $\Lmmf$ is two dimensional, therefore $\VI$
and $\VII$ are linearly independent lines in $\Lmmf$. Take $\H :=
\FSVI$ to be the vector space of complex valued functions on
$\VI$. For an element $ v \in \Lmmf$ define:
\begin{equation} \label{frep}
\Pif(v) = \psi(\half \omega(v_{1},v_{2}))  \bL _{v_{1}} \bM_{v_2}
\end{equation}
where $ v = v_{1} + v_{2}$ is a direct decomposition $v_{1} \in
\VI, \; v_{2} \in \VII $,  $\bL_{v_{1}}$ is the translation
operator defined by $v_{1}$:
$$ \bL_{v_{1}}(f)(x) = f(x+v_{1}), \rev  f \in \FSVI $$
and $\bM_{v_2}$ 
is a notation for the operator of multiplication by the function $
\bM_{v_2}(x) = \psi(\omega(v_{2}, x))$. Next, we show that the
formulas given in (\ref{frep}) satisfy the relations
(\ref{relations}) and thus constitute a representation of the
algebra $\Adf$. Let $u , v \in \Lmmf$. We have to show:
\begin{equation*}     
\Pif(u+v) = \psi(\half \omega(u,v)) \Pif(u) \Pif(v).
\end{equation*}
Compute:
\begin{equation*} 
  \Pif(u+v) = \Pif( (u_{1} + u_{2}) + (v_{1} + v_{2}) )
\end{equation*}
where $u =u_{1} + u_{2}$ and $v = v_{1} + v_{2}$ are
decompositions of $u$ and $v$ correspondingly.
\\
\\
Then:
\begin{equation} \label{eq02_2}
   \Pif( (u_{1} + v_{1}) + (u_{2} + v_{2})) = \psi(\half \omega(u_{1} + v_{1}, u_{2} + v_{2}))
   \bL_{u_{1} + v_{1}} \bM_{u_2+v_2}. 
\end{equation}
This is by definition of $\Pif$ (cf. (\ref{frep})). Now use the
following formulas:
\begin{eqnarray*} 
\bL_{u_{1} + v_{1}} & = & \bL_{u_{1}} \bL_{v_{1}}, \\
\bL_{v_{1}} \bM_{u_2} & = & \bM_{u_2}(v_1) \bL_{v_{1}} 
\end{eqnarray*}
to obtain that the right-hand side of (\ref{eq02_2}) is equal to:
\begin{equation*} 
 \psi(\half \omega(u_{1} + v_{1}, u_{2} + v_{2}) + \omega(u_{2} , v_{1}))
 \bL_{u_{1}} \bM_{u_2} \bL_{v_{1}} \bM_{v_2}. 
\end{equation*}
Now use:
\begin{equation*}  
   \half  \omega(u_{1} + v_{1}, u_{2} + v_{2}) + \omega(u_{2} , v_{1}) = \half \omega(u_{1} + u_{2}, v_{1} + v_{2}) + \half \omega(u_{1},u_{2}) + \half \omega(v_{1},v_{2})
\end{equation*}
to obtain the result:
\begin{equation*} 
  \psi(\half\omega(u, v))  \Pif(u) \Pif(v)
\end{equation*}
which completes the argument.\\\\
As a consequence of constructing $\Pif$ we automatically proved
that $\Adf$ is non-trivial. It is well known that all linear
operators on $\FSVI$ are linear combinations of translation
operators and multiplication by characters, therefore $\Pif: \Adf
\longrightarrow \End(\H)$ is surjective, but $\dim(\Adf) \leq p^2$
therefore $\Pif$ is a bijection. This means that $\Adf$ is
isomorphic to a matrix algebra
$\Adf \simeq M_{p}(\C)$.\\\\
\textbf{Step 4.} Completing the proof of existence. The group
$\Gf$ acts on $\Adf$ therefore it acts on the category of its
representations. But $\Adf$ is isomorphic to a matrix algebra,
therefore it has unique irreducible representation, up to
isomorphism. This is the standard representation being of
dimension $p$. But $\dim(\H) = p$, therefore $\Pif$ is an
irreducible representation and  its isomorphism class is fixed by
$\Gf$ meaning that we have a pair:
\begin{eqnarray*}
\Pif : \Adf & \longrightarrow & \End(\H),
\\
\rhof :\Gf  & \longrightarrow & \PGL(\H)
\end{eqnarray*}
satisfying the Egorov identity:
$$\rhof(B) \Pif(v) \rhof(B^{-1}) = \Pif(Bv)$$
where $B \in \Gf$ and $v \in \Adf$.\\\\
It is a well known general fact (attributed to I. Schur) that the
group $\Gf$, where $p$ is an odd prime, has no non-trivial
projective representations. This means that $\rhof$ can be
linearized\footnote{see Chapter \ref{metaplectique} for an
independent proof based on {\it ``The method of canonical Hilbert
space''}.} to give:
$$ \rhof : \Gf \longrightarrow \GL(\H).$$
Now take:
\begin{eqnarray*}
\Hh & := & \H,
\\
\Pih & := & \Pif \circ q,
\\
\rhoh &  := & \rhof \circ p.
\end{eqnarray*}
Because $q$ intertwines the actions of $\G$ and $\Gf$ (cf. diagram
(\ref{respect})) we see that $\Pih$ and $\rhoh$ are compatible,
namely the Egorov identity is satisfied:
$$\rhoh(B) \Pih(f) \rhoh(B^{-1}) = \Pih(f^B)$$
where $B \in \Gf$ and $f \in \Ad$. Here the notation $\Pih(f^B)$
means to apply any preimage $\overline{B} \in \G$ of $B \in \Gf$
on $f$. In particular, this implies that the isomorphism class of
$\Pih$ is fixed by $\G$. Knowing that such representation $\Pih$
is unique up to an isomorphism, (Theorem \ref {gh}), our desired
object has been obtained. $\EProof$
%
%
%
%
%
\section{Proof of Lemma \ref{factorization}}\label{proofFACT}
{\it Basic set-up}:  let $ (\Pih,\Hh)$ be a representation of
$\Ad$, which is a representative of the unique irreducible class
which is fixed by $\G$ (cf. Theorem \ref{gh}). Let $\rhoh : \Gf
\longrightarrow \GL(\Hh)$ be the associated honest representation
of the quotient group $\Gf$ (see Theorem \ref{GH2} and Proof
\ref{proofGH}). Recall the notation $\Y = \Gf \times \Lm$. We
consider the function $F: \Y \longrightarrow \C$ defined by the
following formula:
\begin{equation} \label{func1}
F(B,\xi) = \Tr(\rhoh(B) \Pih(\xi))
\end{equation}
where $\xi \in \Lm$ and $B \in \Gf$. We want to show that $F$
factors through the quotient set $\YY = \Gf \times \V$:
\[
\qtriangle[\Y`\YY`\C;`F`F]
\]
The proof is immediate, taking into account the construction given
in section \ref{proofGH}. Let $\Pif$ be the unique (up to
isomorphism) representation of the quotient algebra $\Adf$. As was
stated in \ref{proofGH}, $\Pih$ is isomorphic to $\Pif \circ q$,
where $q:\Ad \longrightarrow \Adf$ is the quotient homomorphism
between the algebras. This means that $\Pih(\xi) = \Pif(q(\xi))$
depends only on the image $ q(\xi) \in \Lmmf$, and formula
(\ref{func1}) solves the problem. $\EProof$
%
%
%
%
\section{Proof of the Geometrization Theorem}\label{proofdeligne}
{\it Basic set-up:} in this section we use the notations of
section \ref{proofGH} and Chapter \ref{metaplectique}. Set $\YY :=
{\Sp} \times \V$ and let $\alpha : \Sp \times \YY \lto \YY$
denotes the associated action map. Let $F:\YY \lto \C$ be the
function appearing in the statement of Theorem \ref{deligne},
i.e., $F(B,v) := \Tr(\rhof(B) \Pif(v) )$, where $B \in \Sp$ and $v
\in \V$. We use the notations $\AV$, $\ASp$ and $\AYY$ to denote
the corresponding algebraic varieties. For the convenience of the
reader we repeat here the
formulation of the statement:\\\\
%
%
\textbf{Theorem \ref{deligne} (Geometrization Theorem).} There
exists an object $\SF \in \Db(\AYY) $ satisfying the following
properties:
%
%
\begin{enumerate}
\item \label{prop_d1} (Function) $f^\SF = F$.
\item \label{prop_d2} (Weight) $w(\SF) \leq 0$.
\item \label{prop_d3} (Equivariance)  For every element $S \in
  \ASp$ there exists an isomorphism $\alpha_S ^* \SF \simeq \SF$.
\item \label{prop_d4} (Formula) On introducing coordinates $\AV \simeq \bA^2$
and identifying $\ASp \simeq \AGG$, there exists an isomorphism
$$\SF_{|_{\AT \times \AV}} \simeq \SL_{\psi(\half
\lambda\mu\frac{a+1}{a-1})} \otimes \SL_{\lgn(a)}.$$
Here $\AT = \{\left(\begin{smallmatrix} a & 0 \\
0 & a^{-1} \end{smallmatrix} \right)\}$ stands for the standard
torus and  $(\lambda,\mu)$ are the coordinates on $\AV$.
\end{enumerate}
\textbf{Construction of the sheaf $\SF$}. We use the notations of
Chapter \ref{metaplectique}. Let $\heiz$ be the Heisenberg group.
As a set we have $\heiz= \V \times \Fq$. The group structure is
given by the multiplication rule $(v,\lambda) \cdot (v',\lambda')
:= (v+v',\lambda+\lambda' + \half \ome(v,v')) $. We fix a section
$ s: \V \dashrightarrow \heiz $, $s(v) =(v,0)$. The group of
linear symplectomorphisms $\Sp$ acts on $\heiz$ by the formula $ g
\cdot (v,\lambda) := (g v , \lambda) $. Let $\semi = \Sp \ltimes
\heiz$. We have a map $\YY \lto \semi$ defined using the section
$s$. We use the notations $\Aheiz$ and $\Asemi$ to denote the
corresponding algebraic varieties.
\\
\\
Let $\SK$ be the Weil representation sheaf (see Theorem
\ref{main_thm}). \textit{Define}:
%
%
$$ \SF := \Tr(\SK_{|_\AYY})$$
where $\Tr$ is the {\it trace functor} (\ref{trf}).
\\
\\
We prove that $\SF$ satisfies properties \ref{prop_d1} -
\ref{prop_d4}.
\\
\\
%
%
\textbf{Property \ref{prop_d1}}. The collection of operators $ \{
\Pif(v) \}_{v \in \V}$ extends to a representation of the group
$\heiz$. The representation $\rhof \ltimes \Pif$ of the group
$\semi$ is isomorphic to the analogue representation $\rho \ltimes
\pi$ constructed in Chapter \ref{metaplectique}. Hence:
\begin{equation*}
f^{\SF} = f^{\Tr(\SK_{|_\AYY})} = \Tr(f^{\SK_{|_\AYY}}) =
\Tr(K_{|_\AYY}) =: F.
\end{equation*}
This proves Property 1.
\\
\\
%
%
\textbf{Property \ref{prop_d2}}. The sheaf $\SK$ is of pure weight
0 (Corollary \ref{propofSK}). The restriction to the subvariety
$\AYY$ does not increase weight. The operation of $\Tr$ also does
not increase weight (using Theorem \ref{deligne2}). It is defined
as a composition:
\begin{equation}\label{trf}  
\Tr := {pr_1}_! \Delta_{12}^{{1}^*}
\end{equation}
where $\Delta_{12}^{1} : \bA^1 \longrightarrow \bA^2$ is the
diagonal morphism $\Delta_{12}^{1}(x) = (x,x)$ and  $pr_1$ is
projection on the $\AYY$ coordinate:
%
%
\begin{equation*}
\begin{CD}
   \AYY \times \bA^2   \\
       @A\Delta_{12}^{1}AA           \\
   \AYY  \times \bA^1       \\
       @Vpr_{1}VV          \\
   \AYY
\end{CD}
\end{equation*}
Putting everything together we obtained Property \ref{prop_d2}.
\\
\\
%
%
\textbf{Property \ref{prop_d3}}. Basically follows from the
convolution property of the sheaf $\SK$ (see Theorem
\ref{main_thm} property 2). More precisely, using the convolution
property we obtain the following isomorphism:
$$ \SK_{|_S} * \SK * \SK_{|_{S^{-1}}} \simeq L_S^* R_{S^{-1}}^* \SK $$
where $L_S$, $R_{S^{-1}}$ denotes left multiplication by $S$ and
right multiplication by $S^{-1}$ on the group $\Asemi$
respectively. We write:
$$ \alpha_S^* \SF \simeq \Tr ( \alpha_S ^* \SK_{|_\AYY} ) \simeq \Tr(L_S
^* R_{S^{-1}}^* \SK_{|_\AYY}) \simeq \Tr ( \SK_{|_S} *
\SK_{|_\AYY} * \SK_{|_{S^{-1}}} ).$$
Now use the following facts:
\begin{eqnarray*}
\Tr( \SK_{|_S} * \SK_{|_\AYY} * \SK_{|_{S^{-1}}} ) & \simeq & \Tr
( \SK_{|_{S^{-1}}} * \SK_{|_S} * \SK_{|_\AYY} )
\end{eqnarray*}
and:
\begin{eqnarray*}
\SK_{|_{S^{-1}}} * \SK_{|_S} & \simeq & \SI.
\end{eqnarray*}
The first isomorphism is the basic tracial property. Its proof is
completely formal and we omit it. The second isomorphism is a
consequence of the convolution property of $\SK$. This completes
the proof of Property \ref{prop_d3}.
\\
\\
\textbf{Property \ref{prop_d4}}. It is directly verified using the
explicit formulas appearing in \ref{realization} which are used to
construct the sheaf $\SK$. More-precisely, we need to compute the
formula for the Trace of the Weil representation sheaf restricted
to $\AT \smallsetminus \{\mathrm{I}\} \times \bA^2 \times 0$. We
have:
$$
\SF(a,\lambda,\mu,0) := \Tr(\SK(a,\lambda,\mu,0)) =
\SL_{\psi(\half\lambda\mu
\begin{smallmatrix} \frac{a+1}{a-1}
\end{smallmatrix})}
\otimes \SL_{\lgn(a)}, \rev a\neq 1.
$$
Here $\Slegendre$ is the Legendre character sheaf on $\Gm$.
\\
\\
This completes the proof of the Geometrization Theorem.$\r\EProof$
\\
\\
\section{Computations for the Vanishing Lemma}
\label{compvanishing} In the computations we use some finer
technical tools from the theory of $\ell$-adic cohomology. The
interested reader can find a systematic study of this material in
\cite{K,KW}.
\\
\\
We identify the standard torus $\AT \subset \AGG$ with the group
$\Gm$. Fix a non-trivial character sheaf\footnote{Namely, a {\it
1-dimensional local system} on $\Gm$ that satisfies the property
$m^*\Skummer \iso \Skummer\boxtimes \Skummer$, where $m:\Gm\times
\Gm \to \Gm$ is the multiplication morphism.} $\Skummer$ on $\Gm$.
Consider the variety $\AX := \Gm - \{1\}$, the sheaf:
\begin{eqnarray}\label{SE}
\SE := \SL_{\psi( \half \lambda\mu \frac{a+1}{a-1})} \otimes
\Skummer
\end{eqnarray}
on $\AX$ and the canonical projection $pr: \AX \lto pt$. Note that
$\SE$ is a non-trivial 1-dimensional local system on $\AX$. The
proof of the Lemma will be given in several steps:
\\
\\
\textbf{Step 1.} \textbf{Vanishing}. We want to show that
$\coH^i(pr_!\SE) = 0$ for $i=0,2$.
\\
\\
By definition:
$$
\coH^0(pr_!\SE) := \Gamma(\AYY,j_!\SE)
$$
where $j:\AX \hookrightarrow \AYY$ is the imbedding of $\AX$ into
a compact curve $\AYY$. The statement follows since:
$$
\Gamma(\AYY,j_!\SE) = \Hom(\Qlb,j_!\SE)
$$
and it is easy to see that any non-trivial morphism $\Qlb \to
j_!\SE$ should be an isomorphism, hence $\Hom(\Qlb,j_!\SE) = 0$.
\\
\\
For the second cohomology we have:
$$
\coH^2(pr_!\SE) = \coH^{-2}(D pr_!\SE)^* = \coH^{-2}(pr_\ast
D\SE)^* = \Gamma(\AX,D\SE[-2])^*
$$
where $D$ denotes the Verdier duality functor and $[-2]$ means
translation functor. The first equality follows from the
definition of $D$, the second equality is the Poincar\'{e} duality
and the third equality easily follows from the definitions. Again,
since the sheaf $D\SE[-2]$ is a non-trivial 1-dimensional local
system on $\AX$ then:
$$
\Gamma(\AX,D\SE[-2]) = \Hom(\Qlb,D\SE[-2]) = 0.
$$
\textbf{Step 2.} \textbf{Dimension}. We claim that dim
$\mathrm{H}^1(pr_!\SE) = 2$.
\\
\\
The (topological) \textit{Euler characteristic} $\chi(pr_!\SE)$ of
the sheaf $pr_!\SE$ is the integer defined by the formula:
$$
\chi(pr_!\SE) := \sum_i  (-1)^i \dim \; \coH^i(pr_!\SE).
$$
Hence from the vanishing of cohomologies (Step 1) we deduce:
\\
\\
\textbf{Substep 2.1.} It is enough to show that $\chi(pr_!\SE) =
-2$.
\\
\\
The actual computation of the Euler characteristic $\chi(pr_!\SE)$
is done using the \textit{Ogg-Shafarevich-Grothendieck formula}
\cite{D3}:
\begin{eqnarray}\label{OSG}
\chi(pr_!\SE) - \chi(pr_!\Qlb) = \sum_{y \in \AYY\setminus\AX}
\Swan_y(\SE)
\end{eqnarray}
Where $\Qlb$ denotes the constant sheaf on $\AX$, and $\AYY$ is
some compact curve containing $\AX$. In words, this formula
expresses the difference of $\chi(pr_!\SE)$ from $\chi(pr_!\Qlb)$
as a sum of local contributions.
\\
\\
Next, we take $\AYY := \P1$. Having that $\chi(pr_!\Qlb) = -1$ and
using formula (\ref{OSG}) we get:
\\
\\
\textbf{Substep 2.2.} It is enough to show that: $\Swan_0(\SE) +
\Swan_1(\SE) + \Swan_\infty(\SE) = -1.$
\\
\\
Now, using the explicit formula (\ref{SE}) of the sheaf $\SE$ we
see that:
\begin{eqnarray*}
\Swan_1(\SE) & = & \Swan_\infty (\Sartin),\\
\Swan_\infty (\SE) & = & \Swan_\infty (\Skummer), \\
\Swan_0(\SE) & = & \Swan_0(\Skummer).
\end{eqnarray*}
Applying the Ogg-Shafarevich-Grothendieck formula to the
Artin-Schreier sheaf $\Sartin$ on $\bA^1$ and the projection
$pr:\bA^1 \lto pt$ we find that:
\begin{eqnarray}\label{SwanArtin}
\Swan_\infty (\Sartin) = \chi(pr_!\Sartin) - \chi(pr_!\Qlb) = 0 -
1 = -1.
\end{eqnarray}
Finally we apply the formula (\ref{OSG}) to the sheaf $\Skummer$
on $\Gm$ and the projection $pr:\Gm \lto pt$ and conclude:
\begin{eqnarray}\label{SwanKummer}
\Swan_\infty(\Skummer) + \Swan_0(\Skummer) = \chi(pr_!\Skummer) -
\chi(pr_!\Qlb) = 0 - 0 = 0.
\end{eqnarray}
In (\ref{SwanArtin}) and (\ref{SwanKummer}) we use the fact that
$pr_!\Sartin$ and $pr_!\Skummer$ are the $0-$objects in $\Db(pt)$.
\\
\\
This completes the computations of the Vanishing Lemma. $\EProof$
\end{chapter}

\include{Future}
\backmatter


\addcontentsline{toc}{chapter}{Bibliography}
\begin{thebibliography}{99}
{\small {
%
\bibitem[A]{A} Auslander L., Kostant B. Quantization and representations of
solvable Lie groups. {\it Bull. Amer. Math. Soc. 73} (1967)
692-695.
%
\bibitem[B]{B} Bernstein J., {\it private communication, Max-Planck
  Institute, Bonn, Germany} (August, 2004).
%
\bibitem[BBD]{BBD} Beilinson A. A., Bernstein J. and  Deligne
  P. Faisceaux pervers. {\it Analysis and topology on singular spaces,
  I, Asterisque, 100, Soc. Math. France, Paris}
  (1982), 5--171.
%
\bibitem[BDB]{BDB} Bouzouina A., De Bievre S., Equipartition of the eigenfunctions of quantized ergodic maps on the torus.
{\it Commun.Math.Phys. 178} (1996), 83-105.
%
\bibitem[BL]{BL} Bernstein J. and  Lunts V., Equivariant sheaves and
functors. {\it Lecture Notes in Mathematics, 1578.
Springer-Verlag, Berlin} (1994).
%
\bibitem[CM]{CM} Coxeter H. S. M. and Moser, W. O. J.,
Generators and relations for discrete groups. {\it Ergebnisse der
Mathematik und ihrer Grenzgebiete, 14. Springer-Verlag, Berlin-New
York} (1980).
%
\bibitem[D]{D} Degli Esposti M., Quantization of the orientation preserving
automorphisms of the torus. {\it Ann. Inst. Henri Poincare 58}
(1993), 323-341.
%
\bibitem[D1]{D1} Deligne P., Metaplectique. {\it A letter to Kazhdan} (1982).
%
\bibitem[D2]{D2} Deligne P., La conjecture de Weil II. {\it Publ. Math. I.H.E.S 52}  (1981),  313-428.
%
\bibitem[D3]{D3} Deligne P., Raport sur la Formulae des traces.
{\it Cohomologie \'{e}tale, S\'{e}minaire de G\'{e}om\'{e}trie
Alg\'{e}brique du Bois-Marie SGA 4${1\over 2}$, Springer Lecture
Notes in Mathematics 569} (1977), 76-109.
%
\bibitem[D4]{D4} Deligne P., Note on quantization. {\it Quantum fields
   and strings: a course for mathematicians, Vol. 1, 2 (Princeton, NJ,
   1996/1997),  Amer. Math. Soc. Providence, RI}  (1999)  367--375.
%
\bibitem[DGI]{DGI} Degli Esposti  M., Graffi S. and Isola S. Classical
limit of the quantized hyperbolic toral automorphisms. {\it Comm.
Math. Phys. 167} (1995), no. 3, 471--507.
%
\bibitem[G]{G} Grothendieck A., Formule de Lefschetz et rationalit\'{e} des
fonctions $L$. {\it Seminaire Bourbaki, Vol. 9, Exp. No. 279}
(1964).
%
\bibitem[Ga]{Ga} Gaitsgory D., Informal introduction to geometric Langlands.
{\it An introduction to the Langlands program, Jerusalem 2001 ,
Birkhauser, Boston, MA} (2003) 269-281.
%
\bibitem[Ge]{Ge} G\'{e}rardin P., Weil representations associated to finite
fields. {\it J. Algebra 46} (1977), no. 1, 54-101.
%
\bibitem[GH1]{GH1} Gurevich S.  and  Hadani R., On Hannay-Berry
  Equivariant Quantization of the Torus. {\it arXiv:math-ph/0312039}.
%
\bibitem[GH2]{GH2} Gurevich S.  and  Hadani R., On Hannay-Berry Equivariant Quantization
of Higher-dimesional Tori. {\it arXiv:math-ph/0403036}.
%
\bibitem[GH3]{GH3} Gurevich S.  and  Hadani R., Proof of the Kurlberg-Rudnick Rate
Conjecture. {\it arXiv:math-ph/0404074}.

\bibitem[GH4]{GH4} Gurevich S.  and  Hadani R., The Higher-Dimensional Kurlberg-Rudnick Conjecture.
{\it In preparation}.
%
\bibitem[HB]{HB} Hannay J.H. and Berry M.V., Quantization of linear maps on the
torus - Fresnel diffraction by a periodic grating. {\it Physica
D1} (1980), 267-291.
%
\bibitem[K]{K} Katz N.M., Gauss sums, Kloosterman sums and monodromy
groups. {\it Annals of Mathematics Studies, 116. Princeton
University Press, Princeton, NJ} (1988).
%
\bibitem[Ku]{Ku} Kurlberg P., {\it private communication, Chalmers
  University, Gothenburg, Sweden} (September, 2003).
%
\bibitem[Ka]{Ka} Kazhdan D., {\it private communication, Hebrew
  University, Jerusalem, Israel} (March, 2004).
%
\bibitem[Ki]{Ki} Kirillov A. A., Lectures on the orbit method.
{\it Graduate Studies in Mathematics, 64. American Mathematical
Society, Providence,RI} (2004).
%
\bibitem[Kl]{Kl} Kloosterman H. D., "The Behavior of General Theta Functions
under the Modular Group and the Characters of Binary Modular
Congruence Groups, I." {\it Ann. Math. 47} (1946), 317-375.
%
\bibitem[KL]{KL} Katz  N.M. and  Laumon G., Transformation de Fouri\'{e}r et
majoration de sommes exponenti\`{e}lles. {\it Inst. Hautes Etudes
Sci. Publ. Math. No. 62} (1985), 361-418.
%
\bibitem[Ko]{Ko} Kostant B., Quantization and unitary representations.
{\it Lecture Notes in Math., Vol. 170 Springer, Berlin} (1970)
87-208.
%
\bibitem[KR1]{KR1} Kurlberg P. and Rudnick Z., Hecke theory and
equidistribution for the quantization of linear maps of the torus.
{\it Duke Math. Jour. 103} (2000), 47-78.
%
\bibitem[KR2]{KR2} Kurlberg P. and Rudnick Z., On the distribution of
  matrix elements for the quantum cat map. {\it Annals of
  Mathematics}, to appear.
%
\bibitem[KW]{KW} Kiehl R. and  Weissauer R., Weil conjectures, perverse
sheaves and $\ell$-adic Fourier transform. {\it A Series of Modern
Surveys in Mathematics, 42. Springer-Verlag, Berlin} (2001).
%
\bibitem[LV]{LV} Lion G. and Vergne M., The Weil representation, Maslov
index and theta series. {\it Progress in Mathematics, 6.
Birkh\"{a}user, Boston, Mass.} (1980).
%
\bibitem[M]{M} Milne J.S., Etale cohomology. {\it Princeton Mathematical Series, 33. Princeton University Press, Princeton, N.J.} (1980).
%
\bibitem[Me]{Me} F. Mezzadri, On the multiplicativity of quantum cat maps. {\it Nonlinearity 15} (2002), 905-922.
%
\bibitem[MR]{MR} McConnell J.C. and Robson J.C., Non-commutative Noetherian Rings. {\it Graduate studies in
mathematics, 30. Amer. Math. Soc. Providence, RI} (2001), 343-344.
%
\bibitem[R1]{R1} Rudnick Z., The quantized cat map and quantum
ergodicity. {\it Lecture at the MSRI conference "Random Matrices
and their Applications", Berkeley,  June 7-11, 1999}.
%
\bibitem[R2]{R2} Rudnick Z., On quantum unique ergodicity for linear maps of
the torus. {\it European Congress of Mathematics, Vol. II
(Barcelona, 2000), Progr. Math., 202, Birkh\"{a}user, Basel}
(2001), 429-437.
%
\bibitem[Ri]{Ri} Rieffel M.A., Non-commutative tori---a case study of
non-commutative differentiable manifolds, {\it Contemporary Math.
105} (1990), 191-211.
%
\bibitem[S1]{S1} Sarnak P., {\it private communication, Cambridge
  University, England} (June, 2004).
%
\bibitem[S2]{S2} Sarnak P., Arithmetic quantum chaos. {\it The Schur lectures, 1992, Tel Aviv. Israel Math. Conf. Proc., 8, Bar-Ilan Univ.,
Ramat Gan} (1995) 183-236.
%
\bibitem[So]{So} Souriau J.M., Structure des systemes dynamiques. {\it Dunod, Paris}
(1970).
%
\bibitem[V]{V} Vogan D.A., The method of coadjoint orbits for real
reductive groups. Representation theory of Lie groups (Park City,
UT, 1998) {\it Amer. Math. Soc., Providence, RI} (2000), 179--238.
%
\bibitem[W1]{W1} Weil A., Sur les courbes alg\'{e}briques et les
  vari\'{e}t\'{e}s qui s'en d\'{e}duisent, {\it Hermann et Cie., Paris} (1948).
%
\bibitem[W2]{W2} Weil A., Sur certains groupes d'operateurs unitaires. {\it Acta Math. 111} (1964) 143-211.
}}
\end{thebibliography}
\end{document}